\documentclass[aps,prd,preprint,groupedaddress,nobibnotes,nofootinbib,byrevtex]{revtex4-1}

\usepackage[dvips]{graphicx}
\usepackage{amsmath,amssymb}
\usepackage{mathrsfs}

\begin{document}
\title{Finite-size corrections to Fermi's golden rule: I. Decay rates}
\author{{Kenzo Ishikawa} and {Yutaka Tobita}}
\affiliation{{Department of Physics, Faculty of Science, Hokkaido University, Sapporo 060-0810, Japan}
}
\begin{abstract}
 A quantum mechanical wave of a finite size moves like a classical particle and shows a unique decay probability.
 Because the wave function evolves according to the Schr\"{o}dinger equation,
 it preserves the total energy but not the kinetic energy in the intermediate-time region of a decay process
where those of the parent and daughters overlap. The decay rate computed with Fermi's golden rule requires
corrections that vary with the distance between the initial and final states, and the energy distribution of the 
daughter is distorted from that of plane waves. The corrections have universal properties in relativistically
 invariant systems and reveal macroscopic quantum phenomena for light particles. The implications for precision
experiments in beta decays and various radiative transitions are presented.
\end{abstract}
\preprint{EPHOU-13-002}
\maketitle

\section{Introduction: wave zone vs particle zone}
The wave length of a particle of momentum $\vec{p}$ is given by the Planck constant $h$ as $\hbar/|\vec{p}|$, 
where $\hbar = h/2\pi$ and is of microscopic size.
The momentum eigenstate is a plane wave of uniform density and many free waves of a constant kinetic energy are also
uniform in space and are like free particles. A system of many waves of varying kinetic energy shows non-uniform 
behavior called diffraction. A diffraction pattern normally has a spatial scale comparable to that of the wave
length, but it can become much longer in a system of a space-time symmetry. Diffraction of this kind which depends
on space-time position in many-body scatterings, is studied.

The diffraction gives corrections to transition probabilities computed by Fermi's golden rule. These corrections
are connected with calibrations of detectors and might be known partly to experimentalists. Even so, it is important
and useful to many physicists to clarify them.

In the diffraction of light, electrons or other particles, the potential energy transforms an incoming wave to a sum
of waves of different kinetic energies. Now, a many-body interaction transforms a many-body state to a sum of the 
same kinetic energy, and the waves behave like free particles and do not show diffraction at the asymptotic region,
$t = \infty$. In the non-asymptotic region of a finite $t$, however, the kinetic energy is not constant and takes 
broad values. So the state reveals the diffraction. Since this diffraction is caused by a many-body interaction,
the pattern has universal properties and appears even in vacuum. Furthermore, the diffraction gives peculiar 
corrections to decay rates that depend on the time interval between those of the initial and final states,
which we call a finite-size correction.

\begin{figure}[t]
 \centering{\includegraphics[scale=.3]{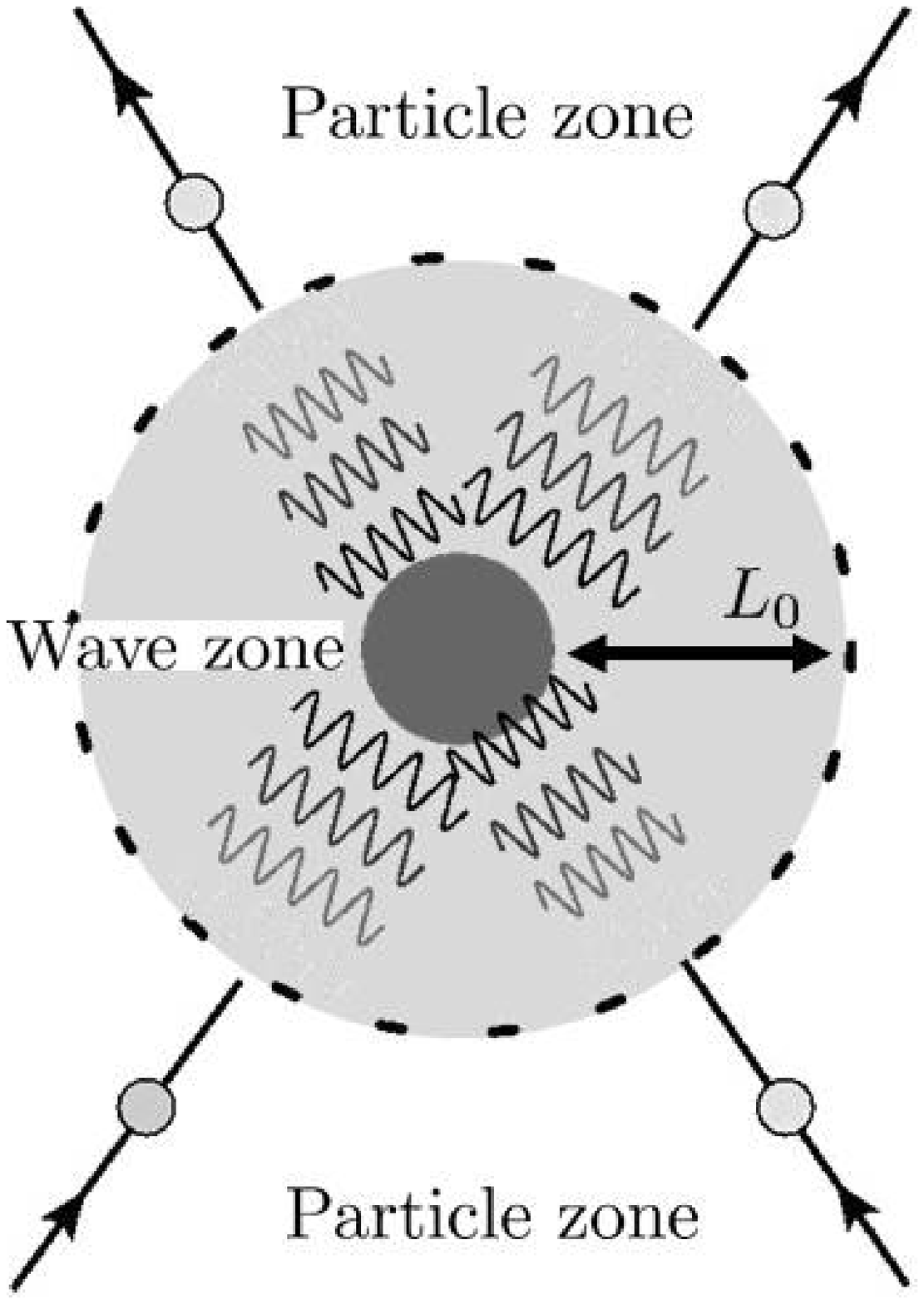}
\caption{In two-body scattering, incoming and outgoing particles in the particle zone behave like classical particles
with constant total kinetic energy, but they behave like waves with non-constant kinetic energy in the wave zone.
The boundary $L_0$ is the coherence length and is normally a microscopic length, but becomes macroscopic in certain
situations, discussed in the present paper.}
\label{fig1}
}
\end{figure}

Scattering processes are defined with initial states prepared at $t = -\infty$ and the final states measured at 
$t=\infty$, where they do not interact with others and have no interaction energy. The initial and final states
 have constant kinetic energy and reveal the particle's nature. Amplitudes and probabilities in the asymptotic
 region have been well studied \cite{1,2,3,4,5}. Near the scattering center, the states overlap and have finite interaction
energy. Thus they retain their wave natures. We call the former region the particle zone, and the latter region 
the wave zone, and the length of the boundary the coherence length. Figure \ref{fig1} shows these for two-body
 scattering. In the particle zone, even at finite $t$, the states behave like particles. In the wave zone, however, 
the state reveals the wave phenomenon that depends on the position and cannot be described with only the 
momentum-dependent distribution function \cite{6}. The coherence length has been considered microscopic in size, 
of the order of de Broglie wave length, which may be true for most cases. Then the phenomena in the wave zone may
be irrelevant to physics and thus unimportant. However, there has been no serious investigation on this length. We 
study problems connected with the wave zone and find that a new length $E\hbar/m^2$, where $m$ and $E$ are the
observed particle's mass and energy, appears for the coherence length and becomes much longer than the de Broglie
wave length in relativistically invariant systems. A space-time-dependent phase of a relativistic wave packet
$(E\left(\vec{p}\,\right)t - \vec{p}\cdot\vec{x})/\hbar $ becomes 
$(E\left(\vec{p}\,\right) - \vec{p}\cdot\vec{v})t/\hbar = m^2t/(\hbar E)$ of the angular velocity,
$m^2/(E\hbar)$ at a position moving with the velocity, $\vec{v}=\vec{p}/E(\vec{p}\,)$, as $\vec{x} =\vec{v}t$. The
angular velocity becomes small for a light particle or at high energy and its inverse gives a new scale of length.
The length even becomes macroscopic for an extremely light particle such as a neutrino. Then the wave zone has a
macroscopic size, and physical phenomena unique to quantum mechanical waves occur in the macroscopic region.
They are natural consequences of the Schr\"{o}dinger equations. Apart from the neutrino, the physics in this region
has not been studied, and is the subject of the present work.

Ordinarily, scattering amplitude is defined in the particle zone and is rigorously formulated with wave packets
 \cite{1,2}; in practical situations, they are approximated well withe plane waves. For scattering processes at a 
finite-time interval, $T$, in the wave zone, the probabilities of detecting particles vary with $T$ and deviate
 from those of an infinite-time interval. We call the deviations finite-size corrections and we study them in 
various processes involving light particles in this paper.

The finite-size corrections of the scattering amplitude and probability have been considered irrelevant to
experiments in high-energy regions. Plane waves with a damping factor $e^{-\epsilon |t|}$ with a positive
and infinitesimal $\epsilon$ in an interaction Hamiltonian often employed for practical calculations
are invariant under translations and are extended in space. This method is powerful for computing the 
asymptotic values but does not supply the finite-size corrections. Because the amplitude in the wave zone
is sensitive to the boundary conditions of the initial and final states, it is dependent on the distance
between them. Hence, the probability has a finite-size correction that has an origin in the boundary
conditions. The correction must, therefore, be included for making a comparison of a theory with an experiment.
An amplitude constructed with wave packets implements manifestly the boundary conditions and supplies the 
finite-size correction.

Previous studies of decay processes at finite-time intervals in the particle zone using an interaction Hamiltonian 
of Damping factor $e^{-\epsilon|t|}$ \cite{7,8,9,10} showed that the time dependences of the decay law of unstable particles
 are modified from simple exponential behaviors due to higher-order effects. These analyses and others of computing
 the decay rates are applicable to kinematical regions where the wave functions of the parent and daughters do not
overlap. As was correctly pointed out in Ref. \cite{8}, the standard method cannot be applied in kinematical regions
where they overlap. The states have wave natures, and the decay rate and other physical quantities in this region
have been thought neither meaningful nor computable since then. This is the region, in fact, where the probability 
of detecting the decay product has a large finite-size correction. One of the main subjects of the present work is 
to develop an $S$-matrix theory that satisfies the boundary conditions of the measuring processes and to find
formulas for the physical quantities in this region. One of our results for decay rate $\Gamma(T)$ at the large
distance $L=cT$, $(T<\tau)$ is 
\begin{align}
 \Gamma(T,\sigma) = \Gamma_0 + N\frac{\sigma}{T}\frac{E}{2m^2}F^2(-\tilde{m}^2),\ \tilde{m}^2 = m_\text{parent}^2
- m^2_\text{daughter},\label{eq1}
\end{align}
where $\Gamma_0$ is the asymptotic value, $\tau$, $\sigma$, $E$, $m$, $m_\text{daughter}$, and $m_\text{parent}$
are the mean life-time, wave packet size, energy, and mass of detected particle, and the mass of daughter and parent,
respectively, $N$ is a numerical constant and $F$ is the form factor. The second term on the right-hand side of 
Eq. \eqref{eq1} is inversely proportional to $T$ and vanishes at $T\to \infty$. So this is the finite-size correction.
From its form, the correction becomes significant at small $m$, large $\sigma$, and $E$, and appears in macroscopic
$T$ for light particles such as photons or neutrinos. This shows
\begin{align}
 \lim_{\sigma \to \infty}\left\{\lim_{T\to \infty} \Gamma(T, \sigma)\right\} = \Gamma_0,\label{eq2}\\
\lim_{T \to \infty}\left\{\lim_{\sigma\to \infty} \Gamma(T, \sigma)\right\} = \infty.\label{eq3}
\end{align}
In Eq. \eqref{eq2}, the rate becomes the asymptotic value, whereas in Eq. \eqref{eq3}, the rate diverges.
The energy distribution also reveals unusual properties even at $T\to \infty$, if particles of large and small 
sizes are involved in one process. They should appear in various situations such as an interface between
two phases, and interesting physics is expected. The implications for particle decay are studied.

The transition probability $P$ composed of many processes in the particle zone is factorized to that of each
microscopic process, $P_i$, as
\begin{align}
 P=\prod_{i}P_i.
\end{align}
Now, the probability for transition processes in the wave zone is not factorized due to the finite-size corrections,
but the whole process is described by the product of wave functions of each microscopic process:
\begin{align}
 \Psi = \prod_{i}\Psi_i,\ P \neq \prod_i P_i.
\end{align}
Because the probability of the whole process is not factorized, the Markov nature of the multiple processes is lost.
The non-Markov nature is related to an EPR correlation \cite{11} and may have various implications.

The decay rates are studied in the present paper and the scattering cross sections will be studied in a subsequent
paper. This paper is organized in the following manner. In Sect. 2, a wave function and $S$-matrix at a finite-time
interval are shown to be different from those of the infinite-time interval. Particles described by wave packets
and their interactions caused by a local Hamiltonian are summarized in Sect. 3. Two-body decays are studied in 
Sect. 4, and radiative decays of atoms, nuclei, and particles are studied in Sect. 5. In Sect. 6, we study the decay
processes and thermodynamics of quantum particle. A summary is given in Sect. 7.

\section{A finite-time interval effect}
In a physical system described by a Hamiltonian $H$ composed of a free term $H_0$ and an interaction term
 $H_\text{int}$,
\begin{align}
 H = H_0 + H_\text{int},
\end{align}
the wave function $|\Psi(t)\rangle$ follows the Schr\"{o}dinger equation
\begin{align}
 i\hbar\frac{\partial}{\partial t}|\Psi(t)\rangle = \left(H_0 + H_\text{int}\right)|\Psi(t)\rangle.
\end{align}
In field theory, the free part $H_0$ is a bi-linear field form and the interaction part $H_\text{int}$ is 
a higher field polynomial. $H_\text{int}$ causes a change in the particle number such as a decay of a pion int a
charged lepton and a neutrino.
\subsection{Finite-size correction to Fermi's golden rule}
The transition rate from an eigenstate of $H_0$, $|\alpha\rangle$ of energy $E_\alpha$, to another, $|\beta\rangle$
of energy $E_\beta$, in a wave zone at a finite-time interval $T$, seems to be computed with the amplitude $f$
and probability $P$ \cite{12,13} in the form,
\begin{align}
f &= \int_0^T dt\langle \beta|H_\text{int}(t)|\alpha\rangle = \int_0^Tdt e^{-i(E_\beta - E_\alpha)t}F_{\alpha,\beta},\\
F_{\alpha,\beta} &= \langle\beta|H_\text{int}(t)|\alpha\rangle,\nonumber\\
P&=|F_{\alpha,\beta}|^2D(E_\beta-E_\alpha;T),\\
D(E_\beta-E_\alpha;T) &=\frac{4\sin^2[(E_\beta-E_\alpha)T/2]}{(E_\beta-E_\alpha)^2},\nonumber
\end{align}
where $F_{\alpha,\beta}$ is the matrix element. In particle decay, the final state constitutes two or more particles
of a continuous energy spectrum and th oscillating function $D(E_\beta-E_\alpha;T)$ approximately agrees with Dirac's
delta function at infinite T \cite{14,15,16},
\begin{align}
 D(E_\beta-E_\alpha;T) = 2\pi T\delta(E_\beta - E_\alpha).
\end{align}
Because the integral of a function $F(E_\beta)$ with weight $D(E_\beta- E_\alpha;T)$ over energy $E_\beta$ is 
computed with a variable $x = (E_\beta - E_\alpha)T$ as
\begin{align}
 P&=\int_{E_\alpha - \Delta_E}^{E_\alpha + \Delta_E}dE_\beta D(E_\beta - E_\alpha;T)\nonumber\\
&=T\int_{-\delta_E T}^{\Delta_E T}dx \left(\frac{\sin(x/2)}{x}\right)^2F(x/T),\label{eq11}\\
F(E_\beta) &= |F_{\alpha,\beta}|^2.\nonumber
\end{align}
The symmetric region of the integration was chosen in Eq. \eqref{eq11}. At large $T$, $F(x/T)$ is replaced with
$F(0)$, and Eq. \eqref{eq11} becomes
\begin{align}
 P=TF(0)\int_{-\Delta_E T}^{\Delta_E T}dx \left(\frac{\sin(x/2)}{x}\right)^2 = 2\pi TF(0).
\end{align}
Thus the transition probability integrated over final states is given by
\begin{align}
 P = 2\pi T\int d\beta\delta(E_\alpha - E_\beta)|F_{\alpha,\beta}|^2,\label{eq13}
\end{align}
and the rate $P/T$ is constant This is Fermi's golden rule.

Now, at finite $T$, expanding $F(x/T)$ in a power series of $x/T$
\begin{align}
 F(x/T) = \sum_lC_l\left(\frac{x}{T}\right)^l,\label{eq14}
\end{align}
we have Eq. \eqref{eq11} in the form 
\begin{align}
 P = \sum_l C_lT^{l-1}\int_{-\Delta_E T}^{\Delta_E T}dx\left(\frac{\sin(x/2)}{x}\right)^2x^l.
\end{align}
The integrals over $x$ are easily evaluated. In a small $|x|$ region, the integral vanishes for $l\geq 1$ and is 
consistent for $l=0$. In a large $x$ region, the integrand behaves as $\frac{x^{l-2}}{2}$. So the above integrals 
becomes, as given in Appendix \ref{appa2},
\begin{align}
 2\pi T C_0 + \sum_{l\geq 1} C_l T^{1-l}\int_{-\Delta_E T}^{\Delta_E T}dx \frac{x^{l-2}}{2} &=
2 \pi T C_0 + \sum_{l\geq 1}C_lT^{1-l}\frac{(\Delta_E T)^{l-1}}{l-1}\nonumber\\
 &= 2\pi T C_0\left\{1 + \sum_{l\geq 2}\frac{C_l}{C_0}\frac{\Delta_E^{l-1}}{2\pi T(l-1)}\right\}.\label{eq16}
\end{align}
The $1/T$ correction is in the second term on the right-and side, which is finite if $\Delta_E$ is finite.
The $1/T$ correction depends on $\Delta_E$ and the eigenvalue distribution and converges if $\Delta_E$ is finite.
Appendix \ref{app1} and \ref{app2} study $1/T$ of various distributions. The value at $T\to \infty$ is then defined uniquely.

In relativistically invariant systems, $\Delta_E = \infty$ and the correction for $l\geq 2$ in Eq. \eqref{eq16}
diverges. The infinite correction emerges due to a large overlap of wave functions in the situation where the 
ordinary scattering theory cannot be applied \cite{8}. The probability at a finite time measured with an
apparatus does not diverge. Hence the amplitude defined according to the boundary conditions of the measurement
process should give the finite value. The boundary condition at $T$ is different from that at $T=\infty$, 
hence the amplitude that satisfies the boundary condition at $T$ is different from that of $T=\infty$.
In the present paper, $S[\infty]$ stands for the standard $S$-matrix, and $S[T]$ stands for the $S$-matrix that
satisfies the boundary conditions at $T$. As is seen later, the function introduced for defining $S[t]$ decrease
rapidly with $x/T$ as $e^{-\sigma(x/T)^2}$ on the right-hand side of Eq. \eqref{eq14}, where $\sigma$ is the size
 of the wave functions determined from the boundary condition, and the coefficients converge.
Since the amplitude at large $x/T$ is determined by the boundary condition, the $1/T$ correction becomes a finite
value that depends on the boundary condition. Nevertheless, they follow a universal relation.
It is important to find the universal properties of the finite-size corrections.
 
The states $|\beta\rangle$ satisfying $E_\beta = E_\alpha$ contribute to the decay rate, Eq. \eqref{eq13}, and 
the states $|\beta\rangle$ of $E_\beta \neq E_\alpha$ contribute to the finite-size correction. Since $E_\beta$ 
is continuous, those states of $E_\beta \approx E_\alpha$ are sensitive to boundary conditions and so is the
finite-size correction. For computation of the probabilities of processes measured in experiments, the wave 
functions for the outgoing waves and incoming waves should be localized around their centers, as has been emphasized
 in textbooks of quantum field theory; see, for instance, Refs. \cite{16,17,18,19,20,21}
\footnote{In Refs. \cite{16,17,18,19,20,21}, $S[\infty]$ was studied with large wave packets. In Ref. \cite{22}, the complete
set of wave packets is constructed with those that have centers of position and momentum and is used. $S[T]$ thus constructed
is studied here.}. 
The wave packets 
satisfy this property and are necessary. They can be replaced with the plane waves in $S[T]$ in the particle zone, 
but, in the wave zone, wave packets in $S[T]$ cannot be replaced with plane waves. We compute the finite-size
corrections to transition probabilities with $S[T]$ expressed by wave packets. $S[T]$ is different from $S[\infty]$,
and has unique properties. Finite $1/T$ corrections are found.

\subsection{Wave function at a finite time}
An initial wave function for $S[T]$ starts from a state at $t=0$ and ends at a final state at $t = T$.
The kinetic energy is not a good quantum number in the wave function at finite $T$. A time-dependent solution
of Eq. \eqref{eqa7} 
in the first order of $H_\text{int}$ that satisfies an initial condition
\begin{align}
 |\Psi(0)\rangle = |\psi^{(0)}\rangle,\ H_0|\psi^{0}\rangle = E_0|\psi^{(0)}\rangle
\end{align}
is 
\begin{align}
 |\Psi(t)\rangle &=e^{-i\frac{E_0}{\hbar}t}\left\{|\psi^{(0)}\rangle + \int d\beta d(\omega,t)|\beta\rangle
\langle\beta|H_\text{int}|\psi^{(0)}\rangle\right\},\\
\omega &= E_\beta - E_0,\ H_0|\beta\rangle = E_\beta|\beta\rangle,\nonumber
\end{align}
where 
\begin{align}
 d(\omega,t) = \frac{e^{-i\omega t} - 1}{\omega} = -2i\frac{\sin(\omega t/2)}{\omega}e^{-\frac{i}{2}\omega t}.
\end{align}
At $t\to \infty$, $d(\omega,t)$ becomes
\begin{align}
 d(\omega,t) = -2\pi i \delta(\omega),\label{eq20}
\end{align}
and the wave function
\begin{align}
 |\Psi(t)_\infty\rangle = e^{-i\frac{E_0}{\hbar}t}\left\{
|\psi^{(0)}\rangle - 2\pi i|\beta\rangle\langle\beta|H_\text{int}|\psi^{(0)}\rangle_{E_\beta = E_0}
\right\}
\end{align}
has the kinetic energy $E_\beta = E_0$. At finite $t$, on the other hand, Eq. \eqref{eq20} is not fulfilled
and the wave function is a superposition of the wide spectrum of the kinetic energy $E_\beta$. An average
of $d(\omega,t)$ over a finite-time interval $\delta t$ satisfying $\omega\delta t \gg 1$ is 
\begin{align}
 d(\omega,t) = -\frac{1}{\omega},
\end{align}
and the average of the wave function over the finite interval is
\begin{align}
 |\Psi(t)_\text{average}\rangle = e^{-i\frac{E_0}{\hbar}t}\left\{
|\psi^{(0)}\rangle - \int d\beta \frac{1}{\omega}|\beta\rangle\langle\beta|H_\text{int}|\psi^{(0)}\rangle
\right\}.
\end{align}

In both cases, the state vectors $|\Psi(t)_\infty\rangle$ and $|\Psi(t)_\text{average}\rangle$ have the frequency
 $E_0/\hbar$ and the total energy $E_0$:
\begin{align}
 H|\Psi(t)_\infty\rangle &= E_0|\Psi(t)_\infty\rangle,\\
H|\Psi(t)_\text{average}\rangle &= E_0|\Psi(t)_\text{average}\rangle.
\end{align}

Thus the wave function at a finite time $t$ is a sum of those of the broad energy spectrum of $H_0$,
whereas that is composed of a discrete spectrum $E_\beta = E_0$ at $t=\infty$. The conservation law of energy
defined with $H$ is reduced to the conservation law of the kinetic energy defined by $H_0$ only at $t=\infty$.

\subsection{Scattering operator at a finite-time interval}
Physical quantities are observed through scattering or decay processes and are computed with $S[T]$,
which is defined from unitary operators
\begin{align}
 U(t) = e^{-iHt},\ U_0 = e^{-iH_0t}.
\end{align}
M{\o}ller operators are defined in the form
\begin{align}
 \Omega_{\pm}(T) =\lim_{t\to\mp T/2}U^\dagger(t)U_0(T),
\end{align}
and satisfy
\begin{align}
 e^{iH\epsilon_t}\Omega_{\mp}(T)=\Omega_{\mp}(T \pm \epsilon_t)e^{iH_0t}.
\end{align}
The scattering operator at a finite $T$ is product
\begin{align}
 S[T] = \Omega_{-}^\dagger(T)\Omega_{+}^\dagger(T),
\end{align}
and satisfies
\begin{align}
 S[T]H_0 = H_0S[T]+i\left\{
\frac{\partial}{\partial T}\Omega_-(T)\right\}^\dagger\Omega_+(T) - i\Omega^\dagger_-(T)\frac{\partial}{\partial T}
\Omega_+(T),
\end{align}
and the commutation relation
\begin{align}
 [S[T],H_0] = i\left\{
\frac{\partial}{\partial T}\Omega_-(T)\right\}^\dagger\Omega_+(T)
-i\Omega_-^\dagger(T)\frac{\partial}{\partial T}\Omega_+(T).\label{eq31}
\end{align}
Thus $S[T]$ does not commute with $H_0$, and the conservation law of kinetic energy is violated at a finite $T$.

From Eq. \eqref{eq31}, the matrix element of $S[T]$ between the eigenstates of $H_0$ has energy-conserving and
non-energy-conserving terms,
\begin{align}
 \langle \beta|S[T]|\alpha = \delta_\epsilon(E_\alpha-E_\beta)f(T) + \delta f,\label{eq32}
\end{align}
where the second term, $\delta f$, vanishes at the energy $E_\beta = E_\alpha$. Since the energy $E_\beta$ of the 
first and second terms is different, the total transition probability is the sum of each probability.
The first term gives a normal constant probability that is also computable by ordinary $S$-matrix of plane waves,
whereas the second term gives a $T$-dependent correction that is not computable by the ordinary $S$-matrix.
In ordinary situations, the non-energy-conserving terms are negligible but they are important in the situations
 studied in the present work.

The magnitude of $\delta f$ and the probability derived from $\delta f$ depend on the dynamics of the system.
When $E_\alpha$ and $E_\beta$ are approximate energies of the states $|\alpha\rangle$ and $|\beta\rangle$, 
we have
\begin{align}
& (E_\alpha - E_\beta)\langle\beta|S[T]|\alpha\rangle = \langle\beta|O(T)|\alpha\rangle,\\
&O(T)=i\left\{\frac{\partial}{\partial T}\Omega_-(T)\right\}^\dagger\Omega_+(T)-i\Omega_-^\dagger
\frac{\partial}{\partial T}\Omega_+(T).\nonumber
\end{align}
Hence
\begin{align}
 \delta f = \frac{\langle \beta |O(T)|\alpha\rangle}{E_\alpha - E_\beta},
\end{align}
and the transition probability for the non-energy-conserving states is given in the form
\begin{align}
\sum_{\beta}\left|\delta f\right|^2 
= \sum_{\beta}\left\{\frac{\langle \beta |O(T)|\alpha\rangle}{E_\alpha - E_\beta}\right\}^2\geq 0,\label{eq35}
\end{align}
where the equality is satisfied at $T\to \infty$. States at ultraviolet energy regions couple in a universal
manner with the operator $O(T)$ and contribute to the probability at the finite-time interval. Since states
of unlimited momentum couple in a Lorentz-invariant manner, they give a universal correction to Eq. \eqref{eq32}.
The finite-size correction appears even in the lowest order of perturbative expansions and is useful for 
probing the physical system in the large momentum region.

Boundary conditions necessary to determine a solution for a wave equation uniquely in scattering or decay processes
are asymptotic boundary conditions \cite{1}. For scattering from an initial state $|\alpha\rangle$ to a final state
$|\beta\rangle$ of a scalar field expressed by $\phi(x)$, the states $|\alpha\rangle$ at $t=-T/2$ are constructed
with free waves $\phi_\text{in}(x)$ and the states $|\beta\rangle$ at $t=T/2$ are constructed with free waves 
$\phi_\text{out}(x)$ and satisfy asymptotic boundary conditions:
\begin{align}
 \lim_{t\to -T/2}\langle \alpha|\phi^f(t)|\beta\rangle = \langle \alpha|\phi^f_\text{in}|\beta\rangle,\\
 \lim_{t\to T/2}\langle \alpha|\phi^f(t)|\beta\rangle = \langle \alpha|\phi^f_\text{out}|\beta\rangle,
\end{align}
where field renormalization $Z^\frac{1}{2}=1$ in the tree levels that we study here. The expansion coefficient
$\phi^f(t)$ is defined by
\begin{align}
 \phi^f(t) = i\int d\vec{x}f^*(\vec{x},t)\overleftrightarrow{\partial}_0\phi(\vec{x},t)\label{eq38}.
\end{align} 
$\phi^f_\text{in}$ and $\phi^f_\text{out}$ are defined in the same way. C-number functions $f(\vec{x},t)$ are
normalized and satisfy the free wave equation. The normalized functions decrease fast in space and form a 
complete set with those functions translated in space. Hence they have central values of position and momentum and 
the state vector is specified by both variables as $|\vec{p},\vec{X}\rangle$. Thus, matrix elements of $S[T]$ are
defined as $\langle \vec{p}_i,\vec{X}_i|S[T]|\vec{p}_j,\vec{X}_j\rangle$ and depend on the position and momentum.
The finite-size corrections are computed with the position dependence of the probability. For normalized functions
to form the complete set, those of different center positions are required \cite{22}. Those of the initial 
state represent the beam and those of the final state represent a detected particle. They are determined by the
experimental apparatus and those of the initial and final states are normally different. Being non-normalizable,
plane waves are not suitable for these functions if the damping factor $e^{-\epsilon|t|}$ is not included.
Instead, wave packets are normalizable and are suitable. $\phi_\text{in}(x)$ and $\phi_\text{out}(x)$  satisfy
the free wave equation and the states $|\alpha\rangle$ and $|\beta\rangle$ are defined with wave packets.
The wave packets, which have finite-spatial sizes and decrease fast at large $|\vec{x} - \vec{x}_0$, ensure
the asymptotic conditions at a finite $T$, where $\vec{x}_0$ is the center position. Hence $S[T]$ is described
by wave packets and the finite-size corrections are studied with $S[T]$. We present several examples where the 
finite-size corrections are non-negligible and give interesting observable effects.

\section{Quantum Particles described by wave packets}
Waves of finite sizes expressed by wave packets used for formulating $S[T]$ exist in various areas.
Wave function at the particle zone lose their wave nature quickly and the time evolution of objects turns
out to be described by the classical equation of motion. Thus a classical mechanical description
is smoothly obtained starting from the quantum mechanical description and physics in this region is understood
well by classical physics, as is explained in most textbooks of quantum mechanics. Now, in the wave zone, the 
phase of the wave function remains and gives physical effects that are different from classical physics.
They could appear in macroscopic space-time regions. Then their universal properties are common in any wave 
functions, and can be studied with Gaussian wave packets. Their sizes are determined from physical processes 
of the particles.

The physics of quantum particles has been neither completely explored nor understood and is becoming
relevant to recent advanced science and technology, especially for precision experiments of light particles.
Various phenomena of neutrinos and photons caused by these unique phases are studied hereafter. A neutrino 
interacts extremely weakly with matter and is not disturbed by the environment; hence, its phase is not washed out,
and consequently the neutrino retains the wave nature even in the macroscopic area and reveals large
finite-size corrections \cite{23,24}. The finite-size correction is observed as a diffraction pattern of the 
neutrino produced in pion decay and in other processes that the neutrino gives rise to.

A photon is massless in vacuum and behaves approximately like a particle of small mass in the high-energy region
in dilute matter. Normally, the quantum mechanical phase of single low-energy photon is washed out and a large
number of these photons behave like a classical electromagnetic wave in macroscopic areas. In various exceptional
situations, its phase is not washed out and photons reveal unusual properties and interact with microscopic objects
as single quanta. The photon is then expressed by a wave packet, and the probability of detecting it at a 
finite distance shows the diffraction behavior of a single quantum. This also leads the photon to have unusual 
thermodynamic properties.

Waves of small sizes move like classical particles \cite{25,26,27,28,29,30,31,32,33,34,35,36,37,38}
 and exhibit wave-like behaviors such as 
anomalous finite-size corrections in scattering cross sections or decay rates, and called quantum particles
in the present paper. Quantum particles of relativistic waves have universal properties.

\subsection{Symmetric wave packets}
The Gaussian wave packet of a relativistic particle of mass $m$ and central momentum $\vec{p}_0$, position 
$\vec{X}_0$, and time $T_0$ is expressed in the momentum representation by
\begin{align}
 \langle t,\vec{p}\,|\vec{p}_0,\vec{X}_0,T_0\rangle = N\sigma^\frac{3}{2}e^{iE(\vec{p}\,)(t-T_0)
-i\vec{p}\cdot\vec{X}_0 - \frac{\sigma}{2}(\vec{p}-\vec{p}_0)},
\end{align}
where $\sigma$ is the spatial size of the wave packet, N is the normalization factor, and the energy is given by
a relativistic form, $E(\vec{p}\,) = \sqrt{\vec{p}^{\,2} + m^2}$. This is a super position of the eigenstates of the
energy and momentum of the widths $\frac{|\vec{p}\,|}{E(\vec{p}\,)\sqrt{\sigma}}$ and $\frac{1}{\sqrt{\sigma}}$,
respectively; it is a simple Gaussian form of $\vec{p}$ at $t=T_0$, and retains its shape afterward. The completeness
of wave packets of the continuous position and momentum, and other important properties, are given in Ref. \cite{22}.
Some of them are summarized in the following for completeness of the present paper. They satisfy
\begin{align}
 \int d\vec{X}\frac{d\vec{p}}{(2\pi)^3}|\vec{p},\vec{X},T\rangle\langle\vec{p},\vec{X},T| = 1,\label{eq40}
\end{align}
and the wave function in the coordinate representation is
\begin{align}
 w(\vec{p}_0,x) = \langle t,\vec{x}|\vec{p}_0,\vec{X}_0,T_0\rangle = \int d\vec{k}\langle \vec{x},\vec{k}
\rangle\langle t,\vec{k}|\vec{p}_0,\vec{X}_0,T_0\rangle;
\end{align}
it also becomes a Gaussian form in $\vec{x}$ around a new center,
\begin{align}
 w(\vec{p}_0,x) &= Ne^{-\frac{1}{2\sigma}\left(\vec{x} - \vec{X}_0 - \vec{v}_0(t-T_0)\right)^2}
e^{-E(\vec{p}_0)(t-T_0) + i\vec{p}_0\cdot(\vec{x} - \vec{X}_0)},\label{eq42}\\
\vec{v}_0 &= \left.\frac{\partial}{\partial p_i}E(\vec{p}\,)\right|_{\vec{p}=\vec{p}_0}\nonumber,
\end{align}
in a small $|t-T_0|$ region. Thus the wave function keeps its shape and moves with a velocity $\vec{v}_0$ and
the modulus is invariant under
\begin{align}
 t \to t+\delta t, \ \vec{x} \to \vec{x} + \vec{v}_0\delta t.\label{eq43}
\end{align}
Since the position of the wave packet moves uniformly with the velocity $\vec{v}_0$ and has the extension $\sigma$,
the wave function becomes finite only inside a narrow strip of this width. Hence the quantum state expressed
by this wave packet behaves like a particle of the extension $\sigma$. At large $|t-T_0|$, the function expands.

The wave function Eq. \eqref{eq42} decrease rapidly with $|\vec{x} - \vec{X}_0 - \vec{v}_0(t-T_0)|$ and 
vanish at $|\vec{x} - \vec{X}_0 - \vec{v}_0(t-T_0)| \to \infty$. Hence they satisfy the asymptotic boundary 
conditions and are appropriate to use as the basis, $f(\vec{x},t)$, of Eq. \eqref{eq38}. The transition process
of the particle prepared at the initial time $T_i$ and of observing the final states at a final time $T_f$ of a 
finite $T=T_f - T_i$ is studied with $S$-matrix at the finite-time interval $S[T]$ thus defined.
Because the $S$-matrix of plane waves defined at $T=\infty$, $S[\infty]$, satisfies the boundary condition at 
$t=\pm\infty$, it is different from $S[T]$ defined at $t=\pm T/2$.
$S[T]$ defined by the wave packets Eq. \eqref{eq42}, and the amplitudes and probabilities obtained from them
are not equivalent to those obtained from $S[\infty]$ generally in the wave zone. Then, the computations 
should be made with $S[T]$. Conversely, if they are equivalent, the computations can be made with either method.
The kinetic energy is strictly conserved in both classical collisions of particles under a force of finite
range and quantum collisions described by $S[\infty]$ of the stationary states of the free Hamiltonian,
whereas the conservation law of kinetic energy is slightly modified in a collision of the finite-time
interval $T$ described by $S[T]$ from the algebra Eq. \eqref{eq31}. The total energy is conserved, but is 
different from the kinetic energy in the space-time region where the interaction Hamiltonian has a finite
expectation value. Hence the kinetic energy is not conserved in this region. The non-conservation of 
the kinetic energy is a unique property of quantum particles described by $S[T]$ and causes unusual behaviors
of the collision or decay probabilities.

The quantum states of finite-spatial extensions are expressed by superpositions of plane waves of different
momenta and energies, and their scatterings re those of the non-stationary states. These non-stationary
wave packets are specified by the values of position, momentum, and complex phase at the center. Even though
its spatial size is so small that it behaves like a point particle, the wave nature represented by the 
phase remains. The phase that depends on dynamical variables gives physical effects that are characteristic of
the quantum particles. 

$\sigma$ in a Gaussian wave packet determines the spatial size of the quantum particle, and depends on the situation.
Because the probability of detecting this particle is unity inside the wave packet, this size is the classical size
 of a quantum particle. So, $\sigma$ for the outgoing state is the size of the unit of the detecting system that gives
a signal, and is the size of the nucleus used in the detector for the neutrino. For a high-energy photon, the signal
is taken from its $e^+e^-$ creation around the electric field of the nucleus used in the detector, hence $\sigma$ 
is about size of the nucleus. $\sigma$ for in-state is also the size of the wave function that expresses this particle.
 This size is infinite for an ideal particle in vacuum, but is finite in matter due to the effects of the environment.
When a particle expressed by a certain wave function interacts with others and both make a transition to other
states, this particle is expressed by one wave function in a finite-time interval between these reactions.
Hence that is determined by the mean free time of this particle. Thus $\sigma$ is determined by the mean free
path for incoming waves. The $\sigma$ values for the pion, kaon, muon, proton, photon, and electron in the initial
states are estimated from their mean free paths in the matter of experiments. Actually, most of them have macroscopic
sizes in high-energy regions. An electron easily loses energy by electromagnetic showers and is exceptional. In
low-energy regions, an electron, negative muon, and negative pion form bound states of microscopic sizes with a nucleus
in matter, and the $\sigma$ values have microscopic sizes. Positive-charged particles such as a positive muon and
 positive pion do not form bound states with a nucleus  and may have larger $\sigma$.

Thus $\sigma$ values of nuclear size, atomic size, or larger size appear depending on the situation. In scattering or
decay of waves with different sizes, the wave functions overlap in the finite and asymmetric region. Consequently,
the conservation laws derived from space-time symmetry are modified.

\subsection{Local interaction}
Characteristic features of quantum particles are connected with the phase factor of wave functions and appear in the
lowest order of interactions of scaler fields. Hence we study the scattering of particles caused by the local 
interaction
\begin{align}
 \mathscr{L}_{\text{int}} = g\prod^{j=N}_{j=1}\varphi_j(x),
\end{align}
in the lowest order of $g$ first. The effects of spin and internal structure will be included later. Interactions of 
$N_1$ incoming and $N_2$ outgoing particles expressed by the wave packets parameterized by 
$(\vec{p}_i,\vec{X}_i,T_i;\sigma_i)$ at a space-time position $(t,\vec{x})$ are given in the form \cite{22}
\begin{align}
 \langle k|\prod_{j=1}^{j=N}\varphi_j(x)|l\rangle &= \prod_{k=1}^{N_2}w^{*}_k(x,\vec{p}_k;\vec{X}_k,T_k,\sigma_k)
\times \prod_{l=1}^{N_1}w_l(x,\vec{p}_l;\vec{X}_l,T_l,\sigma_l)\nonumber\\
&=N_t\exp^{-\frac{1}{2\sigma_S}\left(\vec{x}-\vec{x}_0(t)\right)^2 - \frac{1}{2\sigma_t}\left(t-t_0\right)^2}
\exp^{R+i\phi},\label{eq45}\\
N_t&=\prod_{k,l}N_k^* N_l,\nonumber
\end{align}
where $\sigma_S$ and $\sigma_t$ in the exponent display the extents in $\vec{x}$ and $t$, and are expressed in the form
\begin{align}
 \frac{1}{\sigma_S} = \sum_{j}\frac{1}{\sigma_j},\ \frac{1}{\sigma_t} = \sum_{j}\frac{\vec{v}_j^{\,2}}{\sigma_j} - 
\frac{\vec{v}_0^{\,2}}{\sigma_S},\label{eq46}\\
\vec{v}_0 = \sigma_S\sum_j\frac{\vec{v}_j}{\sigma_j},\ \vec{v}_j = \frac{\vec{p}_j}{E_j}.
\end{align}
Here, $\vec{x}_0(t)$ is the center in $\vec{x}$ and moves with $\vec{v}_0$:
\begin{align}
 \vec{x}_0 &= \vec{v}_0t + \vec{x}_0(0),\label{eq48}\\
\vec{x}_0 &=\sigma_S\left\{\sum_j\frac{\tilde{\vec{X}}_j}{\sigma_j} - i\sum_j(\pm)\vec{p}_j\right\},\nonumber\\
t_0 &= \sigma_t\left\{\frac{\vec{v}_0\cdot\vec{x}_0}{\sigma_S}- \sum_j\frac{\vec{v}_j\cdot\tilde{\vec{X}}_j}{\sigma_j}
 + i\sum_j(\pm)E(\vec{p}_j)\right\},\nonumber\\
\tilde{\vec{X}}_j & = \vec{X}_j - \vec{v}_jT_j.\nonumber
\end{align}
In the above equations and hereafter, $(+)$ and $(-)$ are for incoming and outgoing states, respectively.
The real part of the exponent of Eq. \eqref{eq45}, $R$, determines the magnitude and is composed of position-dependent
and momentum-dependent terms. The former, $R_\text{trajectory}$, and the latter, $R_\text{momentum}$, are expressed by
\begin{align}
 R &= R_\text{trajectory} + R_\text{momentum},\label{eq49}\\
 R_\text{trajectory} &= -\sum_j\frac{\tilde{\vec{X}}_j^2}{2\sigma_j} + 2\sigma_S\left(\sum_j\frac{\tilde{\vec{X}}_j}
{2\sigma_j}\right)^2 + 2\sigma_t\left(\sum_j\frac{(\vec{v}_0 - \vec{v}_j)\cdot\tilde{\vec{X}}_j}{2\sigma_j}\right)^2,
\label{eq50}\\
R_\text{momentum} &= -\frac{\sigma_t}{2}\left\{\sum_j(\pm)(E(\vec{p}_j)- \vec{v}_0\cdot\vec{p}_j)\right\}^2
-\frac{\sigma_S}{2}\left(\sum_j(\pm)\vec{p}_j\right)^2.\label{eq51}
\end{align}
From $R_\text{trajectory}$, particles follow classical orbits and from $R_\text{momentum}$, Eq. \eqref{eq51}, they follow
the approximate energy-momentum conservation. Because the interaction system is invariant under a translation of the
coordinate system, $R_\text{trajectory}$ is invariant under the translation
\begin{align}
 \vec{X}_i \to \vec{X}_i+\vec{d},\ T_i \to T_i + \delta,\label{eq52}
\end{align}
where $(\delta, \vec{d}\,)$ is a constant four vector. From $R_\text{momentum}$, the momentum is approximately conserved
with the uncertainty $1/\sqrt{\sigma_S}$ and the energy of the system moving with $\vec{v}_0$ is approximately conserved
with the uncertainty $1/\sqrt{\sigma_t}$. Since a massless particle has the maximum speed, the moving frame has a large
velocity and the effect becomes significant for a massless or extremely light particle. The product Eq. \eqref{eq45}
also depends on the phase factor
\begin{align}
 \phi &=\phi_0 + \phi_1,\label{eq53}\\
\phi_0 &= \sum_j(\pm)\left(\vec{p}_j\cdot\vec{X}_j - E(\vec{p}_j)T_j\right),\nonumber\\
\phi_1 &= -2\sigma_t\left(\sum_j\frac{(\vec{v}_0 - \vec{v}_j)\cdot\tilde{\vec{X}}_j}{2\sigma_j}\right)\left\{
\sum_j(\pm)\left(\vec{v}_0\cdot\vec{p}_j - E(\vec{p}_j)\right)\right\}\nonumber\\
&-2\sigma_S\left(\sum_j(\pm)\vec{p}_j\right)\cdot\left(\sum_j\frac{\tilde{\vec{X}}_j}{2\sigma_j}\right),\nonumber
\end{align}
where $\phi_0$ agrees with that of a plane wave.

When the values of $\sigma_S$ and $\sigma_t$ are finite, the product Eq. \eqref{eq45} becomes finite in a small region
of $(t,\vec{x})$ and decreases steeply away from this region. Hence the integration over $(t,\vec{x})$ becomes
\begin{align}
 \int d^4x \langle k|\prod_{j=1}^{j=N}\varphi_j(x)|l\rangle = N_t(2\sigma_S\pi)^\frac{3}{2}(2\sigma_t)^\frac{1}{2}
e^{R+i\phi},
\end{align}
and converges fast. The integral over $0\leq t \leq T$ becomes $O(\exp^{-\frac{T^2}{2\sigma_t}})$. Thus the finite-size
correction to the probability is $O(\exp^{-\frac{T^2}{2\sigma_t}})$ with a microscopic $\sigma_t$, and is negligible
 at a macroscopic $T$.

\subsection{Pseudo-Doppler effect}
The first effect caused by the modified conservation law of kinetic energy is the distortion of the energy distribution,
which appears in the amplitude at finite and infinite $T$.

The energy-momentum conservation in n invariant system under the translation
\begin{align}
 x_\mu \to x_\mu + d_\mu,\label{eq55}
\end{align}
where $d_\mu$ is a constant four vector, is derived from the integration for the plane waves
\begin{align}
 \int d^4xe^{i(k_i-k_f)\cdot x}=(2\pi)^4\delta^{(4)}(k_i-k_f),\label{eq56}
\end{align}
where $k_i$ and $k_f$ are the four-dimensional momenta of the initial and final states. In the amplitude of the 
wave packets, the wave functions overlap in a finite space time area and the amplitude is not invariant under 
Eq. \eqref{eq55} generally. However, for a large $\sigma_t$, it is approximately invariant under the transformation
Eq. \eqref{eq43} from Eq. \eqref{eq51}, and the energy in the moving frame is approximately conserved. In a system
of $\sigma_t = \infty$, the invariance is rigorous.

$R_\text{momentum}$ is rewritten as
\begin{align}
R_\text{momentum} &= -\frac{\sigma_t}{2}(\delta \tilde{E})^2 - \frac{\sigma_S}{2}(\delta \vec{p}\,)^2,\label{eq57}\\
\delta\tilde{E} &= \sum_{\text{initial},l}(E^i(\vec{p}_l) - \vec{v}_0\cdot\vec{p}^{\,i}_l)-
\sum_{\text{final},k}(E^f(\vec{p}_k) - \vec{v}_0\cdot\vec{p}^f_k),\nonumber\\
\delta\vec{p} &= \sum_j(\pm)\vec{p}_j.\nonumber
\end{align}
For small $\sigma_S$ and large $\sigma_t$, $|\delta\vec{p}\,|$ becomes large but $|\delta\tilde{E}|$ becomes small,
and the modified conservation law, $\delta\tilde{E} = 0$,
\begin{align}
 \sum_\text{final} E^f(\vec{p}_k) - \sum_\text{initial} E^i(\vec{p}_l) = \sum_\text{final}\vec{v}_0\cdot\vec{p}^f_k
- \sum_\text{initial}\vec{v}_0\cdot\vec{p}^{\,i}_l\label{eq58}
\end{align} 
is fulfilled. The momentum spreading is large and the conservation law for the events of $\delta\vec{p} = 0$ or
$\vec{v}_0 = 0$ takes the form
\begin{align}
 \sum_\text{final}E^f(\vec{p}_k) - \sum_\text{initial}E^i(\vec{p}_l) = 0.
\end{align}
For the events of $\delta \vec{p}\neq 0$ and $\vec{v}_0\neq 0$, the law becomes
\begin{align}
 \sum_\text{final}\gamma_kE^f(\vec{p}_k) - \sum_\text{initial}\gamma_lE^i(\vec{p}_l) = 0,\ \gamma_l = 
\frac{E^i(\vec{p}_l) - \vec{v}_0\cdot\vec{p}^i_l}{E^i(\vec{p}_l)},\label{eq60}
\end{align}
where $\gamma_j$ is the rate of the energies in the moving and rest systems. $\delta \tilde{E}$ is also written in
the high-energy region in the following form:
\begin{align}
 \delta\tilde{E} = \sum_j(\pm)E(\tilde{\vec{p}}_j),\ \tilde{\vec{p}}_j = \vec{p}_j - \frac{\sigma_S}{\sigma_j}\delta
\vec{p}.\label{eq61}
\end{align}
From the momenta and energies of particles in the final state, $\tilde{\vec{p}}_j$ can be computed from Eq. \eqref{eq61},
and $E(\tilde{\vec{p}}_j)$ is calculated. Then Eq. \eqref{eq58} can be verified. The total momenta are distributed with
the width given by $1/\sqrt{\sigma_S}$ but the sum of total energies at $\tilde{\vec{p}}_j$ vanishes at each event.
Even though the detector is at rest and a real Doppler effect is irrelevant, the kinetic energy of the moving frame,
instead of that in the rest system, is conserved. Consequently, the kinetic energy of the final state shifts in magnitude
in events of large $|\delta \vec{p}\,|$. In the Doppler effect, the energy shifts in all events, so the shifts due to 
the wave packet are different and are called the pseudo-Doppler effect.

For Small $\sigma_S$ and $\sigma_t$, $|\delta \vec{p}\,|$ and $|\delta E(\tilde{\vec{p}}_j)|$ become large. For plane
waves, $\sigma_i=\infty$, the velocity $\vec{v}_0$ vanishes and the modified conservation law becomes the standard one.
Thus the energy conservation for the wave packets is different from both that of classical mechanics and that for
the plane waves of quantum mechanics.

The modified law of energy conservation results from $S[\infty]$, which satisfies the commutation relation
$\left[S[\infty], H_0\right] = 0$ duet to the fact that the wave packets are superpositions of states of continuous
eigenvalues of $H_0$. The quantum particle of the momentum $\vec{p}$, kinetic energy $E(\vec{p}\,)$, and size $\sigma$
gives a reaction as a particle of the energy $\gamma E(\vec{p}\,)$, and the modified conservation law, Eq. \eqref{eq60},
is fulfilled. Here $\gamma$ is regarded as the ratio of the time intervals in the moving and rest frames, and
Eq. \eqref{eq58} is understood as that for the average values taken over the time intervals
\begin{align}
 \frac{1}{\sum_j \gamma_j}\left\{\sum_l \gamma_l E(\vec{p}_l) - \sum_k \gamma_k E(\vec{p}_k)\right\}
=\langle E^i\rangle - \langle E^f \rangle =0.\label{eq62}
\end{align}
Thus the conservation law of energy is modified to that for the average values. Because the energy is conjugate to
the time, the equality of average values taken over the time intervals is reasonable. From Eq. \eqref{eq58}, the 
effective action
\begin{align}
 \int \sum_i \left(E_i dt - \vec{p}_i\cdot d\vec{x}_i\right) = \int \sum_i\left(E_i - \vec{p}_i\cdot \vec{v}_i\right)dt
,\ \vec{v}_i = \vec{v}_0
\end{align}
of the initial state coincides with that of the final state in the present reaction.

\subsection{Finite-size correction}
The second effect caused by the modified conservation law is the large finite-size correction. If $\sigma_t$ is 
finite of a microscopic size, the integration over $t$ converges and the amplitude and probability decrease rapidly
due to $R_\text{trajectory}$. In a marginal case of $\sigma_t = \infty$, the modulus of Eq. \eqref{eq45} does not
decrease with $t$ but the wave packets overlap in the infinite-time interval. This happens in various situations.
If all the particles except particle 1 are plane waves,
\begin{align}
 \sigma_1 &\neq \infty,\\
\sigma_j &= \infty ,\ j\neq 1,
\end{align}
the frequency and the real and imaginary parts of the amplitude
\begin{align}
& \prod_kN_k^*\prod_lN_l\exp^{-\frac{1}{2\sigma_S}\left(\vec{x} - \vec{x}_0(t)\right)^2 - i\omega(t-t_0)}e^{R+i\phi},\label{eq66}\\
&\sigma_S = \sigma_1, \ \sigma_t = \infty\nonumber
\end{align}
are 
\begin{align}
 \omega &=\sum_\text{initial} E^i - \sum_\text{final}E^f - \vec{v}_1\cdot\delta\vec{p},\label{eq67}\\
R&=R_\text{momentum} + R_\text{trajectory},\nonumber\\
R_\text{momentum} &= -\frac{\sigma_1}{2}\left(\delta\vec{p}\,\right)^2,\ R_\text{trajectory} = 0,\nonumber\\
\phi &= E(\vec{p}_1)T_1 - \vec{p}_1\cdot\vec{X}_1 - \delta\vec{p}\cdot\vec{X}_1.\nonumber
\end{align}
The modulus of Eq. \eqref{eq66} decreases fast with $|\vec{x} - \vec{x}_0(t)|$ and the total momentum is approximately
conserved, whereas it is constant in $t$. The phase factor has a similar form to that of the plane wave Eq. \eqref{eq42}
but the angular velocity $\omega$ is not identical. $\omega$ in Eq. \eqref{eq67} is the energy in a moving frame with
velocity $\vec{v}_1$, showing the pseudo-Doppler effect, and is
\begin{align}
 \omega &= -E_1(\vec{p}_1) + \vec{v}_1\cdot \vec{p}_1 + \omega_0 = -\left(\sqrt{\vec{p}_1^{\,2} + m_1^2} - |\vec{p}_1|\right)
 + \omega_0\nonumber\\
&= -\frac{m_1^2}{2|\vec{p}_1|^2} + \omega_0,\label{eq68}
\end{align}
where $m_1$ is the mass of particle 1 and $\omega_0$ is independent of $\vec{p}_1$ in the high-energy region.
Hence $\omega$ depends on the momentum of particle 1 in a different wave to the plane wave, and there are more states
satisfying $\omega \approx 0$ than those of the simple plane wave. The amplitude Eq. \eqref{eq66} at a large-time
interval is determined by a state satisfying $\omega=0$ and also the states $\omega\approx0$. From Eq. \eqref{eq68},
$\omega$ is degenerate at $|\vec{p}_1|\to \infty$, and an infinite number of states make a contribution. It will be shown
that the rate derived from this at finite $T$, $\Gamma(T)$, has a large finite-size correction and is described in 
the form
\begin{align}
 \Gamma(T) &= \Gamma_0 + \Gamma_1(T),\\
\Gamma_1(T) &= C_1/T,\nonumber
\end{align}
where $C_1$ is a constant and $\Gamma_0$ is the asymptotic term.

$\sigma_t$ becomes infinite when the right-hand side of Eq. \eqref{eq46} vanishes. This condition is fulfilled in 
particular momenta of the initial and final states. The probability thus has a finite-size correction in this
kinematical region.

\subsection{Asymmetric wave packet}
In some situations, the wave packet is asymmetric in $\vec{k}_L$ and $\vec{k}_T$, which are parallel and perpendicular
to the central momentum, or in $\vec{k}$ and $|\vec{k}|$. A small energy uncertainty, $\delta E\gg |\vec{k}|$, also
often appears. For an asymmetric wave packet or a wave packet with different spreadings in the momentum and energy,
we have 
\begin{align}
 \langle t,\vec{p}\,|\vec{p}_0,\vec{X}_0,T_0\rangle_\text{asy} &= N\sigma^\frac{3}{2}e^{-iE(\vec{p}\,)(t-T_0)
- i\vec{p}_0\cdot\vec{X}_0 - \frac{\sigma_L}{2}\left(\vec{p}_L - \vec{p}_L^{\,0}\right)^2
-\frac{\sigma_T}{2}\left(\vec{p}_T\right)^2},\\
\langle t,\vec{p}\,|\vec{p}_0,\vec{X}_0,T_0\rangle_E &= N\sigma^\frac{3}{2}e^{-iE(\vec{p}\,)(t- T_0)-\vec{p}_0\cdot\vec{X}_0
- \frac{\sigma}{2}\left(\vec{p}- \vec{p}_0\right)^2 - \frac{\sigma_E}{2}\left(E(\vec{p}\,) - E_0\right)^2},
\end{align}
where $\sigma_L$, $\sigma_T$, and $\sigma_E$ are the size in the parallel and perpendicular directions to the center
of momentum, and that in the energy. The functions in the coordinate representation become Gaussian forms in $\vec{x}$
and t:
\begin{align}
 \langle t,\vec{x}\,|\vec{p}_0,\vec{X}_0,T_0\rangle_\text{asy} &= \int d\vec{p}\langle \vec{x}\,|\vec{p}\,\rangle
\langle t,\vec{p}\,|\vec{p}_0,\vec{X}_0,T_0\rangle_\text{asy}\nonumber\\
&=N\sigma^\frac{3}{2}e^{-iE(\vec{p}_0)(t-T_0) - i\vec{p}_0\cdot\vec{X}_0 - \frac{1}{2\sigma_L}\left(
{x} - {X}_L - v(t-T_0)\right)^2 - \frac{1}{2\sigma_T}(\vec{x}_T)^2},\label{eq72}\\
\langle t,\vec{x}\,|\vec{p}_0,\vec{X}_0,T_0\rangle_E &= \int d\vec{p} \langle \vec{x}\,|\vec{p}\,\rangle\langle t,
\vec{p}\,|
\vec{p}_0,\vec{X}_0,T_0\rangle_E\label{eq73}.
\end{align}
In $\sigma_L \approx \sigma_T$ or $\sigma_E \approx \sigma$, the energy spreading is about the same as that of the 
momentum, and the probability of a finite $|\Delta \vec{p}\,|$ around the central momentum $\vec{p}_0$ shows
pseudo-Doppler  and finite-size effects. In $\sigma_L \gg \sigma_T$ or $\sigma_E \gg \sigma$, on the other hand,
 the energy spreading is much smaller than the momentum spreading and Eq. \eqref{eq72} or \eqref{eq73} is applied.
Precision experiments of $\Delta E\approx 0$, $|\Delta\vec{p}\,| = \infty$ of narrow energy levels are studied with
Eq. \eqref{eq73}, and the probability does not show pseudo-Doppler and finite-size effects then.

\section{Two-body decay: $A \to B+C$}
\begin{figure}[t]
 \centering{\includegraphics{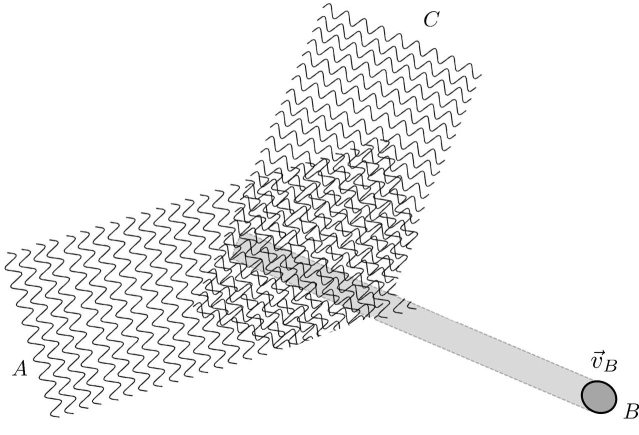}
\caption{The decay amplitudes of $A$ to $B$ and $C$ that are expressed with wave packets of different sizes are
represented. They interact in a region where they overlap. If the size of $B$ is smaller than the others, the region
is mainly determined by $B$.}
\label{Fig:fig2}
}
\end{figure}

The unusual properties of the decay probability at a finite distance are studied in detail for two-body decay here.
The decay rate is computed with $S[T]$ and the finite-size correction to that computed by Fermi's golden rule is found.
The correction depends on the boundary conditions of the experiments and is computed properly with $S[T]$ that satisfies
the boundary condition at $T$, instead of $S[\infty]$.
Two-body decays of a particle $A$ into $B$ and $C$ of masses $m_A$, $m_B$, and $m_C$ satisfying $m_A > m_B+m_C$ and 
governed by a local Lagrangian
\begin{align}
 \mathscr{L} &= \mathscr{L}_0 + \mathscr{L}_\text{int},\label{eq74}\\
\mathscr{L}_0 &= \frac{1}{2}\left[
\left(\partial\varphi_A\right)^2 - m_A^2\varphi_A^2 + \left(\partial\varphi_B\right)^2 - m_B^2\varphi_B^2
+\left(\partial\varphi_C\right)^2 - m_C^2\varphi_C^2
\right],\nonumber\\
\mathscr{L}_\text{int} &= g\varphi_A(x)\varphi_B(x)\varphi_C(x)\nonumber
\end{align}
in the wave zone are studied in the lowest order of coupling constant $g$. The characteristic features of decay
amplitude in the wave packet scattering are seen in Fig. \ref{Fig:fig2}, which shows a space-time picture of the decay
of a large $A$ to a large $B$ and small $C$. Because the interaction occurs in the finite region where these waves 
overlap, the conservation laws of the kinetic energy and momentum are modified from those of plane waves.

\subsection{Average energy in the wave zone}
The kinetic energy of the wave function at finite $t$ is not a constant. The state vector evolves with the 
Schr\"{o}dinger equation of a total Hamiltonian composed of a free part $H_\text{int}$ derived by the Lagrangian,
Eq. \eqref{eq74}:
\begin{align}
 i\hbar\frac{\partial}{\partial t}|\Psi(t)\rangle = (H_0 + H_\text{int})|\Psi(t)\rangle.
\end{align}
From perturbative expansion, a solution satisfying the boundary condition $|\Psi(0)\rangle = |\psi_1\rangle$ is
\begin{align}
 |\Psi(t)\rangle = a_1(t)|\psi_1\rangle + |\psi_2\rangle,
\end{align}
where $|\psi_1\rangle$ is a one-particle state composed of $A$ of a momentum $\vec{p}_A$ and a kinetic energy $E_A$
and $|\psi_2\rangle$ is a two-particle state composed of $B$ and $C$:
\begin{align}
 |\psi_1(t)\rangle &= \exp^{\frac{E_A}{i\hbar}t}|A,\vec{p}_A\rangle,\\
|\psi_2(T)\rangle &= \int_0^tdt'\frac{H_\text{int}(t)}{i\hbar}|\psi_1(t')\rangle = \int_0^tdt'|B,C\rangle\langle B,C|
\frac{H_\text{int}(t')}{i\hbar}|\psi_1(t')\rangle.
\end{align}
In the lowest order in $g$,
\begin{align}
 a_1(t) = 1.
\end{align}
The energy expectation value is 
\begin{align}
 \langle E\rangle &=\frac{\langle\Psi(t)|H|\Psi(t)\rangle}{\langle\Psi(t)|\Psi(t)\rangle},\\
\langle\Psi(t)|H|\Psi(t)\rangle &= |a_1(t)|^2E_A\langle\psi_1|\psi_1\rangle + 2Re[
a_1(t)\langle\psi_2|H_\text{int}|\psi_1\rangle
]
+(E_B+E_C)\langle\psi_2|\psi_1\rangle,\nonumber\\
\langle \Psi(t)|\Psi(t)\rangle &= |a_1(t)|^2\psi_1|\psi_1\rangle + \langle\psi_2|\psi_2\rangle.\nonumber
\end{align}
At infinite $t$,
\begin{align}
 \langle \psi_2|\psi_2\rangle & = 2\pi t \delta (\omega)\left|\langle B,C|\frac{H_\text{int}(0)}{i\hbar}|A\rangle 
\right|^2,\label{eq81}\\
a_1(t) \langle \psi_2|H_\text{int}|\psi_1\rangle &= \frac{i}{\omega}(1 - e^{^i\omega t/\hbar})
\left|\langle B,C|\frac{H_\text{int}(0)}{i\hbar}\right|^2 = O(1),\label{eq82}\\
\omega &= E_A - E_B - E_C,\nonumber
\end{align}
hence the expectation value of the interaction $H_\text{int}(0)$, Eq. \eqref{eq82}, is negligible compared to 
Eq. \eqref{eq81} and the kinetic energy as well as the total energy is $E_A$. For finite $t$, the expectation value
of $H_\text{int}(0)$ is not negligible. An average over a finite-time interval gives
\begin{align}
& 2\text{Aver}(\text{Re}[a_1(t)\langle \psi_2|H_\text{int}|\psi_1\rangle]) =\frac{2}{\omega}\left|
\langle B,C|\frac{H_\text{int}(0)}{i\hbar}|A\rangle\right|^2,\\
&\text{Aver}(\langle \psi_2|\psi_2\rangle) = 2\frac{\hbar^2}{\omega^2}\left|\langle B,C|\frac{H_\text{int}}{i\hbar}
|A\rangle\right|^2,\\
&\text{Aver}(\langle \psi_2|H_0|\psi_2\rangle) = 2(E_B + E_C)\frac{\hbar^2}{\omega^2}\left|
\langle B,C|\frac{H_\text{int}(0)}{i\hbar}|A\rangle\right|^2.
 \end{align}
The expectation value of the total energy becomes
\begin{align}
 \text{Aver}\langle H\rangle = E_A.
\end{align}
Thus the average energy coincides with the initial energy, but the average kinetic energy is 
\begin{align}
 \text{Aver}(\langle H_0\rangle) = \frac{E_A + 2(E_B+E_C)\frac{2\hbar^2}{\omega^2}|\langle
B,C|\frac{H_\text{int}(0)}{i\hbar}|A\rangle|^2}{1 + \frac{2\hbar^2}{\omega^2}|\langle B,C|\frac{H_\text{int}(0)}{i\hbar}
|A\rangle|^2},
\end{align}
and is different from the initial kinetic energy. $H_\text{int}$ causes a transition of $A$ to $B$ and $C$, and is
non-diagonal in the base of eigenvectors defined by $H_0$. Thus $H_\text{int}$ does not contribute to the total energy
 in infinite $t$, but, at finite $t$, the state $|\Psi(t)\rangle$ is superposition of $|A\rangle$ and $|B,C\rangle$ and
$H_\text{int}$ has a finite expectation value. The total energy is always same but the expectation value of 
$H_\text{int}$ is finite in finite $t$. Hence the state becomes a super position of different kinetic energies
and the kinetic energy is not a good quantum number in this region.

$E_A$ is real in the lowest order of $g$ and has an imaginary part in the second order, which represents the life-time
of $A$, $\tau_A$. In $t\ll \tau_A$, the imaginary part of $E_A$ is negligible. For a self-consistent treatment of the 
decay process, we start from $E_A$ of an imaginary part and compute the decay amplitude and probability. The decay
probability is proportional to $T$ in $T\ll\tau_A$ and becomes unity at $T\gg \tau_A$.

\subsection{Transition amplitude and decay probability}
Next we study the transition probability at finite distance. The decay of a particle $A$ at a space-time position
$(\vec{X}_A,T_A)$ into particles $B$ at $(\vec{X}_B,T_B)$ and $C$ at $(\vec{X}_C,T_C)$ in the most general case of
the symmetric wave packets
\begin{align}
 \sigma_A,\ \sigma_B,\ \sigma_C,
\end{align}
of the four-dimensional momenta and masses
\begin{align}
 (E_A,\vec{p}_A;m_A),\ (E_B,\vec{p}_B;m_B),\ (E_C,\vec{p}_C;m_C)
\end{align}
is studied here. The life-time of $A$ expressed with the imaginary part of $E_A$ is assumed negligible in majority of
the present paper. From the interaction Lagrangian Eq. \eqref{eq74}, the transition amplitude is expressed with an 
integral over $(t,\vec{x})$:
\begin{align}
 \mathcal{M}(A\to B+C)&=g\int dt\int d\vec{x}\exp^{-\frac{1}{2\sigma_S}(\vec{x}- \vec{x}_0)^2 - \frac{1}{2\sigma_t}
(t-t_0)^2}e^{R+i\phi}\nonumber\\
&=g(2\pi\sigma_S)^\frac{3}{2}(2\pi \sigma_t)^\frac{1}{2}e^{R+i\phi}\theta(\vec{X}_i,T_i),\label{eq90}
\end{align}
for finite values of $\sigma_S$ and $\sigma_t$, where $\theta(\vec{X}_i,T_i)$ denotes the condition that $t_0$ is
the inside of the time region defined from the boundary conditions; we omit it hereafter. $\sigma_S$ and $\sigma_t$
are given in the expression
\begin{align}
 \frac{1}{\sigma_S} &= \frac{1}{\sigma_A} + \frac{1}{\sigma_B} + \frac{1}{\sigma_C},\label{eq91}\\
\frac{1}{\sigma_t} &=\frac{\vec{v}_A^{\,2}}{\sigma_A} + \frac{\vec{v}_B^{\,2}}{\sigma_B} + \frac{\vec{v}_C^{\,2}}{\sigma_C}
 - \sigma_S\left(\frac{\vec{v}_A}{\sigma_A} + \frac{\vec{v}_B}{\sigma_B} + \frac{\vec{v}_C}{\sigma_C}\right)^2.\label{eq92}
\end{align}
The center position $\vec{x}_0(t)$ is 
\begin{align}
 \vec{x}_0(t) = \vec{x}_0(0) + \vec{v}_0(t-t_0),
\end{align}
of an average velocity $\vec{v}_0$,
\begin{align}
 \vec{v}_0 = \sigma_S\left(\frac{\vec{v}_A}{\sigma_A} + \frac{\vec{v}_B}{\sigma_B} + \frac{\vec{v}_C}{\sigma_C}\right).\label{eq93}
\end{align}
$R$ and $\phi$ in the exponent are obtained from Eqs. \eqref{eq50}, \eqref{eq51}, and \eqref{eq53}, and are given as
\begin{align}
 &R=R_\text{trajectory} + R_\text{momentum},\\
&R_\text{trajectory} = -\sum_j\frac{\tilde{\vec{X}}_j^2}{2\sigma_j} + 2\sigma_S\left(\sum_j\frac{\tilde{\vec{X}}_j}
{2\sigma_j}\right)^2 + 2\sigma_t\left(\sum_j\frac{(\vec{v}_0 - \vec{v}_j)\cdot\tilde{\vec{X}}_j}{2\sigma_j}\right)^2,\label{eq96}\\
&R_\text{momentum}  = -\frac{\sigma_t}{2}\left(\delta E - \vec{v}_0\cdot\delta\vec{p}\,\right)^2 - \frac{\sigma_S}{2}
(\delta \vec{p}\,)^2,\label{eq97}\\
&\delta E = E_A(\vec{p}_A) - E_B(\vec{p}_B) - E_C(\vec{p}_C),\ \delta \vec{p}=\vec{p}_A-\vec{p}_B-\vec{p}_C,
\end{align}
and $\phi$ is a function of the momenta $\vec{p}_j$ and positions $\vec{X}_j$.

Since $R_\text{trajectory}$ is a function of the momenta and coordinates, we write it as 
$R_\text{trajectory}(\vec{X}_A,T_A;\vec{X}_l,T_l)$, where $l$ stands for $B$ or $C$. This is invariant under the 
translation, Eq. \eqref{eq52}:
\begin{align}
 R_\text{trajectory} (\vec{X}_A + \vec{d}, T_A + \delta;\vec{X}_l+\vec{d},T_l+\delta) = 
R_\text{trajectory}(\vec{X}_A,T_A;\vec{X}_l,T_l).
\end{align}
Choosing $\vec{d} = \vec{v}_A\delta$, we have the identity
\begin{align}
 R_\text{trajectory}(\vec{X}_A,T_A;\vec{X}_l+\vec{v}_A\delta - \vec{v}_l\delta,T_l) = R_\text{trajectory}
(\vec{X}_A,T_A;\vec{X}_l,T_l),
\end{align}
and 
\begin{align}
 \frac{\partial}{\partial \delta}R_\text{trajectory}(\vec{X}_A,T_A;\vec{X}_l + \vec{v}_A\delta - \vec{v}_l\delta,T_L)=0.
\label{eq101}
\end{align}

The probability is the integral 
\begin{align}
 P=\int \prod_id\vec{X}_i\frac{d\vec{p}_i}{(2\pi)^3}|\mathcal{M}|^2.\label{eq102}
\end{align}
$|\mathcal{M}|^2$ does not depend on $\delta$ from Eq. \eqref{eq101}, and the phase space is reduced to that in the
$\delta$ component and the orthogonal components, $\vec{X}_T$:
\begin{align}
 \int d\delta \prod_id\vec{X}_{T,i}\frac{d\vec{p}_i}{(2\pi)^3}|\mathcal{M}|^2.\label{eq103}
\end{align}
The parameter $\delta$ is not measured in the ordinary experiment and is integrated. From the integration over $\delta$,
we have
\begin{align}
 \int d\delta \prod_id\vec{X}_{T,i}\frac{d\vec{p}_i}{(2\pi)^3}|\mathcal{M}|^2 = T\int \prod_i d\vec{X}_{T,i}
\frac{d\vec{p}_i}{(2\pi)^3}\frac{\sum_l\left(\vec{v}_l^{\,2} - \vec{v}_l\cdot\vec{v}_A\right)}
{\sqrt{\sum_l\left(\vec{v}_l - \vec{v}_A\right)^2}}|\mathcal{M}|^2.\label{eq104}
\end{align}
Thus the probability in the system of finite $\sigma_S$ and $\sigma_t$ is proportional to time interval, $T$.
Its magnitude is independent of the parameters of the wave packet from the completeness equation, Eq. \eqref{eq40},
and agrees with the value obtained with $S[\infty]$ defined by plane waves combined with $i\epsilon$ prescription.

\begin{figure}[t]
\centering{
 \includegraphics[scale=.5,angle=-90]{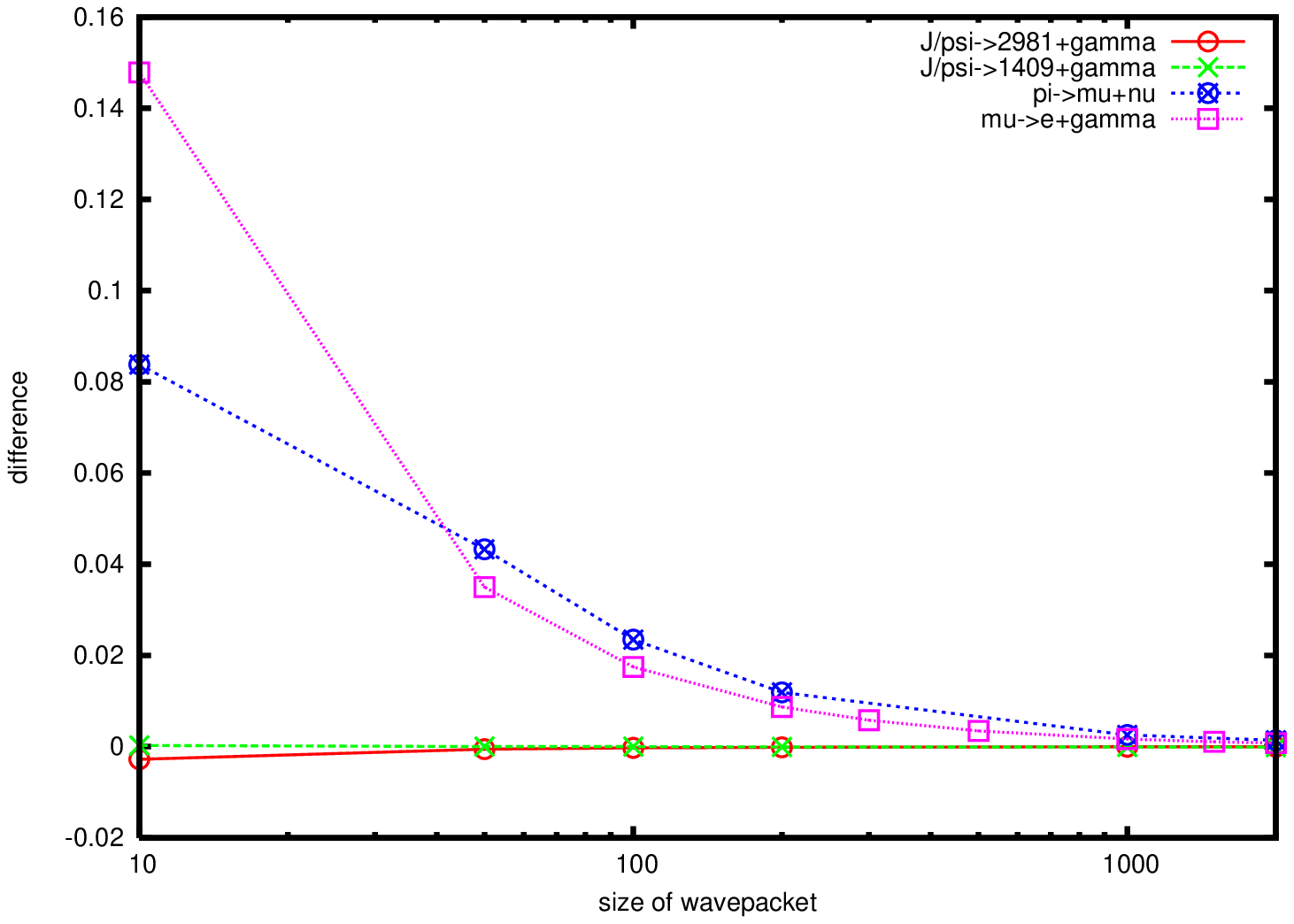}
\caption{The probability of decays at rest (Eq. \eqref{eq104}) for the wave packets in two-body decays:
$J/\Psi \to M(2981) + \gamma$ (red solid), $J/\Psi \to \eta(1409) + \gamma$ (green dot), $\pi \to \mu + \nu$ (blue dot),
and $\mu \to e + \gamma$ (magenta dot). The wave packet of another daughter is $\infty$ and that of the parent is 
$\sigma_\text{parent}m_\pi^2 = 10000$. The horizontal axis shows the size of the wave packets of the light particle
in units of $\sigma m_\pi^2$ and the vertical axis shows the deviations of the rates for wave packets over the rates 
for the plane waves, $1 - \frac{P_\text{wave packet}}{P_\text{plane wave}}$. Errors for $\pi$ and $\mu$ are slightly
larger than others due to numerical uncertainty.
 }
\label{fig3}}
\end{figure}

In Fig. \ref{fig3}, the rates computed with wave packets of various sizes are compared with those of the plane waves
in various decays, $J/\Psi \to M(2981) + \gamma$, $J/\Psi \to \eta(1409) + \gamma$, $\pi \to \mu + \gamma$, and
$\mu \to e + \gamma$, which will be discussed later. The wave packet of another daughter is $\infty$ and that of
the parent is $\sigma_\text{parent}m_\pi^2 = 10000$. The value is the same for all processes.
Within small errors, they agree.

\subsection{Various cases of wave packets}
We study the amplitude and probability of the systems (1) $\sigma_S=\text{finite}$, $\sigma_t=\text{finite}$,
(2) $\sigma_S=\text{finite}$, $\sigma_t=\infty$, (3) $\sigma_S = \infty$, $\sigma_t=\infty$ in the following.

\subsubsection{Finite $\sigma_S$ and finite $\sigma_t$}
When $\sigma_A$, $\sigma_B$ and $\sigma_C$ are finite, $\sigma_S$ and $\sigma_t$ are also finite;
the integrand in Eq. \eqref{eq90} decreases fast at $t\to \infty$ and $|\vec{x}|\to \infty$, the integrals over
$t$ and $\vec{x}$ converge fast, and the results of Eqs. \eqref{eq50}, \eqref{eq51}, and \eqref{eq53} are applied. 
The total probability is obtained by integration the momentum and position of Eq. \eqref{eq102}, and does not
have a finite-size correction at a macroscopic $T$.

When $\sigma_A$ and $\sigma_B$ are finite and $\sigma_C = \infty$, $\sigma_t$ and $\sigma_S$ are finite generally.
We have 
\begin{align}
 \frac{1}{\sigma_S} &= \frac{1}{\sigma_A} + \frac{1}{\sigma_B},\label{eq105}\\
\frac{1}{\sigma_t} &=\frac{\left(\vec{v}_A - \vec{v}_B\right)^2}{\sigma_A + \sigma_B},\label{eq106}\\
\vec{v}_0 &= \sigma_S\left(\frac{\vec{v}_A}{\sigma_A} + \frac{\vec{v}_B}{\sigma_B}\right).\label{eq107}
\end{align}
The integrand of Eq. \eqref{eq90} decreases fast at $|\vec{x} - \vec{x}_0| \to \infty$ and the integral over $\vec{x}$
converges fast. $\sigma_t$ is finite when $\vec{v}_A \neq \vec{v}_B$, the integrand decreases fast at 
$|t-t_0|\to \infty$, and the integral over $t$ converges fast. We have
\begin{align}
 R_\text{trajectory} & = -\frac{\left\{\left(\tilde{\vec{X}}_B - \tilde{\vec{X}}_A\right)_T\right\}^2}
{2\left(\sigma_A + \sigma_B\right)},\label{eq108}\\
\left(\tilde{\vec{X}}_B - \tilde{\vec{X}}_A\right)_T &= \left(\tilde{\vec{X}}_B - \tilde{\vec{X}}_A\right)
 - \frac{\left(\vec{v}_B-\vec{v}_A\right)}{|\vec{v}_B - \vec{v}_A|}\cdot
\left(\tilde{\vec{X}}_B - \tilde{\vec{X}}_A\right)\frac{\left(\vec{v}_B - \vec{v}_A\right)}{|\vec{v}_B - \vec{v}_A|}.
\nonumber
\end{align}
Thus the probability depends on the transversal components of coordinate $\vec{X}_B - \vec{X}_A$ but not on the
longitudinal component. The coordinate of $B$ is integrated over the transversal and longitudinal components
\begin{align}
 \int d\vec{X}_Be^{2R_\text{trajectory}} = \int d\left(\tilde{\vec{X}}_B-\tilde{\vec{X}}_A\right)_T
d\left(\tilde{\vec{X}}_B - \tilde{\vec{X}}_A\right)_Le^{2R_\text{trajectory}},\label{eq109}
\end{align}
where the former variables are integrated in the form:
\begin{align}
 \int d\left(\tilde{\vec{X}}_B-\tilde{\vec{X}}_A\right)_Te^{2R_\text{trajectory}} = \pi\left(\sigma_A+\sigma_B\right),
\label{eq110}
\end{align}
and the latter variable is integrated using $\theta(\vec{X}_i,T_i)$ in Eq. \eqref{eq90} as
\begin{align}
 \int d\left(\tilde{\vec{X}}_B - \tilde{\vec{X}}_A\right)L = |\vec{v}_B - \vec{v}_A|\int d\left(T_B-T_A\right)
= |\vec{v}_B - \vec{v}_A|T.\label{eq111}
\end{align}
Thus the probability is proportional to $T$, and does not have a finite-size correction. $R_\text{momentum}$ is
expressed with Eq. \eqref{eq97} or with the energies of the momenta
\begin{align}
 \tilde{\vec{p}}_A &= \vec{p}_A - \frac{\sigma_B}{\sigma_A + \sigma_B} \left(\vec{p}_A - \vec{p}_B-\vec{p}_C\right),
\label{eq112}\\
 \tilde{\vec{p}}_B &= \vec{p}_B + \frac{\sigma_A}{\sigma_A + \sigma_B} \left(\vec{p}_A - \vec{p}_B-\vec{p}_C\right),
\label{eq113}\\
\tilde{\vec{p}}_C &= \vec{p}_C.\label{eq114}
\end{align}

Other cases with two wave packets and one plane wave are equivalent to the previous case.

In the $\vec{v}_B = \vec{v}_A$ case, in the limit $\vec{v}_B \to \vec{v}_A$, $\sigma_t$ diverges and many cause a
large diffraction effect. Nuclei trapped in matter have momenta $\vec{p}_A = \vec{p}_B = 0$ and M\"{o}ssbauer effect
is a phenomenon that occurs through absorption of a gamma ray by a nucleus.

\subsubsection{Finite $\sigma_S$ and infinite $\sigma_t$}
In finite $\sigma_S$ and infinite $\sigma_t$, the wave functions of initial and final states overlap in a long strip 
region; accordingly, the probability shows unusual finite-size corrections. \\
\textbf{A: Small mass}

We study next the situation where the particles $A$ and $C$ are described by plane waves,
\begin{align}
 \sigma_A = \infty,\ \sigma_C = \infty,\label{eq115}
\end{align}
of the momenta $\vec{p}_A$ and $\vec{p}_C$ and $B$ is described by a wave packet of the size $\sigma_B$ and momentum
$\vec{p}_B$. $B$ is assumed to have a small mass $m_B$. $A$ is prepared at $T_A$ and $B$ is detected at the space-time
position $(\vec{X}_B,T_B)$. Obviously, the parameters of Eq. \eqref{eq46} become
\begin{align}
 \sigma_S &= \sigma_B,\label{eq116}\\
\vec{v}_0 &= \vec{v}_B,\label{eq117}\\
\frac{1}{\sigma_t} &= \frac{\vec{v}_B^{\,2}}{\sigma_B} - \frac{\vec{v}_0^{\,2}}{\sigma_S} = 0.\label{eq118}
\end{align} 
Since $\sigma_t = \infty$, the integrand in the probability does not decrease with $t$ and may receive a finite-size
correction.

\begin{figure}[t]
\centering{\includegraphics{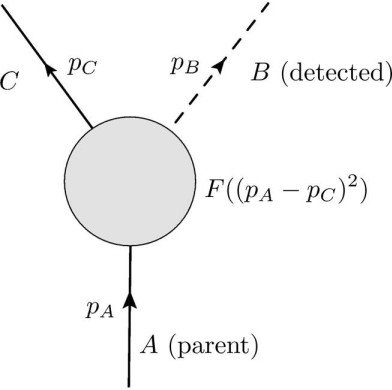}}
\caption{Form factor in $\langle A|J_B(0)|C\rangle$.}
\label{fig4}
\end{figure}

The transition amplitude is expressed in the form
\begin{align}
\mathcal{M} = \int d^4xN_1w_B(\vec{p},\vec{x})e^{-ip_A\cdot x + ip_C\cdot x}F((p_A-p_C)^2),\label{eq119}
\end{align}
where $N_1 = ig/(2E_B2E_C(2\pi)^6)^{\frac{1}{2}}$ and the coefficient $N_B$ in $w_B(\vec{p}_B,\vec{x})$ is 
$N_B = (\pi\sigma_B)^{-\frac{3}{4}}$, $F((p_A-p_C)^2)$ is the form factor shown in Fig. \ref{fig4}, and the time $t$ is integrated
over the region $T_A\leq t\leq T_B$. $\sigma_B$ is estimated using the size of a constituent object in a target that $B$ interacts
with. The coordinate $\vec{x}$ is integrated next and the amplitude finally becomes
\begin{align}
 \mathcal{M} = & N_1N_B(2\pi\sigma_B)^\frac{3}{2}e^{-i(E_BT_B - \vec{p}_B\cdot\vec{X}_B)}
e^{-\frac{\sigma_B}{2}\left(\vec{p}_A-\vec{p}_B-\vec{p}_C\right)^2}\nonumber\\
&\times \int_0^Tdt e^{-i(E_A-E_C-E_B-(\vec{p}_A-\vec{p}_B-\vec{p}_C)\cdot\vec{v}_B)t}F((p_A-p_C)^2)\nonumber\\
=&N_1N_B(2\pi\sigma_B)^\frac{3}{2}e^{-i(E_BT_B - \vec{p}_B\cdot\vec{X}_B)}
e^{-\frac{\sigma_B}{2}(\vec{p}_A-\vec{p}_B-\vec{p}_C)^2}\nonumber\\
&\times F((p_A-p_C)^2)\frac{\sin(\omega T/2)}{\omega}e^{i\omega T/2},\label{eq120}
\end{align}
where $\omega$ is 
 \begin{align}
  \omega = E_A -E_B - E_C - (\vec{p}_A - \vec{p}_B-\vec{p}_C)\cdot\vec{v}_B.\label{eq121}
 \end{align}
Because the magnitude is inversely proportional to $\omega$, $\mathcal{M}$ receives contributions from small and large $\omega$
regions. The amplitude receives a large contribution at large $T$ from the region 
\begin{align}
 \omega \approx 0.
\end{align}
A normal root satisfying 
\begin{align}
 E_A - E_B - E_C \approx 0,\ \vec{p}_A - \vec{p}_B - \vec{p}_C \approx 0
\end{align}
and a new root satisfying
\begin{align}
 E_A - E_B - E_C \neq 0,\ \vec{p}_A - \vec{p}_B - \vec{p}_C \neq 0
\end{align}
exists. Because the kinetic energy and momentum are different from those of the initial state, the secondary root gives a 
finite-size correction due to the diffraction. The dependence of the amplitude on $\vec{p}_B$ is determined by the root of
$\omega = 0$ and its slope $\frac{\partial\omega}{\partial \vec{p}_B}$. 

Assuming that $|\vec{p}_A - \vec{p}_B - \vec{p}_C|$ is small, we have
\begin{align}
 E_B(\vec{p}_B) + (\vec{p}_A - \vec{p}_B - \vec{p}_C)\cdot\vec{v}_B = E(\vec{p}_A - \vec{p}_C)\label{eq125}
\end{align}
and 
\begin{align}
 \omega = E_A(\vec{p}_A) - E_C(\vec{p}_C) - E(\vec{p}_A - \vec{p}_C).\label{eq126}
\end{align} 
The probability integrated over $\vec{p}_C$ becomes 
\begin{align}
 |N_1|^2N_B^2(2\pi\sigma_B)^3\int \frac{d\vec{p}_C}{(2\pi)^3}e^{-\sigma_B(\vec{p}_A-\vec{p}_B-\vec{p}_C)^2}
\left(\frac{\sin(\omega T/2)}{\omega}\right)^2F(({p}_A-{p}_C)^2)\nonumber\\
=|N_1|^2N_B^2(2\pi\sigma_B)^3\int d\omega\left(\frac{\sin(\omega T/2)}{\omega}\right)^2\rho(\omega),\label{eq127}
\end{align}
where the spectrum density $\rho(\omega)$ is
\begin{align}
 \rho(\omega) = \int \frac{d\vec{p}_C}{(2\pi)^3}e^{-\sigma_B(\vec{p}_A-\vec{p}_B-\vec{p}_C)^2}F(({p}_A-{p}_C)^2)
\delta(\omega-E_A(\vec{p}_A)+E_C(\vec{p}_C)+E_B(\vec{p}_A-\vec{p}_C))\label{eq128}.
\end{align}
Because $\rho(0)$ is finite and $\rho(\omega)$ deceases rapidly in the large $\omega$ region, as shown in Appendix A,
the following integral converges at finite $T$;
\begin{align}
 \int d\omega\left(\frac{\sin(\omega T/2)}{\omega}\right)^2\rho(\omega) = T\left\{2\pi\rho(0) + \frac{1}{T}\zeta\right\},
\label{eq129}
\end{align}
where $\zeta$ is equal to $C(T)$ in Appendix A. Thus the finite-size correction becomes finite.

The finite-size correction to the total probability integrated over the whole momentum region of $\vec{p}_C$ is easily obtained
with the correlation function \cite{26,27}
\begin{align}
 \int \frac{d\vec{p}_C}{(2\pi)^3}|\mathcal{M}|^2 = \frac{N_2}{E_B}\int d^4x_1d^4x_2e^{-\frac{1}{2\sigma_B}
\sum_i(\vec{x}_i - \vec{x}_i^{\,0})^2}\Delta_{A,C}(\delta x)e^{i\phi(\delta x)},\label{eq130}
\end{align}
where $N_2 = g^2(4\pi/\sigma_B)^\frac{3}{2}((2\pi)^32E_A V)^{-1}$, $V$ is the normalization volume for the initial state $A$,
$\vec{x}_i^{\,0} = \vec{X}_B + \vec{v}_B(t_i - T_B)$, $\delta x = x_1 - x_2$, $\phi(\delta x) = p_B\cdot \delta x$, and
\begin{align}
 \Delta_{A,C}(\delta x) = \frac{1}{(2\pi)^3}\int \frac{d\vec{p}_C}{E(\vec{p}_C)} e^{-i(\vec{p}_A - \vec{p}_C)\cdot \delta x}
F((p_A - p_C)^2).\label{eq131}
\end{align}
On the right-hand side of Eq. \eqref{eq131}, the integration region of the momentum $\vec{p}_C$ is that of the complete set and
is reduced to the smaller one if the integrand $|\mathcal{M}|^2$ vanishes in some kinematical region. This happens for the
amplitude of plane waves at the asymptotic region $T=\infty$, which includes the delta function, $\delta^{(4)}(\delta p)$, from
the integration over $x$, reflecting the conservation law of kinetic energy and momentum. The phase space of the final state then
becomes proportional to the initial energy. On the right-hand side of Eq. \eqref{eq131}, the coordinates are fixed and are not 
integrated. Thus the correlation function $\Delta_{A,C}(\delta x)$ does not include $\delta^{(4)}(\delta p)$, and $\vec{p}_C$ is 
integrated over the whole region.

Because the probability is finite, integration variables can be interchanged. For $\tilde{m}^2 = m_A^2-m_C^2\geq 0$  and a 
real $p_A$, \cite{23,24,26,27}, and from Appendix C,
\begin{align}
& \Delta_{A,C}(\delta x) = 2i\left[F(-\tilde{m}^2) D_{\tilde{m}}\left(-i\frac{\partial}{\partial \delta x}
\right)\left(\frac{\epsilon(\delta t)}{4\pi}\delta(\lambda)+f_\text{short}\right) + I_2
\right],\label{eq132}\\
&D_{\tilde{m}}\left(-i\frac{\partial}{\partial\delta x}\right)
=\sum_l\frac{1}{l!}\left(2p_\pi\cdot\left(-i\frac{\partial}
{\partial\delta x}\frac{\partial}{\partial\tilde{m}^2}
\right)\right),\nonumber\\
&f_\text{short}=-\frac{i\tilde{m}^2}{8\pi\xi}\theta(-\lambda)\left\{
N_1(\xi)-i\epsilon(\delta t)J_1(\xi)
\right\}
-\frac{i\tilde{m}^2}{4\pi^2\xi}\theta(\lambda)K_1(\xi),\nonumber
\end{align}
where $\epsilon(\delta t)$ is equal to $+1$ or $-1$ for positive or negative $\delta t$, respectively, $\lambda = (\delta x)^2$,
$\xi = \tilde{m}\sqrt{\lambda}$, and $N_1$, $J_1$, and $K_1$ are Bessel functions. $f_\text{short}$ has a singularity of the form
$1/\lambda$ around $\lambda = 0$ and decreases as $e^{-\tilde{m}\sqrt{|\lambda|}}$ or oscillates as $e^{i\tilde{m}\sqrt{|\lambda|}}$ at large $|\lambda|$. The condition for the convergence of the series will be studied later. The formula for $A$ 
with a finite life-time is obtained later. The last term
\begin{align}
 I_2 = \frac{1}{(2\pi)^3}\int\frac{d\vec{p}_C}{E(\vec{p}_C)} F((p_A-p_C)^2)\theta(p_A^0-p_C^0)e^{-i(p_A-p_C)\cdot\delta x}.
\label{eq1330}
\end{align}
For $\tilde{m}^2 = m_A^2-m_C^2<0$, 
\begin{align}
 \Delta_{A,C}(\delta x) = 0.\label{eq134}
\end{align}

Thus $\Delta_{A,C}(\delta x)$ is composed of the light-cone singularity $\delta(\lambda)\epsilon(\delta t)$ \cite{23,40,41},
regular terms given by Bessel functions, and $I_2$. The former two terms come from the integration from $E_A \leq E_C$,
and are finite in finite $T$. Therefore, using this expression, the finite $T$ correction, which is unobtainable with 
standard calculations of plane waves, can be found. Because the integration region for this is outside of the kinematical
region conserving energy and momentum, this integral vanishes at $T=\infty$. $I_2$, on the other hand, comes from the 
region $E_C\leq E_A$, which is the kinematical region satisfying the energy and momentum conservation, and determines the 
quantities at $T=\infty$, This expression giving the probability with the light-cone singularity converges and is valid
in the kinematical region $2p_A\cdot p_B\leq \tilde{m}_C^2$, where $\tilde{m}_C^2 = m_A^2-m_C^2$.

Substituting the expression of $\Delta_{A,C}(\delta x)$ into Eq. \eqref{eq130} and integration over $\vec{x}_1$ and 
$\vec{x}_2$, we have
\begin{align}
 J_{\delta(\lambda)} &= \int d\vec{x}_1d\vec{x}_2e^{i\phi(\delta x)}e^{-\frac{1}{2\sigma_B}\sum_i\left(
\vec{x}_i-\vec{X}_B -\vec{v}_B(t_i - T_B)
\right)^2}\frac{1}{4\pi}\delta(\lambda)\epsilon(\delta t)\nonumber\\
&\approx (\sigma_B\pi)^\frac{3}{2}\frac{\sigma_B}{2}\frac{\epsilon(\delta t)}{|\delta t|}e^{i\bar{\phi}_C(\delta t)}
\label{eq135}
\end{align}
for the leading singular part and
\begin{align}
 J_{1/\lambda} &= \int d\vec{x}_1d\vec{x}_2e^{i\phi(\delta x)}e^{-\frac{1}{2\sigma_B}\sum_i\left(
\vec{x}_i-\vec{X}_B -\vec{v}_B(t_i - T_B)
\right)^2}\frac{i}{4\pi^2\lambda}\nonumber\\
&\approx (\sigma_B\pi)^\frac{3}{2}\frac{\sigma_B}{2}\left(\frac{1}{\pi\sigma_B|\vec{p_B}|^2}\right)^\frac{1}{2}
e^{-\sigma_B|\vec{p}_B|^2}\frac{1}{|\delta t|}e^{i\bar{\phi}_C(\delta t)}
\label{eq136}
\end{align}
for the next term of the form $1/\lambda$.

Finally, we integrate $t_1$ and $t_2$ over the finite region $T=T_B - T_A$, and we have the slowly decreasing term
$\tilde{g}(\omega_B T)$,
\begin{align}
& i\int_0^Tdt_1dt_2\frac{\epsilon(\delta t)}{|\delta t|}e^{i\omega_B\delta t} = T(\tilde{g}(\omega_B T)-\pi),\label{eq137}\\
& \omega_B = E_B - |\vec{p}_B| = \frac{m_B^2}{2E_B},\nonumber
\end{align}
and the normal term $G_0$. $\tilde{g}(\omega_BT)$ is generated from the light-cone singularity and related term, satisfies
\begin{align}
 &\tilde{g}(0) = \pi,\label{eq138}\\
&\tilde{g}(\omega_BT)\to \frac{2}{\omega_BT};\ T\to\infty,\label{eq139}
\end{align}
and vanishes at $T=\infty$. $G_0$ is from the rest
\begin{align}
 G_0= 2\sqrt{\frac{\sigma_B}{\pi}}t\int \frac{d\vec{p}_C}{E(\vec{p}_C)}\delta(E_A-E_B-E_C(\vec{p}_C))
e^{-\sigma_B(\vec{p}_A-\vec{p}_B-\vec{p}_C)^2}\theta(E_A-E_C(\vec{p}_C)),\label{eq140}
\end{align}
approximately conserves the kinetic energy and momentum,
\begin{align}
 p_A-p_C = p_B,\label{eq141}
\end{align}
and gives the asymptotic value. Due to the rapid oscillation in $\delta t$, $G_0$ receives a contribution only from the 
microscopic $|\delta t|$ region and is constant in $T$. Integration of this term does not depend on $\sigma_B$ and agrees
with the normal probability obtained with the standard method of using plane waves. In the region 
$2p_A\cdot p_B > \tilde{m}^2_C$, $\Delta_{A,C}(\delta x)$ does not have the light-cone singularity and diffraction term 
exists only in the kinematical region $2p_A\cdot p_B\leq \tilde{m}_C^2$. 

We have 
\begin{align}
 \int \frac{d\vec{p}_C}{(2\pi)^3}|\mathcal{M}|^2 = \frac{N_3}{E_B}\left\{
F(-\tilde{m}^2)T\tilde{g}(\omega_BT) + F(\tilde{m}_B^2)G_0
\right\},\label{eq142}
\end{align}
where $N_3 = 4g^2\pi^3\sigma_BV^{-1}$. The form factor gives different corrections to the diffraction and normal terms.
They are evaluated later.

\textbf{B: Massless particle $m_B=0$}

For a massless $B$, the leading singularity $\delta(\lambda)\epsilon(\delta t)$ cancels on integrating over the times, $t_1$
and $t_2$, and the next term proportional to $1/\lambda$ gives a dominant contribution. The integral of this term
\begin{align}
 J_{1/\lambda} &= \int d\vec{x}_1d\vec{x}_2e^{i\phi(\delta x)}e^{-\frac{1}{2\sigma_B}\sum_i\left(
\vec{x}_i-\vec{X}_B -\vec{v}_B(t_i - T_B)
\right)^2}\frac{i}{4\pi^2\lambda}\nonumber\\
&\approx (\sigma_B\pi)^\frac{3}{2}\frac{\sigma_B}{2}\left(\frac{1}{\pi\sigma_B|\vec{p_B}|^2}\right)^\frac{1}{2}
e^{-\sigma_B|\vec{p}_B|^2}\frac{1}{|\delta t|}e^{i\bar{\phi}_C(\delta t)}
\label{eq143} 
\end{align}
leads to 
\begin{align}
 J_{1/\lambda} \approx (\sigma_B\pi)^\frac{3}{2}\frac{\sigma_B}{2}\left(\frac{1}{\pi\sigma_B|\vec{p}_B|^2}
\right)^\frac{1}{2}e^{-\sigma_B|\vec{p}_B|^2}\frac{1}{|\delta t|}.\label{eq144}
\end{align}
This term also has universal dependence on $|\delta t|$ and its integration over the times becomes
\begin{align}
 \int dt_1dt_2J_{1/\lambda} = (\sigma_B\pi)^\frac{3}{2}\frac{\sigma_B}{2}\left(
\frac{1}{\pi\sigma_B|\vec{p}_B|^2}\right)^\frac{1}{2}e^{-\sigma_B|\vec{p}_B|^2}
\int dt_1dt_2\frac{1}{|t_1-t_2|}.\label{eq145}
\end{align}
The integration over the times in a finite region from $\epsilon$ to $T$ is
\begin{align}
 \int_\epsilon^T dt_1dt_2 \frac{1}{|t_1-t_2|}=T\left(2\log\frac{T}{\epsilon} - 1\right)\label{eq146}
\end{align}
and
\begin{align}
 \int dt_1dt_2J_{1/\lambda} = (\sigma_B\pi)^\frac{3}{2}\frac{\sigma_B}{2}\left(
\frac{1}{\pi\sigma_B|\vec{p}_B|^2}\right)^\frac{1}{2}e^{-\sigma_B|\vec{p}_B|^2}
T\left(2\log\frac{T}{\epsilon} - 1\right).\label{eq147}
\end{align}
This term gives the probability 
\begin{align}
 P_\text{diffraction} = N_4\int d\vec{p}_Be^{-\sigma_B|\vec{p}_B|^2},\label{eq148}
\end{align}
where
\begin{align}
 N_4 = 8Tg^2\left(\frac{\sigma_B^2}{4}\right)\left(2\log\frac{T}{\epsilon} - 1\right).\label{eq149}
\end{align}

\textbf{Large time: $T>\tau_A$}

If $T$ is larger than the life-time of $A$, $\tau_A$, Eqs. \eqref{eq137} and \eqref{eq146} are replaced with
\begin{align}
& i\int_0^Td t_1dt_2\frac{\epsilon{\delta t}}{|\delta t|}e^{i\omega_B\delta t}e^{-\frac{t_1+t_2}{\tau_A}}
=\tilde{g}(\omega_B,T;\tau_A) - \tilde{g}(\omega_B,\infty;\tau_A),\label{eq150}
\end{align}
and
\begin{align}
 \int_\epsilon^Tdt_1dt_2 \frac{1}{|t_1-t_2|}e^{-\frac{t_1+t_2}{\tau_A}}\label{eq151}.
\end{align}
$N_4$ becomes approximately
\begin{align}
 N_4 = 8\tau_Ag^2\frac{\sigma_B^2}{4}\left(2\log\frac{\tau_A}{\epsilon}-1\right).\label{eq152}
\end{align}
Thus, the system of $\sigma_t=\infty$ has a finite-size correction of the form Eq. \eqref{eq142} for $T\ll\tau_A$, and
$T\tilde{g}(\omega_BT)$ in Eq. \eqref{eq142} is replaced with $\tilde{g}(\omega_B,T;\tau_A)$ in $T\approx \tau_A$.
The correction depends on $T$ in the universal manner and on the size of wave packet in magnitude. At $\sigma_B = \infty$,
the correction becomes infinite.

We compute the total probability next. From the integration over $\vec{X}_B$, the total volume $V$ is obtained and canceled 
with the normalization of $A$. The total probability thus becomes the integral of the sum of $G_0$ and 
$\tilde{g}(\omega_BT)$,
\begin{align}
 P=\begin{cases}
    N_3\int\frac{d\vec{p}_B}{(2\pi)^3E_B}\left[
    F(-\tilde{m}^2)T\tilde{g}(\omega_BT) + F(m_B^2)G_0
    \right],\ \text{for }T\ll\tau_A,\\
\ \\
    N_4\int\frac{d\vec{p}_B}{(2\pi)^3E_B}\left[
    F(-\tilde{m}^2)\tilde{g}(\omega_B,T;\tau_A)+F(m_B^2)G_0
\right],\ \text{for }\tau_A\leq T,
   \end{cases}\label{eq153}
\end{align}
The second terms, $P_\text{normal}$, on the right-hand sides of Eq. \eqref{eq153} are independent of $T$ and $\sigma_B$,
and agree with the standard value computed with the plane waves. $\tilde{g}(\omega_BT)$ and $\tilde{g}(\omega_B,T;\tau_A)$
in the first terms depend on $\omega_B$ and $T$, and are corrections due to the finite distance between the initial and
final states. The magnitudes of the first terms, $P_\text{diffraction}$, at $T\to \infty$ are proportional to
\begin{align}
 P_\text{diffraction} = \tilde{N}F(-\tilde{m}^2)\frac{\sigma_B}{\omega_BT} = \tilde{N}F(-\tilde{m}^2)\frac{\sigma_B E_B}
{m_B^2T}\label{eq154}
\end{align}
where $\tilde{N}$ is constant. $P_\text{diffraction}$ becomes significant for large $(\sigma_BE_B)/m_B^2$, i.e., small mass
or large wave packet.
\subsubsection{Infinite $\sigma_S$ and infinite $\sigma_t$}
When three particles are plane waves, $\sigma_S=\sigma_t = \infty$, the scattering amplitude and cross section are the 
standard ones if $H_\text{int}(t)e^{-\epsilon|t|}$ is used. The space-time coordinates $(t,\vec{x})$ are integrated over 
the whole region, and the energy and momentum are strictly conserved. The asymptotic values thus obtained with $S[\infty]$,
\begin{align}
& \mathcal{M} = (2\pi)^4g\delta^{(4)}(p_A-p_B-p_C)f,\label{eq155}\\
& P = g^2|f|^2\times (\text{phase space}),\label{eq156}
\end{align}
agree with the asymptotic values obtained with $S[T]$. If the convergence factor $e^{-\epsilon|t|}$ is absent, the limit
$T\to \infty $ is not unique and is consistent with the diverging correction in $\sigma_B\to \infty$ of the previous case.
\subsubsection{Coherence length}
The coherence length found from the amplitude of the initial and final states expressed with wave packets is finite. From
Eq. \eqref{eq91}, the integral in $\vec{x}$ converges for a finite $\sigma_S$ and that in $t$ converges for a finite 
$\sigma_t$. $\sigma_t$ becomes infinite with $\vec{v}_A = \vec{v}_B$ and $\sigma_C=\infty$, or 
$\sigma_A = \sigma_B = \infty$. In the latter case, the coherence length is $\hbar E_C/(m_C^2c^3)$.
\subsubsection{Asymmetric wave packets}
For asymmetric wave packets, the integral over $(t,\vec{x})$ is expressed by
\begin{align}
 \int dt\int d\vec{x}e^{-\frac{1}{2\sigma_S^L}\left(\vec{x}_L - \vec{x}_0^L\right)^2
-\frac{1}{2\sigma_S^T}\left(\vec{x}_T - \vec{x}_0^T\right)^2 - \frac{1}{2\sigma_t}\left(t-t_0\right)^2},\label{eq157}
\end{align}
where the sizes of the Gaussian exponents and other parameters are given by complicated expressions. Experiments on
$\delta E\ll|\delta\vec{p}|$ are studied with asymmetric wave packets.

\section{Emission and absorption of light}
Radiative transitions of particles
\begin{align}
 \begin{cases}
  A\to C+\gamma,\\
A + \gamma \to C
 \end{cases}
\label{eq158}
\end{align}
expressed with wave packets are studied in various parameter regions. Electromagnetic interaction is expressed with
\begin{align}
 H_\text{int} = e\int d\vec{x}J_\mu(x)\mathcal{A}^\mu(x),\label{eq159}
\end{align}
where $\mathcal{A}_\mu(x)$ is the photon field and $J_\mu(x)$ is the electromagnetic current. The matrix element of the
current between eigenstates of energy and momentum is written as
\begin{align}
 \langle C;p_C|J_\mu(x)|A;p_A\rangle = e^{i(p_A-p_C)\cdot x}\langle C;p_C|J_\mu(0)|A;p_A\rangle,\label{eq160}
\end{align}
where 
\begin{align}
 \langle C;p_C|J_\mu(0)|A;p_A\rangle = \Gamma_\mu F((p_A-p_C)^2),\label{eq161}
\end{align}
with the form factor $F((p_A-p_C)^2)$ and the spin-dependent factor $\Gamma_\mu$. We assume one form factor for simplicity,
but it is straightforward to extend to a case with many form factors. In the normal term of the radiative transition, the 
energy momentum is conserved, and 
\begin{align}
 F((p_A-p_C)^2) = F(k_\gamma^2)=F(0),\label{eq162}
\end{align}
hence the coupling strength is determined by $F(0)$.

In detectors, the fundamental processes of a photon are the photo-electric effect, Compton effect, or $e^+e^-$ pair 
production. The wave packet sizes of the photon, $\sigma_\gamma$, are nuclear sizes for pair production due to the 
nuclear electric field, or atomic sizes or larger for the photo-electric and Compton effects, depending on the energy.

\subsection{Universal background}
The transition probabilities of radiative processes receive finite-size corrections under certain situations and their
energy spectra are modified by pseudo-Doppler effects. Since the finite-size correction is caused by states that violate
the conservation law of kinetic energy and momentum, the corresponding events look like backgrounds even though they are
produced dynamically. They have universal properties and magnitudes that depend on the experimental apparatus.

\subsubsection{Universal background}
The universal background derived from the finite-size correction resulting from
\begin{align}
 \left|\frac{e^{i\omega T}-1}{\omega} - 2\pi\delta(\omega)\right|\neq 0 \label{eq163}
\end{align}
 is an inevitable consequence of the Schr\"{o}dinger equation. Since it is generated by states with kinetic energies
different from that of the initial state, it is positive semi-definite from Eq. \eqref{eq35} and is added to the 
normal component in the wave zone. Its magnitude is computed rigorously in relativistic systems, Eq. \eqref{eq153}.
The correction vanishes in the particle zone. The energy spectrum for wave packets is distorted in both the particle
and wave zones due to the pseudo-Doppler effects, even though the total probability agrees with the normal value.

\subsubsection{Form factor}
The nucleus, atom, and molecule are composite states and have internal structures. Therefore, they have finite extensions
and interact with photons or neutrinos non-locally. This non-locality is negligible if the size $R$ and the photon momentum
$k_\gamma$ satisfy $k_\gamma R \ll 1$, where multi-pole expansions are applicable.

For X-rays of atoms, they are approximately 
\begin{align}
 k_\gamma R = 10^{-3}; \ k_\gamma \sim \text{keV},\ R=10^{-11}\text{m},\label{eq164}
\end{align}
and for transitions of the nucleus
\begin{align}
 k_\gamma R = 10^{-1};\ k_\gamma\sim\text{MeV},\ R = 10^{-15}\text{m}.\label{eq165}
\end{align}
Since $k_\gamma R$ is small,
\begin{align}
 F((p_A-p_C)^2)=F(0).\label{eq166}
\end{align}

\subsubsection{Life-time effect}
If the parent $A$ has a finite life-time, $\tau_A$, it modifies the results. In a region
\begin{align}
 c\tau_A \leq \sqrt{\sigma_A},\label{eq167}
\end{align}
the integral over the times in the transition probability receives a dominant contribution from the region
\begin{align}
 t\leq \tau_A.\label{eq168}
\end{align}
Then the effect of the wave packet is diminished and the pseudo-Doppler effect becomes negligible. If the life-time 
satisfies
\begin{align}
 c\tau_A\geq \sqrt{\sigma_A},\label{eq169}
\end{align}
the integral over the times in the transition probability receives a dominant contribution from the region
\begin{align}
 t\leq \frac{\sqrt{\sigma_A}}{c},\label{eq170}
\end{align}
and the pseudo-Doppler effect is prominent.
\subsubsection{Photon effective mass}
A photon is massless in vacuum but its properties are modified in matter due to the dielectric constant. In high-energy
regions, the refraction constant behaves with frequency as
\begin{align}
 n = 1 - \frac{\omega_p^2}{\omega},\label{eq171}
\end{align}
where $\omega_p$ is the plasma frequency and is given as
\begin{align}
 \omega_p = \frac{NZe^2}{\epsilon_0 m_e};\label{eq172}
\end{align}
it depends on the material, density, and other parameters. The wave vector satisfies
\begin{align}
 (ck)^2 = \omega^2 - \omega_p^2.\label{eq173}
\end{align}
and the energy dispersion becomes 
\begin{align}
 \vec{p}^{\,2} = E^2(\vec{p}) - (\hbar\omega_p)^2.\label{eq174}
\end{align}
Thus the photon has an effective mass
\begin{align}
 m_\text{eff} = \hbar\sqrt{\frac{NZe^2}{\epsilon_0m_e}},\label{eq175}
\end{align}
where $N$ and $Z$ are the number density and atomic number of the gas, and $m_e$ is the electron's  mass. $m_\text{eff}$ 
depends upon the density of matter and is variable. A high-energy photon behaves like a massive particle.
\subsubsection{Light-cone singularity for general systems}
For particles $A$ and $C$ of internal structures, Eq. \eqref{eq161} is substituted for Eq. \eqref{eq131}. As is shown in
Appendix C, the singular part of the correlation function is written in the form
\begin{align}
 \Delta_{A,C}^{\text{light-cone}}(\delta x) = F(-m_A^2+m_C^2)\Delta_{A,C}^{(0),\text{light-cone}}(\delta x),\label{eq176}
\end{align}
where $\Delta_{A,C}^{(0)}(\delta x)$ is that of the point particle. Thus the form factor
\begin{align}
 F(m_C^2-m_A^2) \label{eq177}
\end{align}
determines the strength of the singularity and is given in Appendix C as
\begin{align}
 F(m_C^2-m_A^2)/F(0) = \begin{cases}
			O(1);\ \text{hadron, positronium, light nucleus,}\\
			O(10^{-1});\ \mu N\text{ atom, heavy nucleus,}\\
			O(10^{-5});\ \mu e,\text{ K-electron,}\\
			O(10^{-10})\ \text{atom}
		       \end{cases}\label{eq178}
\end{align}
Thus the form factors do not modify the magnitude of the light-cone singularity for hadrons, light nuclei, or positronium, 
and reduce to $1/10$ for the $\mu N$ atom and heavy molecules. For $\mu e$, the K-electron, and atoms, the magnitudes
 become extremely small. Equation \eqref{eq177} is almost the same as on-shell coupling, Eq. \eqref{eq162}, in the former
 but much smaller in the latter. The singularity is caused by waves of translational motion, which retain their relativistic
invariance even for particles with internal structure, but the magnitude depends on their sizes.

\subsection{Emission of light}
\subsubsection{Decay in flight in vacuum}
\textbf{1. Finite $\sigma_A$ and $\sigma_\gamma$: pseudo-Doppler effect}\\
The amplitudes of the momenta, positions, and wave packet sizes for the radiative decay of $A$ to $C$ and a photon $\gamma$,
\begin{align}
 A:(\vec{X}_A,E_A,\vec{p}_A,\sigma_A),\nonumber\\
\gamma:(\vec{X}_\gamma,E_\gamma,\vec{p}_\gamma,\sigma_\gamma),\nonumber\\
C:(\vec{p}_C,E_C,\sigma_C=\infty),\label{eq179}
\end{align}
is expressed with the matrix element of the current operator and the photon field:
\begin{align}
 \mathcal{M} = &\int d^4x \langle C|J_\mu(x)|A\rangle\langle\gamma|\mathcal{A}^\mu(x)|0\rangle\nonumber\\
= &\int d^4xe^{i(p_A - p_C - p_\gamma)\cdot x} F_{AB}e^{ip_\gamma\cdot X_\gamma - \frac{1}{2\sigma_\gamma}
\left(\vec{x}- \vec{X}_\gamma - \vec{v}_\gamma (t-T_\gamma )\right)^2}\nonumber\\
 &\times e^{-ip_A\cdot X_A - \frac{1}{2\sigma_A}\left(\vec{x} - \vec{X}_A - \vec{v}_A(t- T_A)\right)^2}\nonumber\\
=&e^{R+i\phi},\label{eq180}\\
F_{AB} = & \langle C|J_\mu(0)|A\rangle \epsilon^\mu(\vec{p}_\gamma),\nonumber
\end{align}
where $\epsilon^\mu(\vec{p}_\gamma)$ is the polarization vector of the photon. We have $|\mathcal{M}|^2$ in the form
\begin{align}
 |\mathcal{M}|^2 = &N^2\int d^4x_1d^4x_2 e^{i(p_A-p_C-p_\gamma)\cdot(x_1-x_2) - \frac{1}{2\sigma_\gamma}
\sum_i\left(\vec{x}_i - \vec{X}_\gamma - \vec{v}_\gamma(t_i - T_\gamma)\right)^2}\nonumber\\
&\times e^{-\frac{1}{2\sigma_A}\sum_i\left(\vec{x}_i - \vec{X}_A - \vec{v}_A(t_i - T_A)\right)^2}W_{i,j}(p_A,p_C)
\left(\delta^{i,j} - \frac{p_\gamma^ip_\gamma^j}{\vec{p}_\gamma^{\,2}}\right),\label{eq181}
\end{align}
in the Coulomb gauge,
\begin{align}
 \mathcal{A}_0(x) = 0, \ \vec{\nabla}\cdot\vec{\mathcal{A}}=0,\label{eq182}
\end{align}
where $N$ is the normalization factor, and
\begin{align}
& W_{i,j}(p_A,p_C) = \langle C|J_i(0)|A\rangle(\langle C|J_j(0)|A\rangle )^*,\label{eq183}\\
&\left(\delta^{i,j} - \frac{p_\gamma^ip_\gamma^j}{\vec{p}_\gamma^{\,2}}\right) = \sum\epsilon^i(\vec{k}_\gamma)
(\epsilon^j(\vec{k}_\gamma))^{*}.\label{eq184}
\end{align}

$R$ in Eq. \eqref{eq180} is composed of the momentum-dependent part $R_\text{momentum}$ and the coordinate-dependent
part $R_\text{trajectory}$. The former is
\begin{align}
 R_\text{momentum} &= -\frac{\sigma_t}{2}\left(E_A(\vec{p}_A) - E_C(\vec{p}_C) - E_\gamma(\tilde{\vec{p}}_\gamma)\right)^2
-\frac{\sigma_S}{2}\left(\vec{p}_A - \vec{p}_C - \vec{p}_\gamma\right)^2,\nonumber\\
\tilde{\vec{p}}_\gamma &= \vec{p}_\gamma + \frac{\sigma_S}{\sigma_\gamma}\left(\vec{p}_A-\vec{p}_C-\vec{p}_\gamma\right),
\label{eq185}
\end{align}
where $\sigma_S$, $\sigma_t$, and $\vec{v}_0$ are
\begin{align}
 \sigma_S &= \frac{\sigma_A\sigma_\gamma}{\sigma_A+\sigma_\gamma},\label{eq186}\\
\vec{v}_0 &= \frac{\sigma_A}{\sigma_A+\sigma_\gamma}\vec{v}_\gamma,\label{eq187}\\
\frac{1}{\sigma_t}&=\frac{\vec{v}_\gamma^{\,2}}{\sigma_\gamma} - \frac{\vec{v}_0^{\,2}}{\sigma_S} 
= \frac{\vec{v}_\gamma^{\,2}}{\sigma_A+\sigma_\gamma}.\label{eq188}
\end{align}
Thus, the energy momentum satisfies the modified conservation law. The momentum is conserved approximately around the center
$\delta \vec{p}=0$, whereas the photon's energy at the momentum $\tilde{\vec{p}}_\gamma$ fulfills the approximate 
conservation law. the implications of this will be studied in detail shortly.

The position-dependent exponent is written in the form
\begin{align}
 R_\text{trajectory} &= -\frac{\vec{X}_A^{\,2}}{2\sigma_A} - \frac{\tilde{\vec{X}}_\gamma^{\,2}}{2\sigma_\gamma}
+2\sigma_S\left(\frac{\vec{X}_A}{2\sigma_A} + \frac{\tilde{\vec{X}}_\gamma}{2\sigma_\gamma}\right)^2
+ 2\sigma_t\left(\frac{\vec{v}_0\cdot\vec{X}_A}{2\sigma_A} + \frac{(\vec{v}_0-\vec{v}_\gamma)\cdot\tilde{\vec{X}}_\gamma}
{2\sigma_\gamma}
\right)\nonumber\\
&=-\frac{1}{2(\sigma_A+\sigma_\gamma)}\left[\left(\vec{X}_A - \tilde{\vec{X}}_\gamma\right)^2
-\frac{1}{\vec{v}_\gamma^{\,2}}\left(\vec{v}_\gamma\cdot(\vec{X}_A-\tilde{\vec{X}}_A)\right)^2\right].\label{eq189}
\end{align}
The probability is expressed as 
\begin{align}
 P = &N^2e^{2R_\text{momentum} + 2R_\text{trajectory}}|F_{A,B}|^2\nonumber\\
=&N^2|F_{A,B}|^2e^{-\sigma_t(E_A-E_C-E_\gamma(\tilde{\vec{p}}_\gamma))^2 - \sigma_S(\vec{p}_A-\vec{p}_C-\vec{p}_\gamma)^2}
\nonumber\\
&\times e^{-\frac{(\vec{X}_A-\tilde{\vec{X}}_\gamma)_T^2}{\sigma_A+\sigma_\gamma}},\label{eq190}
\end{align}
and has no finite-size correction. Thus the total probability agrees with that of plane waves. Nevertheless, the energy
spectrum of Eq. \eqref{eq190} is distorted due to the pseudo-Doppler effect. The photon's momentum is distributed around a 
center $\vec{p}_A - \vec{p}_C$ and the photon's energy at the momentum $\tilde{\vec{p}}_\gamma$ is distributed around
$E_A-E_C$. If $\sigma_S$ is small and $\sigma_t$ is large, the momentum distribution is wide but the energy 
$E(\tilde{\vec{p}}_\gamma)$ almost coincides with $E_A-E_C$. The observed photon's energy is $E_\gamma(\vec{p}_\gamma)$
and is given from Eq. \eqref{eq185}:
\begin{align}
 E_\gamma(\vec{p}_\gamma) &= E_\gamma(\tilde{\vec{p}}_\gamma - \frac{\sigma_S}{\sigma_\gamma}(\vec{p}_A - \vec{p}_C
-\vec{p}_\gamma))\nonumber\\
&=E_A - E_C - \frac{\sigma_S}{\sigma_\gamma}\vec{v}_\gamma\cdot(\vec{p}_A-\vec{p}_C-\vec{p}_\gamma).\label{eq191}
\end{align}
Thus $E_\gamma(\vec{p}_\gamma)$ is very different from $E_A - E_C$.

The photon is on the mass shell and satisfies
\begin{align}
 E^2(\vec{p}_\gamma) - \vec{p}_\gamma^{\,2} = 0. \label{eq192}
\end{align}
In an event where the energy momenta $(E_A,\vec{p}_A)$, $(E_C,\vec{p}_C)$, and $(E_\gamma,\vec{p}_\gamma)$ are measured,
and momenta satisfy
\begin{align}
 \vec{p}_\gamma \neq \vec{p}_A-\vec{p}_C,\label{eq193}
\end{align}
the photon's energy at the momentum $\tilde{\vec{p}}_\gamma$ satisfies
\begin{align}
 E_A - E_C = E_\gamma(\tilde{\vec{p}}_\gamma).\label{eq194}
\end{align}
Consequently, the mass shell condition at $\tilde{\vec{p}}_\gamma$,
\begin{align}
 E_\gamma^2(\tilde{\vec{p}}_\gamma) - \tilde{\vec{p}}_\gamma^{\,2} = (E_A - E_C)^2 - \tilde{\vec{p}}_\gamma^{\,2} = 0,
\label{eq195}
\end{align}
is satisfied. Substituting $\tilde{\vec{p}}_\gamma$, we have
\begin{align}
 (E_A - E_C)^2 - \left(\vec{p}_\gamma + \frac{\sigma_S}{\sigma_\gamma}(\vec{p}_A-\vec{p}_C-\vec{p}_\gamma)\right)^2=0,
\label{eq196}
\end{align}
which gives the relation between the energies and momenta with the ratio $\sigma_S/\sigma_\gamma$. Measuring the energies
and momenta, the ratio $\sigma_S/\sigma_\gamma$ will be determined.

In a situation with
\begin{align}
 \sigma_\gamma\ll\sigma_A,\label{eq197}
\end{align}
we have 
\begin{align}
& \sigma_S = \sigma_\gamma,\ \sigma_t = \sigma_A,\ \sigma_S\ll\sigma_t,\label{eq198}\\
&\vec{v}_0 = \frac{\sigma_A}{\sigma_A + \sigma_\gamma}\vec{v}_\gamma,\ \tilde{\vec{p}}_\gamma = \frac{\sigma_\gamma}
{\sigma_A+\sigma_\gamma}\vec{p}_\gamma + \vec{p}_A-\vec{p}_C.\label{eq199}
\end{align}
The central values of energies and momenta satisfy
\begin{align}
 \langle E_\gamma(\tilde{\vec{p}}_\gamma)\rangle &= \langle E_A-E_C\rangle,\label{eq200}\\
\langle \vec{p}_\gamma\rangle &= \langle \vec{p}_A - \vec{p}_C\rangle\label{eq201},
\end{align}
with variations 
\begin{align}
 \delta E&=\frac{1}{\sqrt{\sigma_t}},\label{eq202}\\
|\delta\vec{p}|&=\frac{1}{\sqrt{\sigma_S}}.\label{eq203}
\end{align}
The energy spreading is narrower than the momentum spreading,
\begin{align}
 \delta E \ll |\delta \vec{p}\,|,\label{eq204}
\end{align}
hence the constraint to the energy is more stringent than that of the momentum.

\textbf{Heavy $A$ and $C$ (pseudo-Doppler effect combined with M\"{o}ssbauer effect)}

If $C$ and $A$ are a ground state and an excited state of a heavy atom, which are bound together to become massive
objects, the correlation function of Eq. \eqref{eq183} does not only vanish at the same momenta,
\begin{align}
 \vec{p}_A = \vec{p}_C,\label{eq205}
\end{align}
like those of the M\"{o}ssbauer effect. We study the photon's energy spectrum when this condition is satisfied in a large
wave packet $\sigma_A = \sigma_C$. The reduced momentum becomes
\begin{align}
 \tilde{\vec{p}}_\gamma = \frac{\sigma_\gamma}{\sigma_A + \sigma_\gamma}\vec{p}_\gamma.\label{eq206}
\end{align}
from Eq. \eqref{eq198}. The energy of the massless particle is proportional to the momentum and 
\begin{align}
 \langle E_\gamma(\tilde{\vec{p}}_\gamma)\rangle = \frac{\sigma_\gamma}{\sigma_A+\sigma_\gamma}\langle 
E_\gamma(\vec{p}_\gamma)\rangle.\label{eq207}
\end{align}
Substituting Eq. \eqref{eq200}, we have the expectation value of $E_\gamma$:
\begin{align}
 \langle E_\gamma(\vec{p}_\gamma)\rangle &= \kappa \Delta E_\text{electron},\label{eq208}\\
\kappa = \frac{\sigma_A + \sigma_\gamma}{\sigma_\gamma},\ \Delta E_\text{electron} = E_A - E_C,\label{eq209}
\end{align}
which is much larger than the energy difference $E_A - E_C$. Thus the product of average energy with the time interval
for the photon is equal to that for the atom:
\begin{align}
 \sigma_\gamma\langle E_\gamma(\vec{p}_\gamma)\rangle = (\sigma_A + \sigma_\gamma)(E_A-E_C).\label{eq210}
\end{align}
Now, $\sigma_\gamma$ is the size of the particle with which the photon interacts and $\sigma_A$ is that of the atom;
they are proportional to the average-time intervals of their reactions. Thus the conservation law Eq. \eqref{eq62}
for the energy is satisfied for the average value. This unusual phenomenon occurs because the electromagnetic interaction
tales place in a narrow space-time region where the wave functions of $A$, $C$ and the photon overlap. When $\sigma_\gamma$
is much smaller than $\sigma_A$, the region has the area $\sigma_\gamma$ and also moves with the velocity $\vec{v}_\gamma$.
Hence the energy is conserved in this moving frame where the photon has the effective energy 
$E_\gamma(\tilde{\vec{p}}_\gamma)$, which is much smaller than $E_\gamma(\vec{p}_\gamma)$. Hence the average energy of
$\gamma$ becomes much larger than the energy difference between $A$ and $C$. This is the pseudo-Doppler effect caused by
the wave packets. 

The condition Eq. \eqref{eq197} is fulfilled in various situations. A molecule in a gas propagates almost freely and an atom
is bound in a solid. The wave packet size of a molecule in a gas is given by the square of the mean free path and is of the
order of $10^{-14}$ m${}^2$, whereas that is the atomic distance in solid of the order of $10^{-20}$ m${}^2$. Hence we
have 
\begin{align}
 \kappa = \frac{\sigma_A}{\sigma_\gamma} = 10^6.\label{eq211}
\end{align}
Consequently, the photon in this situation interacts with the atom in a solid with the energy 
$\kappa\Delta E_\text{electron}$. for $\Delta E_\text{electron} = 0.1$ eV, $E_\gamma$ can be as large as 100 keV. Some
anomalous X-ray or $\gamma$-ray luminescence \cite{42,43,44,45,46,47,48} may be connected with this energy enhancement.

For a photon produced from an excited atom in a solid and interacting with a nucleus in a solid, we have $10^{-20}$ m${}^2$
for the former size and $10^{-28}$ m${}^2$ for the latter size, and 
\begin{align}
 \kappa = \frac{\sigma_A}{\sigma_\gamma} \approx O(10^8).\label{eq2120}
\end{align} 
Consequently, the photon produced from excited atoms interacts with a nucleus with much larger energy than the energy 
difference $\Delta E_\text{electron} =E_A-E_C$. Because the photon-nucleus cross section is much smaller than that of
the photon-atom scattering, the probability of this event is extremely small.

A similar phenomenon is expected when charged particles propagate in a magnetic field. A plane wave with charge $q$ and mass
$M$ in the magnetic field $\vec{\mathcal{B}}$,
\begin{align}
 e^{-i(E(\vec{p}_0) +\frac{q\vec{x}\times\vec{\mathcal{B}}}{2M}\cdot\vec{p}_0)(t-T_0) + i\vec{p}_0\cdot\vec{x} },
\label{eq213}
\end{align}
has a phase proportional to the cyclotron frequency
\begin{align}
 \omega = \frac{q|\vec{\mathcal{B}}|}{M}.\label{eq214}
\end{align}
These waves behave like plane waves in a time region less than $T_i = \frac{2\pi}{\omega_i}$. $T_i$ for the electron and 
proton is 
\begin{align}
 T_i = \frac{2\pi}{\omega_i}=\frac{M_i}{q|\vec{\mathcal{B}}|},\ i=e,\ p.\label{eq215} 
\end{align}
Thus the waves have different sizes, the ratio of which is
\begin{align}
 \frac{T_e}{T_p} = \frac{m_e}{m_p} = \frac{1}{2000}.\label{eq216}
\end{align}
Thus, the photon emitted from the atom interacting with the electron in a magnetic field can reveal the same energy 
enhancement.

The anomalous enhancement of the photon's energy results from the overlap of wave functions of different sizes. This 
occurs when the photon's wave packets, which are the sizes of the wave functions with which the photons interact, are 
much smaller than the parent's wave functions, Hence, the rate of these events may be quite low.

\textbf{2. Infinite $\sigma_A$ and finite $\sigma_\gamma$: finite-size correction}

The amplitude of the momenta, positions, and wave packet sizes of the radiative decay of $A$ to $C$ of plane waves and a 
$\gamma$,
\begin{align}
& A:\ (\vec{X}_A,E_A,\vec{p}_A,\sigma_A=\infty),\nonumber\\
&\gamma:\ (\vec{X}_\gamma,E_\gamma,\vec{p}_\gamma,\sigma_\gamma),\ E_\gamma^2-\vec{p}_\gamma^{\,2}=0,\nonumber\\
&C:\ (\vec{p}_C,E_C,\sigma_C=\infty)\label{eq217}
\end{align}
 is expressed with the matrix element of the current operator and the photon field:
\begin{align}
 \mathcal{M} &=e\int d^4x \langle C|J_\mu(x)|A\rangle\langle \gamma|\mathcal{A}^\mu(x)|0\rangle\nonumber\\
&=e\int d^4xe^{i(p_A-p_C-p_\gamma)\cdot x}\langle C|J_\mu(0)|A\rangle \epsilon^\mu(\vec{p}_\gamma)
e^{i{p}_\gamma\cdot X_\gamma - \frac{1}{2\sigma_\gamma}\left(\vec{x}-\vec{X}_\gamma -\vec{v}_\gamma(t-T_\gamma)\right)^2}
\nonumber\\
&=e^{R+i\phi}.\label{eq218}
\end{align}
We have $|\mathcal{M}|^2$ in the form 
\begin{align}
 |\mathcal{M}|^2 =&N^2\int d^4x_1d^4x_2e^{i(p_A-p_C-p_\gamma)\cdot(x_1-x_2)-\frac{1}{2\sigma_\gamma}
\sum_i(\vec{x}_i-\vec{X}_\gamma -\vec{v}_\gamma(t_i-T_\gamma))^2}\nonumber\\
&\times W_{i,j}(p_A,p_C)\left(\delta^{i,j}-\frac{p_\gamma^ip_\gamma^j}{\vec{p}_\gamma^{\,2}}\right).\label{eq219}
\end{align}
Integrating over $\vec{p}_C$ with a variable $r={p}_A-p_C$, we have
\begin{align}
 &\int \frac{d\vec{p}_C}{(2\pi)^3E_C}e^{i(p_A-p_C)\cdot(x_1-x_2)}W_{i,j}(p_A-p_C)\nonumber\\
&= \int \frac{d^4r}{(2\pi)^3}\text{Im}\left[
\frac{1}{r^2-2p_A\cdot r + m_A^2 - m_C^2 - i\epsilon}
\right]
W_{i,j}(p_A,p_A-r)e^{ir\cdot(x_1-x_2)},\label{eq220}
\end{align}
which has the light-cone singularity
\begin{align}
 \frac{i}{2\pi}\delta(\lambda)\epsilon(t_1-t_2)W_{i,j}(p_A,p_A-r)|_{r^2=m_C^2-m_A^2},\label{eq221}
\end{align}
from the integration over the momentum $r=(r^0,\vec{r}\,)$ of the region
\begin{align}
 (r^0)^2-\vec{r}^{\,2} = m_C^2-m_A^2 <0,\ r^0\leq 0.\label{eq222}
\end{align}
It is noted that $|(p_A-p_C)^2| = |m_C^2-m_A^2|$ is small and $|W_{i,j}(p_A,p_A-r)|$ is almost same as the on-shell matrix
element of the radiative transition. Equation \eqref{eq220} also has regular terms; one of them is generated from the 
above kinematical region and the others are from the region $0\leq r^0\leq p_A^0$. The latter coincides with the normal
term of the decay probability. Thus we have 
\begin{align}
 \int \frac{d\vec{p}_C}{E_C} |\mathcal{M}|^2 &= P_\text{normal} + P_\text{diffraction},\label{eq223}\\
P_\text{diffraction } &= \mathcal{C}\tilde{g}(\omega_\gamma T),\ \omega_\gamma = \frac{m_\text{eff}^2}{2E_\gamma},\nonumber
\end{align}
where $\mathcal{C}$ is determined by the wave packet size. From the convergence condition in the expansion 
Eq. \eqref{eq220}, the light-cone singularity exists in the momentum region 
\begin{align}
 2p_A\cdot p_\gamma \leq m_A^2-m_C^2.\label{eq224}
\end{align}

\subsubsection{Decay at rest in a solid}
A decay of $A$ in a solid to $C$ and a photon, $\gamma$, which have the following momenta, positions, and wave packet sizes:
\begin{align}
 A:\ (\vec{X}_A,E_A,\vec{p}_A=0,\sigma_A),\nonumber\\
\gamma:\ (\vec{X}_\gamma,E_\gamma,\vec{p}_\gamma,\sigma_\gamma=\infty),\nonumber\\
C:\ (\vec{X}_A,\vec{p}_C=0,E_C,\sigma_A)\label{eq225}
\end{align}
is a kinematical region of the M\"{o}ssbauer effect. The amplitude
\begin{align}
 \mathcal{M}(A\to C+\gamma) = g\int d^4xw_A(\vec{p}_A=0)w_C(\vec{p}_C=0)e^{ip_\gamma\cdot x},\label{eq226}
\end{align}
where
\begin{align}
 w_A &= N_A\left(\frac{2\pi}{\sigma_A}\right)^\frac{3}{2}e^{-\frac{1}{2\sigma_A}\left(\vec{x}-\vec{x}_0^A\right)^2
-i\phi_0^A},\label{eq227}\\
w_C &=N_C\left(\frac{2\pi}{\sigma_A}\right)^\frac{3}{2}e^{-\frac{1}{2\sigma_A}\left(\vec{x}-\vec{x}_0^C\right)^2-i\phi_0^C},
\label{eq228}\\
\vec{x}^A_0 &= \vec{X}_A,\ \phi_0^A = m_A(t-T_A),\ \vec{x}^C_0 = \vec{X}_A,\ \phi_0^C = m_C(t-T_C)\nonumber
\end{align}
is given as
\begin{align}
 \mathcal{M}(A\to C + \gamma) &=Ne^{im_AT_A-m_CT_C}\int_0^Tdte^{-i(m_A-m_C-E_\gamma)t}\int d\vec{x}e^{-\frac{1}{\sigma_A}
(\vec{x}-\vec{X}_A)^2+i\vec{p}_\gamma\cdot\vec{x}}\nonumber\\
&=Ne^{i\Phi_0}\frac{2\sin[(m_A-m_C-E_\gamma)T/2]}{m_A-m_C-E_\gamma}e^{-\frac{\sigma_A}{4}\vec{p}_\gamma^{\,2}}.\label{eq229}
\end{align}
In the above equation, $N$ and $\Phi_0$ are constants. The square of the modulus of $\mathcal{M}$ is expressed in the form
\begin{align}
 \int_0^T dt_1dt_2\frac{d\vec{p}_\gamma}{E_\gamma}e^{-i(m_A-m_C-E_\gamma)(t_1-t_2)}
e^{-\frac{\sigma_A}{2}\vec{p}_\gamma^{\,2}},\label{eq230}
\end{align}
where the integral 
\begin{align}
 \int \frac{d\vec{p}_\gamma}{E_\gamma} e^{-(m_A-m_C-E_\gamma)(t_1-t_2)}e^{-\frac{\sigma_A}{2}\vec{p}_\gamma^{\,2}}
\label{eq231}
\end{align}
is a smooth and short-range function of $t_1-t_2$. Hence, the total probability is proportional to $T$ and has no
finite-size correction.

Particles in a liquid are also described with wave packets and the probabilities of their reactions are studied in the same
way.
\subsubsection{Decay in flight in a dilute gas}
A photon has an effective mass in the X-ray or $\gamma$-ray region in a dilute gas and the rate is modified by the large
finite-size correction. The radiative decay of $A$ in flight in a gas to $C$ in flight and a photon, $\gamma$, which have
the following momenta, positions, and wave packet sizes:
\begin{align}
& A:\ (E_A,\vec{p}_A,\sigma_A=\infty),\nonumber\\
&\gamma :\ (\vec{X}_\gamma,E_\gamma,\vec{p}_\gamma,\sigma_\gamma),\nonumber\\
&C:\ (E_C,\vec{p}_C,\sigma_C=\infty),\label{eq232}
\end{align}
is studied in a similar manner. Since $\sigma_A = \sigma_C = \infty$, the amplitude is expressed in the form of
 Eq. \eqref{eq218} with the effective mass of the high-energy photon in the X-ray or $\gamma$-ray regions, 
Eq. \eqref{eq175}. The probability of detecting this photon is given in Eq. \eqref{eq153} for the finite-size correction.
The frequency that determines the finite-size correction for this photon with energy $E_\gamma$ is
\begin{align}
 \omega=\frac{m_\text{eff}^2}{2E_\gamma},\label{Eq233}
\end{align}
which gives a macroscopic distance.

\subsection{Absorption}
The absorption of $\gamma$ is studied in a similar manner to the decay process. The changes in $A$, $C$, and $\gamma$ in
terms of the parameters
\begin{align}
& A:\ (\vec{X}_A,E_A,\vec{p}_A,\sigma_A),\nonumber\\
&\gamma:\ (\vec{X}_\gamma,-E_\gamma,-\vec{p}_\gamma,\sigma_\gamma),\nonumber\\
&C:\ (\vec{p}_C,E_C,\sigma_C),\label{eq234}
\end{align}
is described by replacing the sign of the photon's momentum in the previous amplitudes, Eq. \eqref{eq180} or 
Eq. \eqref{eq218}. The distribution function deviates and the central value of the photon's energy 
$E_\gamma(\vec{p}_\gamma)$ becomes different from $E_A-E_C$ with the pseudo-Doppler effect, and the probability receives 
large finite-size corrections in certain parameter regions.
\begin{figure}[t]\centering{
 \includegraphics[angle=-90,scale=0.5]{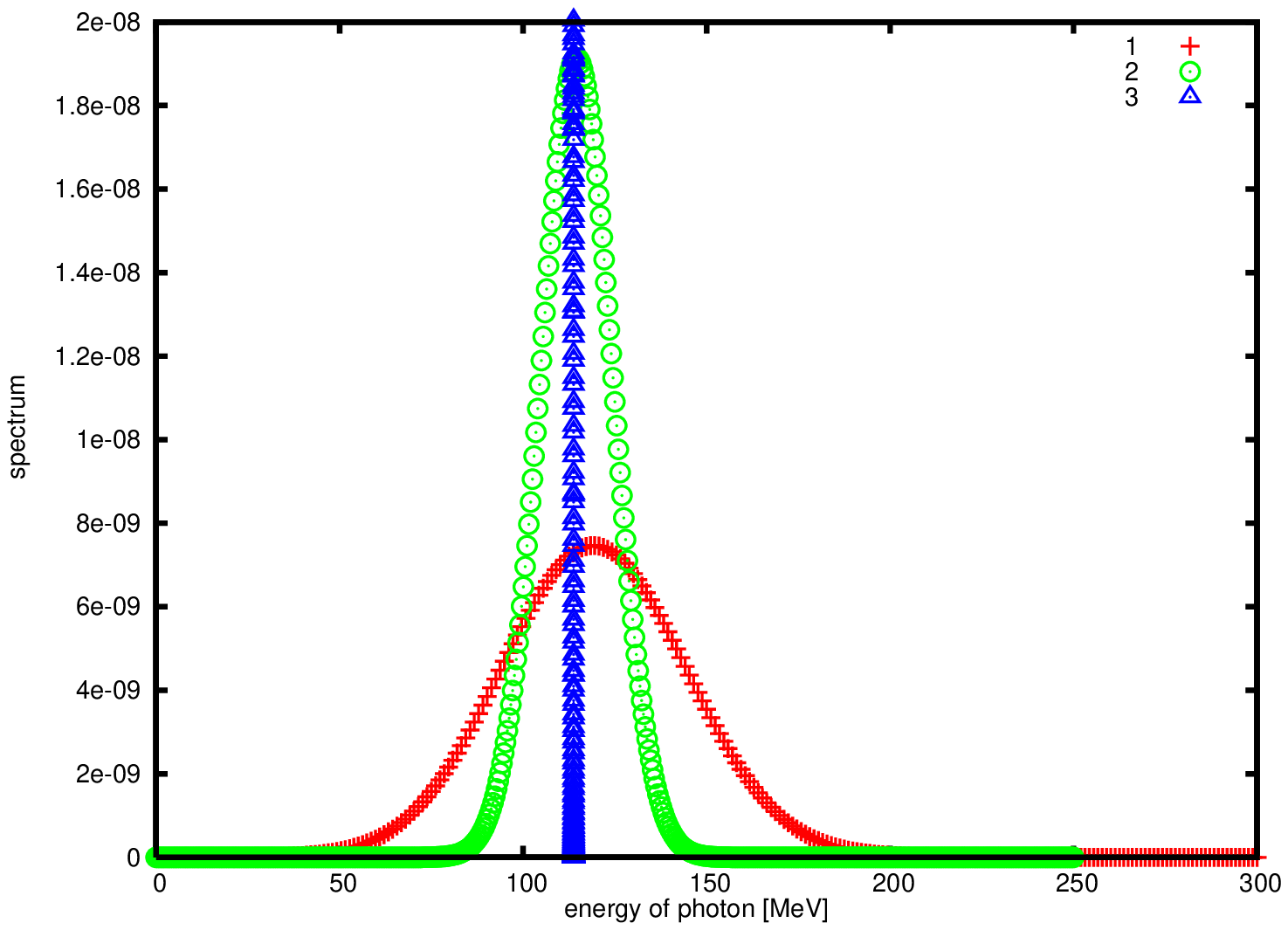}
\caption{The energy spectrum of $\gamma$ in $J/\Psi$ decay at rest to $M(2981)$ and $\gamma$ at $T=\infty$ is shown. The 
horizontal axis shows the energy of $\gamma$ in MeV for $\sigma_\gamma m_\pi^2 =14.6$ (red crosses), 100 (green circles),
and $\infty$ (blue triangles), and the vertical axis shows the probability. Wave packets of another daughter and parent are
$\infty$ and $\sigma_\text{parent}m_\pi^2=10000$. The probability is the same for a wide region of the parent's wave 
packets. The spectrum is sharp for the plane wave and broad for the wave packets. The position of the peak shifts for the
small wave packet due to the pseudo-Doppler effect.}
\label{fig5}}
\end{figure}
 \begin{figure}[t]
 \centering{
\includegraphics[angle=-90,scale=.4]{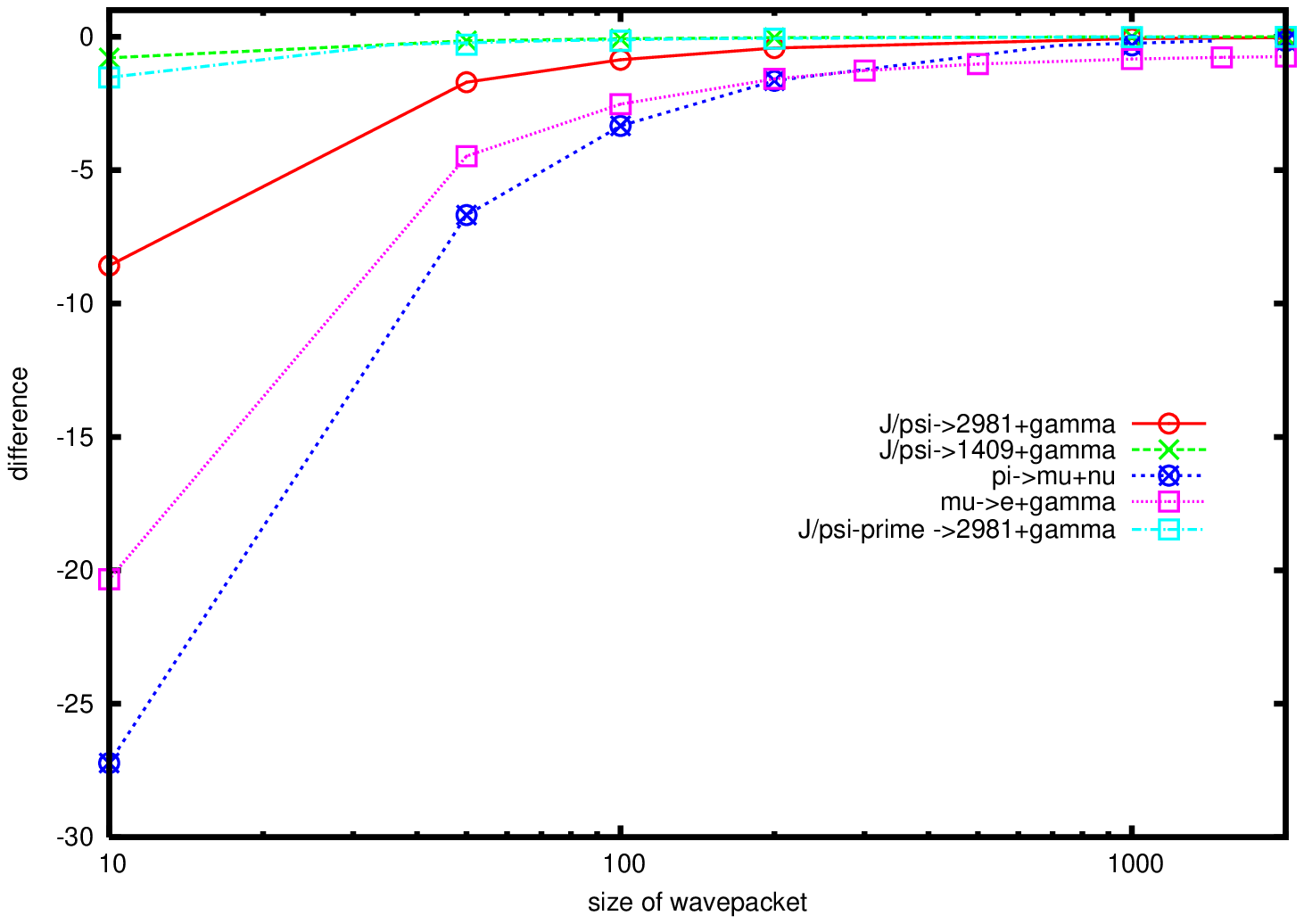}
\caption{The variance of final energy, $\Delta E = \sqrt{\langle(E_B+E_C)^2\rangle - \langle E_B+E_C\rangle^2}$, in particle
decays at rest measured at $T=\infty$ for $J/\Psi \to M(2981) + \gamma$ (solid red line), $J/\Psi \to \eta(1409) + \gamma$
(green dots), $\pi \to \mu + \nu$ (blue dots), and $mu \to e + \gamma$ (magenta dots). The horizontal axis shows the size
of wave packets in units of $\sigma m_\pi^2$ and the vertical axis shows the variance, $\Delta E$. The wave packets of 
another daughter and parent are $\infty$ and $\sigma_\text{parent}m_\pi^2 = 10000$. The curves are almost on one line.}
\label{fig6}
}
\end{figure}
\begin{figure}[t]
 \centering{
\includegraphics[angle=-90,scale=.4]{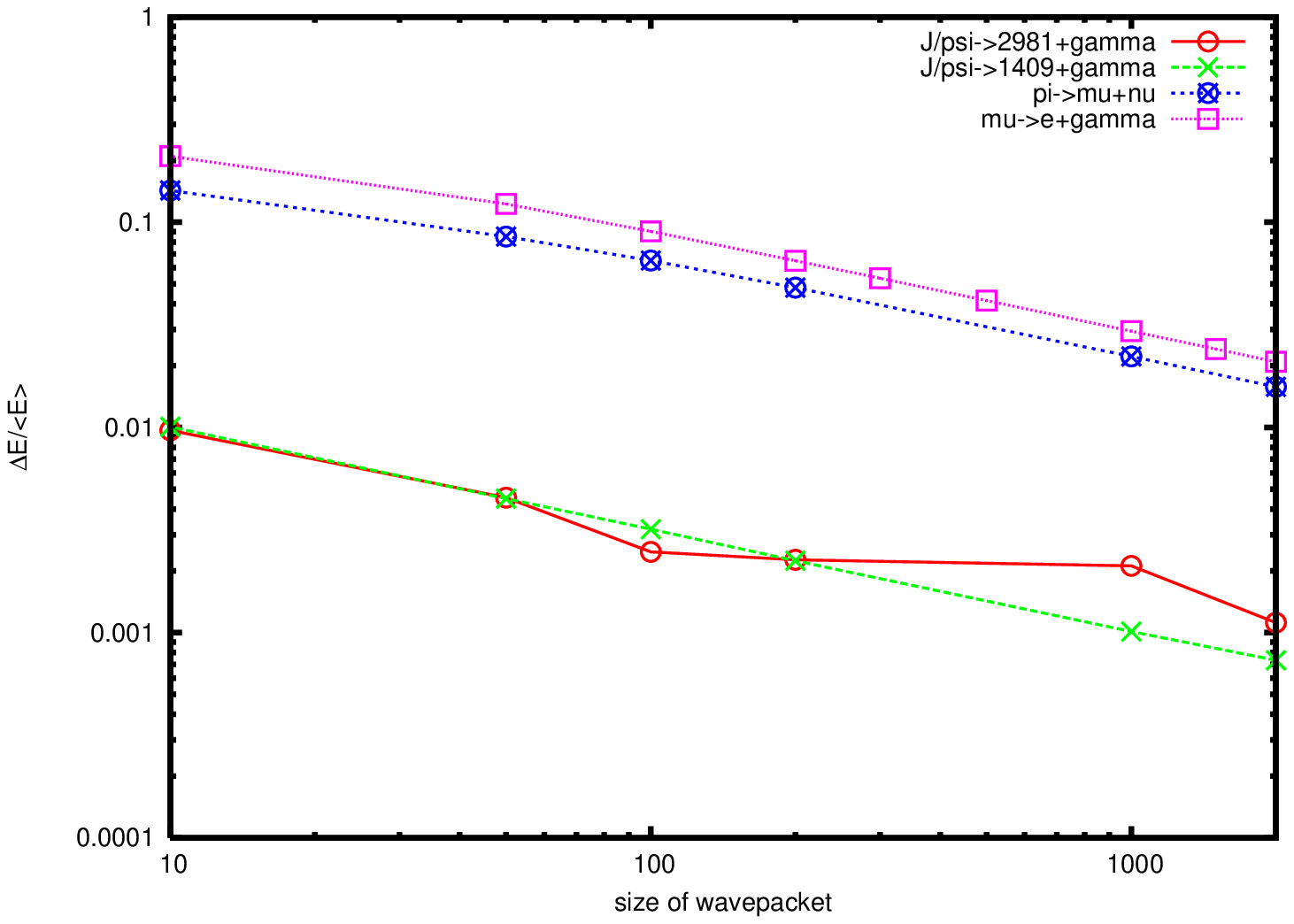}
\caption{The variance of energy of the final state over the average energy of the final state, $\Delta E/
\langle E_B + E_C\rangle$, in particle decays at rest measured at $T=\infty$ for $J/\Psi \to M(2981) + \gamma$
 (solid red line), $J/\Psi \to \eta(1409) + \gamma$ (green dots), $\pi \to \mu + \nu$ (blue dots), and $\mu \to e+\gamma$
(magenta dots). The horizontal axis shows the size of wave packets in units $\sigma m_\pi^2$ and the vertical axis shows the
ratio. The wave packets of another daughter and parent are $\infty$ and $\sigma_\text{parent}m_\pi^2
= 10000$. $\Delta E / \langle E_B + E_C\rangle$ is proportional to $(\sigma m_\pi^2)^{-1/2}$.}
\label{fig7}
}
\end{figure}

\section{Implications in particle decays}
The implications of the probabilities modified by the finite-size correction or the pseudo-Doppler effect are studied in
decay experiments. The former correction depends on the mass, energy, life-time, and time interval in a universal manner and
its  magnitude depends on the wave packet sizes and the internal structures. The effect of the internal wave function
on the light-0cone singularity is analyzed in Appendix C, and it is shown that, for hadrons, nucleus, and positronium, the
internal wave function does not modify the magnitude, but, for atoms,it does. If the initial wave packets are small or 
$T\gg \tau$, the overlap of the wave functions becomes negligible, whereas it becomes large if the initial wave packets
are large and $T\leq \tau$. In particular, the probability reveals various unusual behaviors for $\sigma_\text{initial}
\gg\sigma_\text{final}$. The finite-size effect is easily observed directly with measurements made with a detector located
at various $L$. Conversely, the latter correction becomes large for small wave packets, and the energy spectrum modified
due to the pseudo-Doppler effect is easily observed with the detector if the energy resolutions and other properties of
the detector are well understood. If these are unknown, the parameters of the detector are determined by comparing the 
theoretical values with the experimental data obtained from a standard sample. Calibration of the measuring apparatus
may be used for this purpose.

We study various decay processes and present magnitudes of the finite-size and pseudo-Doppler effects for the parents of 
plane waves and the detecting particles of wave packets. The life-times of the parents are included, and spin-independent
components are studied.

\subsection{Pseudo-Doppler effect}
The energy spectrum is modified by the pseudo-Doppler effect over a wide area and the distortion must be known not only for
 a precise analysis of experimental data but also to understand physical phenomena. A comparison of the rates computed for
plane waves and wave packets of various sizes is given for $J/\Psi \to M(2981) + \gamma$ in Fig. \ref{fig5}. The total
rates integrated over the final states agree but the energy spectra differ depending upon the wave packet size. The 
distributions and the shifts become wider and larger in smaller wave packets.

The broadening and shift of energies in other processes such as $J/\Psi \to M(2981) + \gamma$, $J/\Psi \to \eta(1409) + 
\gamma$, $\pi\to\mu+\nu$, and $\mu\to e+\gamma$ are compared. They are sensitive to the wave packet size, as shown in
Figs. \ref{fig6}-\ref{fig8}. Figure \ref{fig6} shows the variance of the final energy of the various processes. The curves
are almost on one line. Hence, the wave packet size can be found from the variance of the energy of the final state. In
Fig. \ref{fig7}, the normalized variance of the energies of the final states are presented. Those of heavy particles are
different from those of light particles. In Fig. \ref{fig8}, the average energies of the final states are compared with the 
initial energies. The deviations are clearly seen, and the total energies of the final states become larger than those of 
the initial states.

\subsubsection{Radiative transitions of atoms and positronium}
An atom is a bound state for a nucleus and electrons and is heavy. Radiative transitions of atoms from an excited state to
lower energy state, emitting a photon, are examples of two-body decays. Electrons bound to a nucleus have sizes of about
$10^{-10}$ m and energies of about 10 eV or less. The photon is detected through its interaction with matter in a detector.
Among the various reactions, The photo-electronic effect is the most important, where an electron is emitted from the 
photon interacting with electrons. We assume here that electron with which the photon interacts is a bound electron in the 
atom at rest. The size of its wave function is about $10^{-10}$ m. So $\sigma_\gamma$ has this size. For the initial 
particle $A$, $\sigma_A$ is either (1) about the same size, $10^{-10}$ m, for $A$ in matter, or (2) larger than $10^{-10}$
m, for $A$ in vacuum or a dilute gas. In exceptional situations, (3) $\sigma_A$ is smaller than $10^{-10}$ m. In experiments
of $\delta E \approx |\delta\vec{p}\,|$ in the following three cases of wave packet sizes
\begin{align*}
& \textbf{1}:\ \sigma_A\approx \sigma_\gamma,\\
& \textbf{2}:\ \sigma_A\gg \sigma_\gamma,\\
& \textbf{3}:\ \sigma_A\ll \sigma_\gamma,\\
\end{align*} 
the energy spectra are modified differently.

Positronium is a bound state of an electron and its anti-particle, a positron. Positronium of positive charge conjugation
decays to two gammas and that of negative charge conjugation decays to three gammas. The former is a second-order QED
process and the latter is a third-order QED process, and the phase spaces are also different. Hence their decay rates are 
very different.

\subsubsection{$J/\Psi$ radiative decay}
Photons produced in the decay 
\begin{align}
 J/\Psi \to M + \gamma\label{eq235}
\end{align}
have energies in GeV region and may receive the pseudo-Doppler effect. $J/\Psi$ is produced in the $e^+e^-$ reaction and
has a size determined by beam sizes, and the meson $M$ is detected by its decay products, which are stable hadrons such
as pions, kaons, and others. These charged particles have semi-microscopic sizes and $\sigma_M$ has the same size. 
$\sigma_\gamma$ is of the order of the nuclear size.

\begin{figure}[t]
 \centering{
\includegraphics[angle=-90,scale=.4]{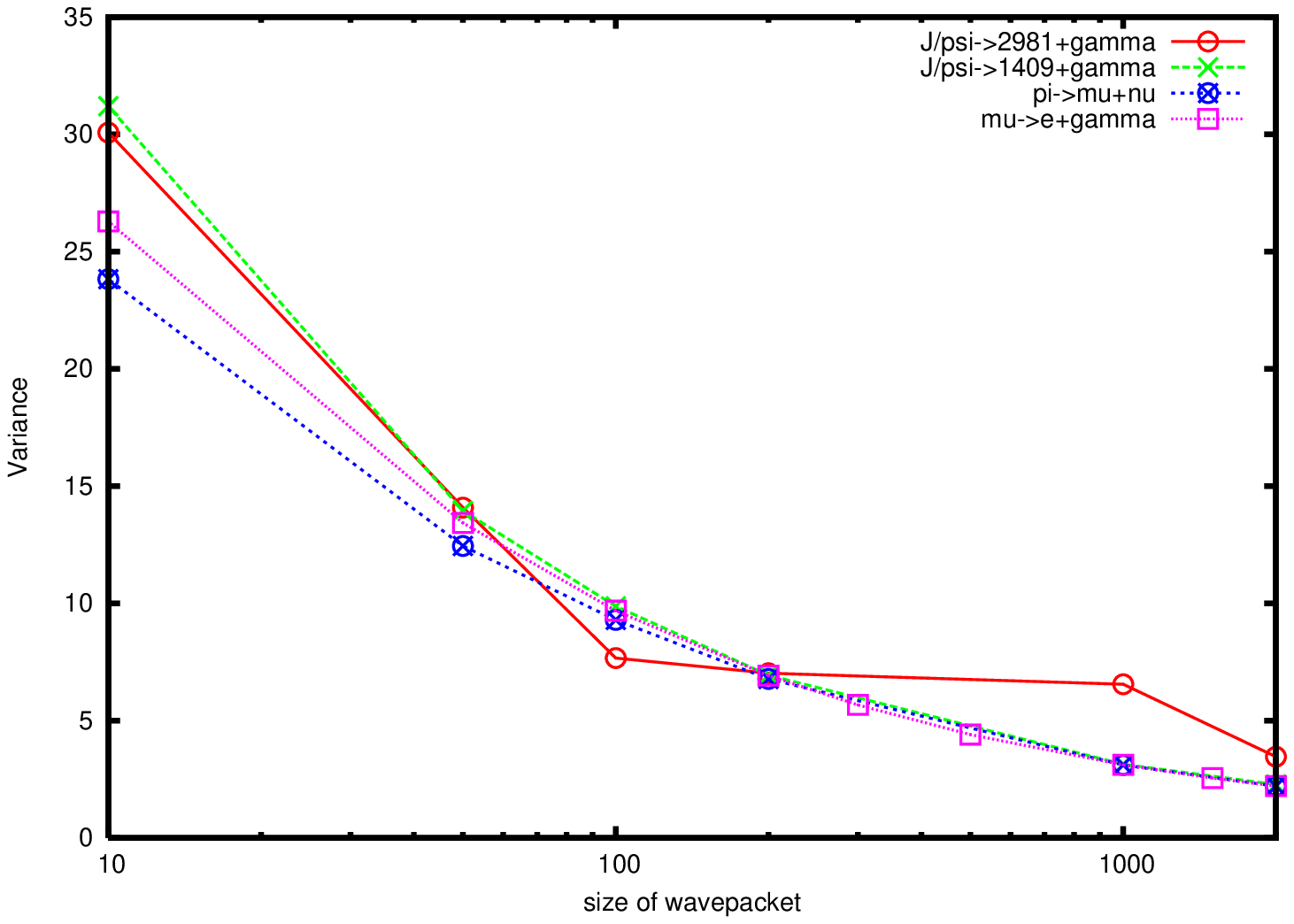}
\caption{The deviation of the average energy of the final state from the initial mass, $m_A - \langle E_B-E_C$
(MeV), due to the pseudo-Doppler effect is shown. There are finite deviations in various decays at rest measured at 
$T=\infty$ for $J/\Psi \to M(2981) + \gamma$ (solid red line), $J/\Psi \to \eta(1409) + \gamma$ (green dots),
$\pi \to \mu + \nu$ (blue dots), $\mu \to e + \gamma$ (magenta dots), and $\psi' \to M(2981) + \gamma$ (light-blue dots).
The horizontal axis shows the size of wave packets in units of $\sigma m_\pi^2$ and vertical axis shows the deviation of
the energies. The wave packets of another daughter and parent are $\infty$ and $\sigma_\text{parent}m_\pi^2 =10000$.}
\label{fig8}
}
\end{figure}

These processes are important for quantum chromodynamics (QCD) dynamics for the $M=c\bar{c}$ state (see section on 
charmonium in Ref. \cite{50}) or for the glue ball $M=\text{glueball}$ \cite{51,52} (see also particle data summary on
$\eta(1409)$ \cite{50}). The magnitudes of the corrections to the probabilities are not negligible as shown in Figs. 
\ref{fig6}--\ref{fig8}, The decay
\begin{align}
 \psi' \to M + \gamma\label{eq236}
\end{align}
is almost equivalent to Eq. \eqref{eq235}, except for the phase space and the fact that it has a smaller pseudo-Doppler
effect due to the large $\gamma$ energy. Experiments show a difference between Eqs. \eqref{eq236} and \eqref{eq235}
(see, e.g., Ref. \cite{49}).

\subsubsection{Two gamma decays of heavy scalar particles}
Positronium, neutral pions, charmonium P-states, and Higgs scalars decay to two photons. They are identified by the 
reconstructed photon's energies and momenta. Detection of photons is done with photo-electric or Compton effects in
low-energy regions and with $e^+e^-$ pair production at high energies. The bound electrons of atoms in the insulator
have a size of $10^{-10}$ m so $\sigma_\gamma$ for the former processes are of this size. The $e^+e^-$ pair is
produced by an electric field around a heavy nucleus, which is of nuclear size. Hence the wave packet size for the
latter process is approximately the nuclear size in high-energy regions. Hence, the wave packet sizes of $\gamma$ vary
over a wide range. They have short mean life-times and pseudo-Doppler effects may appear in 
\begin{align}
 M\to 2\gamma.\label{eq237}
\end{align}

\subsection{Finite-size correction}
\begin{figure}[t]
 \centering{
\includegraphics[angle=-90,scale=.4]{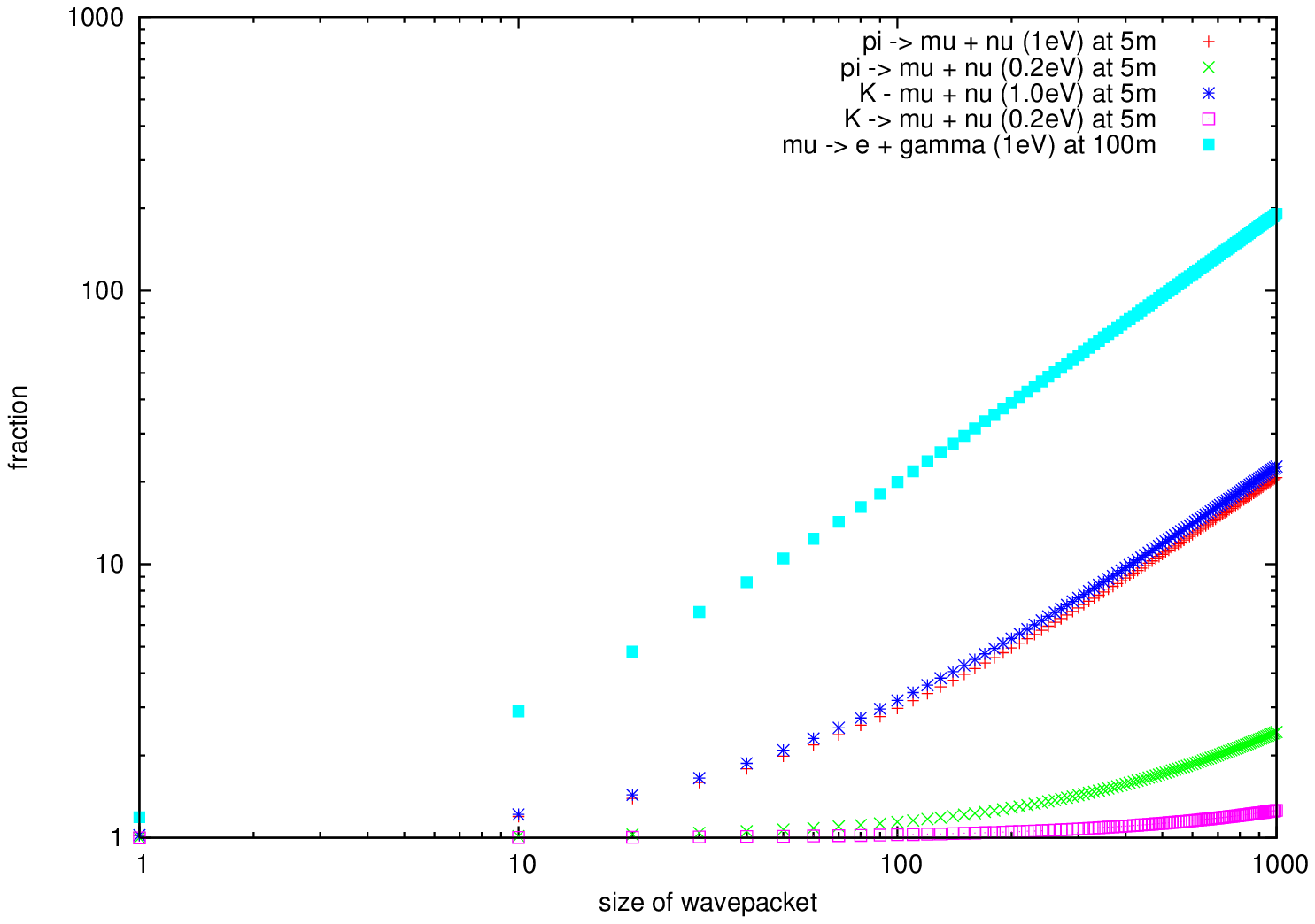}
\caption{The magnitudes of the probabilities in radiative and weak decays at rest measured at $T=0.33\times10^{-6}$ ($\mu$)
and $T=1.7\times 10^{-8}$ ($\pi$, $K$) (s) are shown. Parents and unobserved daughters are plane waves and observed 
particles  are wave packets. The horizontal axis shows the size of wave packets in units of $\sigma m_\pi^2$ and the 
vertical axis shows the enhancement factors at finite distance, i.e., the ratios of the sum of the normal and diffraction
components over the normal component. The decays of a pion to a muon and neutrino (red and green crosses), of a kaon to a
muon and neutrino (blue circles and magenta boxes), and of a muon to an electron and photon (light-blue boxes) are shown.
The masses are $m_\nu = 0.2$ and 1.0 eV/$c^2$, and $m_\gamma^\text{eff} = 1.0$ eV/$c^2$.}
\label{fig9}
}
\end{figure}

The finite-size correction becomes large in the situation where the wave functions of the initial and final states overlap
over a wide area. This is realized at $T\leq \tau$ and is important in slow decays of particles, such as weak decays and
some gamma decays. Figure \ref{fig9} shows the enhancement factors at finite distance, i.e., ratios of the total 
probabilities over the normal probabilities of the asymptotic region in various weak and radiative decays. For large wave
packets, the values become large. In this figure, the initial states are plane waves and the size of the wave packet for
the neutrino or photon is expressed in units of $1/m_\pi^2$ and is shown on the horizontal axis. The ratios
$\displaystyle{\frac{P_\text{normal} + P_\text{diffraction}}{P_\text{normal}}}$ are shown on the vertical axis. 
$P_\text{diffraction}$ is large in the region $\sigma m_\pi^2 > 10$. Thus, the finite-size corrections are non-negligible
and important.

In the region $T\gg\tau$, the finite-size corrections vanish, and the decay rates are expressed by the standard formula.
In this region, the number of parents decreases as $N_0e^{-T/\tau}$ and that of daughters becomes constant.
\subsubsection{Slow gamma decays of the nucleus}
Photons produced from radioactive nuclei are measured through their interactions with nuclei in targets with finite sizes. 
Hence $S[T]$ expressed by wave packets describes the amplitudes of the process,
\begin{align}
 N \to N' + \gamma. \label{eq238}
\end{align}
From Appendix C, the magnitude of the light-cone singularity and the diffraction component are almost equivalent to those
of point particles, and the total probabilities are modified by $P_\text{diffraction}$.

\begin{figure}[t]
 \centering{
\includegraphics[angle=-90,scale=.4]{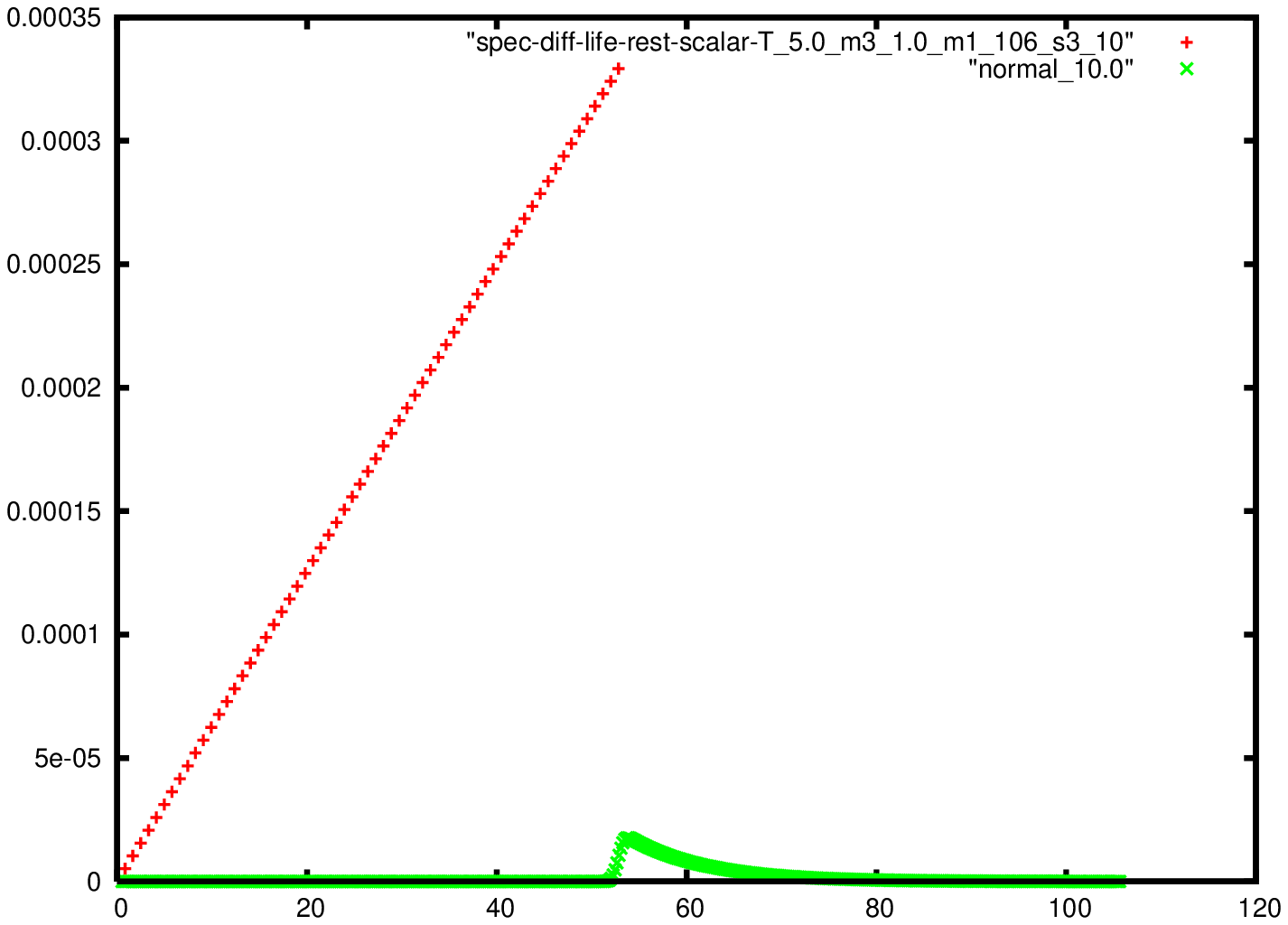}
\caption{The energy spectrum of photons in muon decays is shown at $T=1.7\times 10^{-8}$ s and $m_\gamma^\text{eff} = 1.0$
eV/$c^2$. The horizontal axis shows the energy of the photon in MeV at $\sigma_\gamma m_\pi^2 = 100$, $\sigma_\mu=
\sigma_e=\infty$ and the vertical axis shows the probability. The normal component (green) has a sharp peak around 54 MeV,
and the diffraction component (red) spreads over a wide region on the lower energy side and resembles the background.}
\label{fig10}
}
\end{figure}
\subsubsection{Muon decay to an electron and gamma}
A muon decays to an electron and a photon,
\begin{align}
 \mu \to e + \gamma,\label{eq239}
\end{align}
where the photon's energy is about 50 MeV, if the lepton number is violated. The lepton number violation has been observed
in neutrino oscillation phenomena but not in charged leptons. Precision measurements have been made and a new experiment
has started \cite{53}. Since the rate of this transition process is extremely small, it is important to know the corrections
due to the pseudo-Doppler effect and finite-size correction. Those for the plane wave muon at rest are studied here. From
Fig. \ref{fig6}, the average energy of the final state is larger than the initial energy.

Figure \ref{fig9} reveals the enhancement of the rates due to the diffraction for plane waves of the muon and electron and 
the wave packet for gamma. Figure \ref{fig10} shows the energy spectrum of $\gamma$ in the normal and diffraction components
for $\sigma_\gamma m_\pi^2 = 100$, $\sigma_\mu=\sigma_e=\infty$ in the muon decay of $\vec{p}_\mu=0$ at 
$T=1.7\times 10^{-8}$ s. The normal component has a sharp peak around $E_\gamma\approx 54$ MeV, whereas the diffraction
component spreads over a wide region. Moreover the latter is much larger than the former in these parameters. Thus the
corrections become important if the initial muon is a plane wave. The wave packet size of gamma can be determined from the 
spectrum at the higher-energy region of known process, and is used for the calculation of the diffraction component
of the present process.
\subsubsection{Weak decays}
\begin{figure}[t]
 \centering{
\includegraphics[angle=-90,scale=.4]{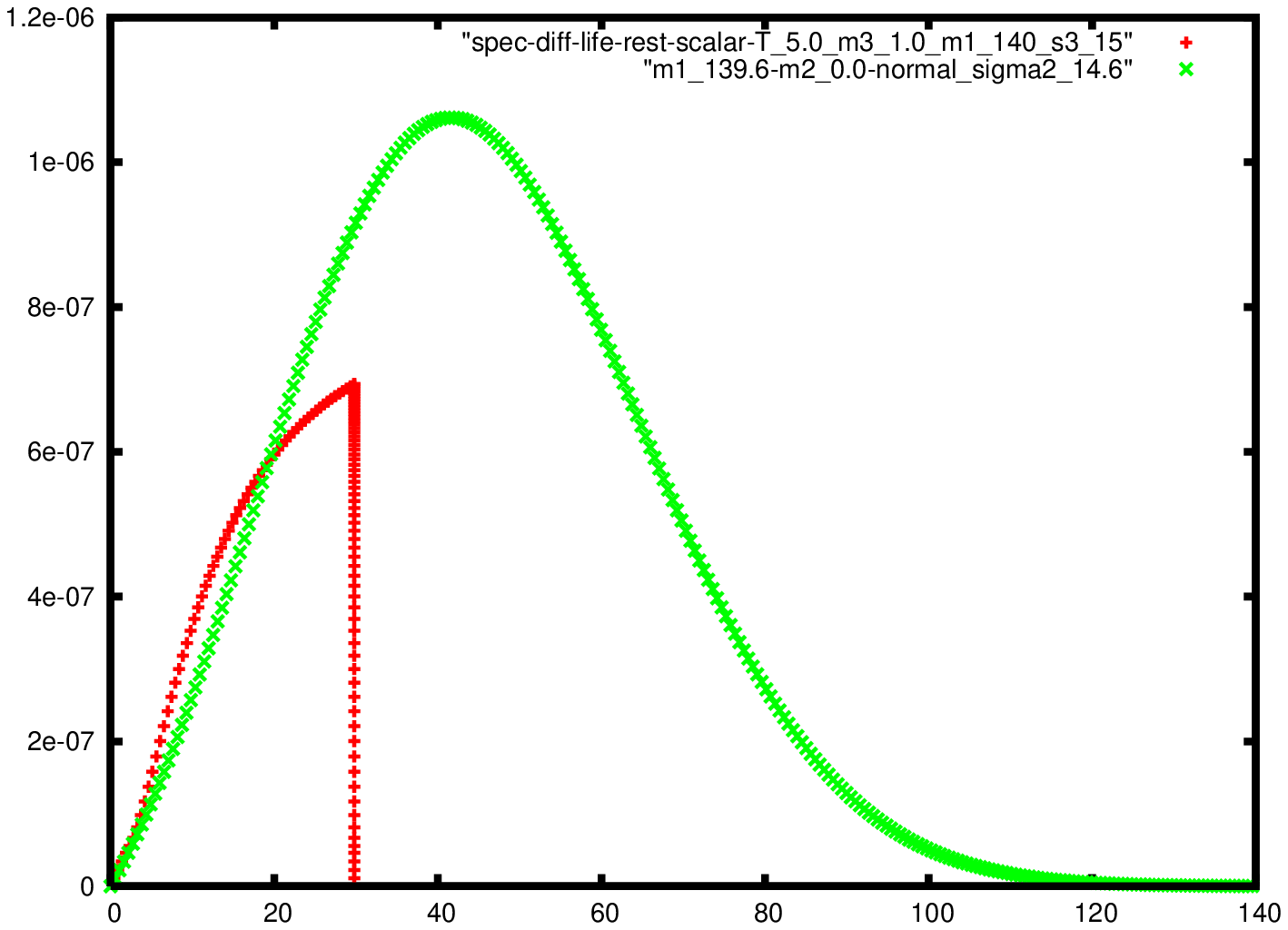}
\caption{The energy spectrum of neutrinos in pion decay at rest at $T=1.7\times 10^{-8}$, s is shown. The horizontal
axis shows the energy in MeV at $\sigma_\nu m_\pi^2 = 14.6$, which corresponds to ${}^{56}$Fe and the vertical axis
shows the probability. Wave packets of another daughter and parent are $\infty$. The neutrino mass is $m_\nu=1.0$ eV/$c^2$.
The normal component (green) has a broad peak, and the diffraction component (red) spread over the low-energy region.}
\label{fig11}
}
\end{figure}

A neutrino measured through its interaction with a nucleus has the same wave packet size as the nucleus. Hence the process 
of nucleus
\begin{align}
 A\to A' + \nu,\label{eq240}
\end{align}
is described by $S[T]$. Pion decay has been discussed in previous papers \cite{54,55,56}, and the neutrino's energy 
distribution is given in Fig. \ref{fig11}. From Appendix C, the magnitude of the light-cone singularity and the diffraction
component are about the same as point particles. The spectrum of the diffraction component that gives the finite-size 
correction is distributed in the low-energy region and that of the normal component is wide and has a peak. The peak is
slightly shifted from that of plane waves due to the pseudo-Doppler effect. From the shift and width of the normal 
component, the wave packet size can be determined and is used for the theoretical calculation of the diffraction component.

\subsection{Proton decay}
The proton is unstable and decays in grand unified theory (GUT). In $SU(5)$ GUT, a main decay mode is
\begin{align}
 \text{proton} \to \pi^0 + e^+.\label{eq241}
\end{align}
The initial proton is in matter in ground experiments and final states are detected through wave packets. For the large wave
functions of a proton, neutral pion, and positron, they overlap over a wide area. For small wave functions, they overlap
over small area. General cases with the symmetric wave packets
\begin{align}
 \sigma_p,\sigma_{\pi^0},\sigma_{e^+}\label{eq242}
\end{align}
of the four-dimensional momenta at positions
\begin{align}
 (p_p^0,\vec{p}_p;\vec{X}_p,T_p),\ (p_{\pi^0}^0,\vec{p}_{\pi^0};\vec{X}_{\pi^0},T_{\pi^0}),\ (p_{e^+}^0,\vec{p}_{e^+};
\vec{X}_{e^+},T_{e^+})\label{eq243}
\end{align}
are studied in the following. They are governed by an interaction Lagrangian
\begin{align}
 \mathscr{L}_{\text{int}} = g\bar{\phi}_p(x) e(x)\varphi(x).\label{eq244}
\end{align}
The transition amplitude is an integral over $(t,\vec{x})$:
\begin{align}
 \mathcal{M}(p\to \pi^0 + e^+) &=g\int dt\int d\vec{x}e^{-\frac{1}{2\sigma_S}(\vec{x}- \vec{x}_0)^2
-\frac{1}{2\sigma_t}(t-t_0)^2}e^{R+i\phi}\tilde{\mathcal{M}}\nonumber\\
&=g(2\pi\sigma_S)^\frac{3}{2}(2\pi\sigma_t)^\frac{1}{2}e^{R+i\phi}\tilde{\mathcal{M}},\label{eq245}
\end{align}
for finite values of $\sigma_S$ and $\sigma_t$. $\tilde{\mathcal{M}}$ includes the spinors. $\sigma_S$ and $\sigma_t$ are
given in the expressions
\begin{align}
 \frac{1}{\sigma_S} &= \frac{1}{\sigma_p}+\frac{1}{\sigma_{\pi^0}} + \frac{1}{\sigma_{e^+}},\label{eq246}\\
 \frac{1}{\sigma_t} &= \frac{v_p^2}{\sigma_p}+\frac{v_{\pi^0}^2}{\sigma_{\pi^0}}+\frac{v_{e^+}^2}{\sigma_{e^+}}
-\sigma_S\left(\frac{\vec{v}_p}{{\sigma_p}}+\frac{\vec{v}_{\pi^0}}{\sigma_{\pi^0}}+\frac{\vec{v}_{e^+}}{\sigma_{e^+}}
\right)^2\label{eq247}.
\end{align}
$t_0$ and $\vec{x}_0(t)$ are given in the form of Eq. \eqref{eq48} of an average velocity $\vec{v}_0$,
\begin{align}
 \vec{v}_0 = \sigma_S\left(\frac{\vec{v}_p}{\sigma_p}+\frac{\vec{v}_{\pi^0}}{\sigma_{\pi^0}}+\frac{\vec{v}_{e^+}}
{\sigma_{e^+}}\right).\label{eq248}
\end{align}
$R$ and $\phi$ in the exponent are obtained from Eqs. \eqref{eq50}, \eqref{eq51}, and \eqref{eq53} as
\begin{align}
 R &= R_\text{trajectory} + R_\text{momentum},\label{eq249}\\
 R_\text{trajectory} &= -\sum_j\frac{\tilde{vec{X}}_j^2}{2\sigma_j} + 2\sigma_S\left(\sum_j\frac{\tilde{\vec{X}}_j}
{2\sigma_j}
\right)^2 + 2\sigma_t\left(\sum_j\frac{(\vec{v}_0-\vec{v}_j)\cdot\tilde{\vec{X}}_j}{2\sigma_j}\right)^2,\nonumber\\
R_\text{momentum} &= -\frac{\sigma_t}{2}\left(
E_p(\tilde{\vec{p}}_p)-E_{\pi^0}(\tilde{\vec{p}}_{\pi^0})-E_{e^+}(\tilde{\vec{p}}_{e^+})\right)^2
-\frac{\sigma_S}{2}\left(\vec{p}_p-\vec{p}_{\pi^0} - \vec{p}_{e^+}\right),\nonumber
\end{align}
where
\begin{align}
 \tilde{\vec{p}}_p &= \vec{p}_p - \frac{\sigma_S}{\sigma_p}\left(\vec{p}_p - \vec{p}_{\pi^0} - \vec{p}_{e^+}\right),
\label{eq250}\\
 \tilde{\vec{p}}_{\pi^0} &= \vec{p}_{\pi^0} + \frac{\sigma_S}{\sigma_p}
\left(\vec{p}_p - \vec{p}_{\pi^0} - \vec{p}_{e^+}\right),\label{eq251}\\
 \tilde{\vec{p}}_{e^+} &= \vec{p}_{e^+} + \frac{\sigma_S}{\sigma_p}\left(\vec{p}_p - \vec{p}_{\pi^0} - \vec{p}_{e^+}\right),
\label{eq252}
\end{align}
and $\phi$ is a function of the momentum $\vec{p}_j$ and positions $\vec{X}_j$.

\subsubsection{Proton at rest}
The proton  in a solid is at rest and is expressed with a small wave packet. In the remaining proton system, the reduced 
momenta of Eq. \eqref{eq250} are
\begin{align}
 \tilde{\vec{p}}_p &= \frac{\sigma_S}{\sigma_p}\left(\vec{p}_{\pi^0} + \vec{p}_{e^+}\right),\label{eq253}\\
 \tilde{\vec{p}}_{\pi^0} &= \vec{p}_{\pi^0} - \frac{\sigma_S}{\sigma_p}\left(\vec{p}_{\pi^0} + \vec{p}_{e^+}\right),
\label{eq254}\\
 \tilde{\vec{p}}_{e^+} &= \vec{p}_{e^+} - \frac{\sigma_S}{\sigma_p}\left(\vec{p}_{\pi^0} + \vec{p}_{e^+}\right).
\label{eq255}
\end{align}
If the wave packet size of the positron is much larger than the others:
\begin{align}
 \sigma_{e^+}\gg\sigma_p,\ \sigma_{\pi^0},\ v_{e^+}\approx v_{\pi^0},\label{eq256}
\end{align}
then
\begin{align}
\sigma_S&= \frac{\sigma_p\sigma_{\pi^0}}{\sigma_p+\sigma_{\pi^0}},\label{eq257}\\
\frac{1}{\sigma_t}&=\frac{v_{\pi^0}^2}{\sigma_p+\sigma_{\pi^0}} + \frac{v_{e^+}}{\sigma_{e^+}}-\frac{2\sigma_p}
{\sigma_{e^+}(\sigma_p+\sigma_{\pi^0})}(\vec{v}_{e^+}\cdot\vec{v}_{\pi^0})\nonumber\\
\end{align}
and we have 
\begin{align}
 R_\text{momentum} &= -\frac{\sigma_t}{2}\left(E_p(\tilde{\vec{p}}_p) - E_{\pi^0}(\tilde{\vec{p}}_{\pi^0})
-E_{e^+}(\tilde{\vec{p}}_{e^+})\right)^2 - \frac{\sigma_S}{2}\left(\vec{p}_{\pi^0}-\vec{p}_{e^+}\right)^2,\label{eq259}\\
\tilde{\vec{p}}_p &= \frac{\sigma_S}{\sigma_p}\left(\vec{p}_{\pi^0} + \vec{p}_{e^+}\right),\ \tilde{\vec{p}}_{\pi^0}
= \vec{p}_{\pi^0},\ \tilde{\vec{p}}_{e^+} = \tilde{p}_{e^+}.\nonumber
\end{align}
Thus, in a region of large $\sigma_S$ and $\sigma_t$, the conservation law of the momentum and energy is the same as
that of plane waves, but at small $\sigma_S$ the momenta are spread over a wide region and the energy conservation law
is modified. For large $\sigma_t$, in the event of 
\begin{align}
 \vec{p}_{\pi^0} + \vec{p}_{e^+}\neq 0,\label{eq260}
\end{align}
the energy conservation tales the form
\begin{align}
 E_p(\tilde{\vec{p}}_p) - E_{\pi^0}(\vec{p}_{\pi^0}) - E_{e^+}\approx 0.\label{eq261}
\end{align}
$\tilde{\vec{p}}_p$ could be very different from $\vec{p}_p=0$, hence the modified conservation law derived from the 
pseudo-Doppler effect should be taken into account for the experimental analysis in this region.

Since $\sigma_t$ is finite, the decay probability is proportional to $T$ in the region $T\ll\tau_\text{proton}$, and 
the decay rate is constant over a wide range of $T$, despite the fact that the spectrum is distorted, where 
$\tau_\text{proton}$ is the average life-time. Thus a proton at rest decays at a constant rate even at small $T$,
and the proton decay experiment is feasible if the life-time is less than $10^{34}$--$10^{35}$ years.

\subsection{Other decay processes}
Three-body decays such as $\mu\to e+\bar{\nu} + \nu$, $n\to p+e+\bar{\nu}$ and others have light particles in the final 
states and are modified by the pseudo-Doppler effect and finite-size corrections. They will be presented in a separate paper
(K. Ishikawa and Y. Tobita, manuscript in preparation).

\subsection{Thermodynamics of small quantum particles}
When an excited state of a heavy atom of the large wave packet size makes a transition without changing the momentum and
emits a photon, it follows the modified energy conservation law. If the atoms are in thermodynamic equilibrium with a 
temperature T, the state of the energy $E$ follows the distribution
\begin{align}
 \rho(E,\beta) = N(-\beta E)
\end{align}
where $\beta$ is inversely proportional to the temperature, and $N(-\beta E)$ becomes the Planck distribution for bosons
and the Fermi-Dirac distribution for fermions.

In the situation where the wave packet size of the atoms is much larger than the wave packet size of a photon and the atoms
are bound together strongly, similar to the M\"{o}ssbauer effect, the photon distribution receives the pseudo-Doppler
effect and a M\"{o}ssbauer-like effect. Then the temperature of the photons begins to deviate from that of the atoms.

From Eq. \eqref{eq57}, in the situation
\begin{align}
 \sigma_\gamma \ll \sigma_A,\label{eq263}
\end{align}
the velocity $\vec{v}_0$ agrees with the velocity of the photon. The photon's energy is given by Eq. \eqref{eq208},
hence the energy distribution of the photon emitted from the atoms is
\begin{align}
 \rho(E_\gamma,\beta) = N\frac{1}{e^{\tilde{\beta}E_\gamma}-1},\ \tilde{\beta} = \frac{\beta}{\kappa}.\label{eq264}
\end{align}
Thus the effective temperature of the photon is $\kappa$ times that of the atoms.

\section{Summary and implications}
We have developed a theory for the diffraction induced by many-body interactions and computed the finite-size corrections
to the rates of slow transitions caused by electromagnetic and weak interactions. Large corrections to Fermi's golden rule,
Eq. \eqref{eq13}, were found in certain processes.

Fermi's golden rule is applicable to the rates in the particle zone where the initial and final states are completely
separate and their wave functions do not overlap. The rates are not subject to the $1/T$ correction. In the wave zone,
however, the $1/T$ corrections are found using the wave functions that satisfy the boundary conditions. Because they have
universal properties, they are observable in scattering experiments. The finite-size correction reveals the diffraction
pattern of single quantum interference. The intermediate-time region of particle decays when the parent and daughters
co-exist with a finite over lap of wave functions is an example of a wave zone. This state is a superposition of the
parent and daughters and has a finite expectation value of $H_\text{int}$. Thus the kinetic energy varies here;
the decay rates and scattering cross sections are different from their asymptotic values.

The finite-size corrections are inevitable consequences of the boundary conditions at $T$ and have a magnitude that depends
on the sizes of $\sigma_S$. The size of $\sigma_S$ can be controlled, and the finite-size corrections will be verified in
experiments. In particular, if the coherence length, $\hbar E/(m^2c^3)$, is a macroscopic size much larger than de Broglie
wave length $\hbar/|\vec{p}\,|$, they would be revealed in macroscopic scales. For neutrinos or photons of the effective mass
of the order (eV/$c^2$), $\hbar E/(m^2c^3)$ becomes a macroscopic size. Hence the finite-size correction may become visible
in a macroscopic distance.

If $\sigma_S = \text{finite}$, $\sigma_t = \text{finite}$, the wave functions overlap only in microscopic regions. Waves in 
experiments at macroscopic distance are in the particle zone, and the finite-size correction disappears. Nevertheless, the 
probability receives a pseudo-Doppler effect even in this region. The distortion of the energy distribution becomes stronger
with smaller wave functions and may become visible. It becomes drastic if both the M\"{o}ssbauer and pseudo-Doppler effects
are combined as in Eq. \eqref{eq208}. The final state of huge kinetic energy, much larger than the initial one, is formed with
a small probability. The conservation of kinetic energy is violated in each event but is satisfied for the average value over
the classical time interval. Despite the fact that $S[\infty]$ conserves the kinetic energy, the modified energy 
$\delta \tilde{E}$ conserves it approximately. In fact, these behaviors could have been considered as artifacts of the
detectors and absorbed in calibrations of the detector. The present results might help toward a complete understanding
in this direction.

In slow radiative and weak decays of particles $A'$ and $N'$ of plane waves,
\begin{align}
A' \to A + \gamma,\ N' \to N + \nu,\label{eq265}
\end{align}
the rates and energy distributions of $\gamma$ and $\nu$ in the asymptotic region,
\begin{align}
 \Gamma_\text{total} &=\Gamma^{(0)},\label{eq266}\\
\mathcal{P}_\text{total}(E) &= \mathcal{P}^{(0)}(E),\label{eq267}
\end{align}
are computed with plane waves and the $i\epsilon$ prescription, where $E$ is the energy of the observed particle. Our results
for $\gamma$ or $\nu$ measured at $L$ reveal the finite-size corrections and pseudo-Doppler effects and are expressed in the
form 
\begin{align}
 \Gamma_\text{total} &= \Gamma^{(0)}+\Gamma^{(\text{diff})}(L;\sigma),\label{eq268}\\
\mathcal{P}_\text{total} &= \mathcal{P}^{(n)}(E;\sigma) + \mathcal{P}^{(\text{diff})}(L,E;\sigma).\label{eq269}
\end{align}
The diffraction components have magnitudes summarized in Eqs. \eqref{eq154} and \eqref{eq178}, and are important in various
processes of leptons, hadrons, nuclei, and positronium, but negligible in ordinary atoms. At $L\gg L_0$,
\begin{align}
& \Gamma^{(\text{diff})} (L;\sigma) \to 0,\label{eq270}\\
&\mathcal{P}^{(\text{diff})}(L,E;\sigma) \to 0,\label{eq271}
\end{align}
where $L_0$ is the minimum value of the mean life-time and coherence length,
\begin{align}
 L_0 = \text{min}\left\{c\tau,\frac{\hbar E}{m^2c^3}\right\}.\label{eq272}
\end{align}
As $\sigma\to 0$,
\begin{align}
& |\mathcal{P}^{(n)}(E;\sigma) - \mathcal{P}^{(0)}(E)| \to \text{large},\label{eq273}\\
&\Gamma^{(\text{diff})}(L;\sigma),\ \mathcal{P}^{(\text{diff})}(L,E;\sigma) \to 0,\label{eq274}
\end{align}
whereas as $\sigma \to \text{large}$,
\begin{align}
& |\mathcal{P}^{(n)}(E;\sigma) - \mathcal{P}(E)| \to 0,\label{eq275}\\
& \Gamma^{(\text{diff})}(L;\sigma),\ \mathcal{P}^{(\text{diff})}(L,E;\sigma)\to \text{large}.\label{eq276}
\end{align}
The normal and diffraction components behave differently with $\sigma$, and Figs. \ref{fig6} and \ref{fig9} show them. From
Eqs. \eqref{eq273} and \eqref{eq275}, $\sigma$ as an effect on observables regardless of its magnitude and can be determined
experimentally from the energy spectrum, Figs. \ref{fig5}--\ref{fig8}, of the normal components $\mathcal{P}^{(n)}(E)$. The 
values computed by Fermi's golden rule have large corrections, Figs. \ref{fig5}--\ref{fig11}, and the theoretical values are
 not consistent with the experiments without corrections. Hence it would be easy to verify the finite-size corrections 
or pseudo-Doppler effect. The magnitudes of $\sigma$ depend on the processes and were not studied in the present paper. Those
values in neutrino experiments are given in Refs. \cite{54,55,56}. For outgoing states, they are determined by the sizes 
of microscopic objects with which they interact in the detector, and are of nuclear or atomic sizes fro neutrinos, $\gamma$-rays,
and charged particles, respectively. Thus they are around (5--10)$\times \frac{1}{m_\pi^2}$ or $(10^{-10})^2$, a few times 
(5--10)$\times \frac{1}{m_\pi^2}$, and $\geq (10^{-10}\text{m})^2$. For incoming states, they are determined by beam sizes,
or mean free paths. They are around 0.5--1.0 m, for the high-energy proton, pion, and muon. Those of other situations will be
given elsewhere (K. Ishikawa and Y. Tobita, manuscript in preparation). Because Fermi's golden rule is applied to many
problems in a wide area, it is important to confirm the corrections.

The decay rate of protons in matter is constant in $T\ll \tau_\text{proton}$, if the final states are measured due to 
the boundary conditions. This agrees with the standard one, and the proton decay will be detected if GUT is correct. Finally,
unusual luminescence and thermodynamic properties of quantum particles caused by overlap of wave functions will be verified 
in experiments.

Constituent particles such as molecules, atoms, nuclei, or elementary particles have small intrinsic sizes and are expressed
with wave functions of finite sizes in certain situations. Consequently, their reactions may be affected by finite-size 
corrections or pseudo-Doppler effects, even though no measurements are made. Physical systems may show unusual behavior.
such as the non-conservation of kinetic energy. Nevertheless, the average energy over a long period recovers the conservation
law. Hence the phenomena may appear in non-stationary and time-dependent processes \cite{42,43,44,45,46,47,48}. Macroscopic
quantum phenomena in this situation have been barely studied and will be discussed in subsequent works.

\begin{acknowledgments}
The present work was partially supported by a Grant-in-Aid for Scientific Research (Grant No. 24340043). The authors thank
Dr Kobayashi, Dr Maruyama, Dr Nakaya, Dr Nishikawa, and Dr Suekane for useful discussion on neutrino experiments. Dr Asai,
Dr Kinoshita, Dr Kobayashi, Dr Minowa, Dr Mori, Dr Nio, and Dr Yamada for useful discussions on interferences, and Dr Sato,
Dr Sorai, Dr Takesada, and Dr Watanabe for useful discussions on wide physical phenomena.
\end{acknowledgments}

\appendix

\section{Finite-size correction to Fermi's golden rule}\label{app1}
\subsection{Approximation with Dirac's delta function}
Integrals over the finite time interval
\begin{align}
 \int_0^T dte^{i\omega t}&=e^{i\omega T/2}\frac{\sin(\omega T/2)}{\omega/2},\label{eqa1}\\
\int_0^Tdt_1dt_2 e^{i\omega(t_1-t_2)} &=\left(\frac{\sin(\omega T/2)}{\omega/2}\right)^2\label{eqa2}
\end{align}
are normally approximated with 
\begin{align}
 \int_0^T dt e^{i\omega t} &= 2\pi\delta(\omega),\label{eqa3}\\
\int_0^Tdt_1dt_2e^{i\omega(t_1-t_2)} &= 2\pi T\delta(\omega)\label{eqa4}
\end{align}
for large $T$. They have been applied in computing the decay rate and cross section and are explained in most textbooks.

The finite-size correction to this formula depends on $\omega$. If $\omega$ is discrete,
\begin{align}
 \int_0^Tdte^{i\omega t}&=\begin{cases}
			  T;\ \omega=0,\\
			  \frac{2\sin(\omega T/2)}{\omega}e^{i\omega T/2};\ \omega\neq 0,
			 \end{cases}\label{eqa5}\\
\int_0^Tdt_1dt_2e^{i\omega(t_1-t_2)} &= \begin{cases}
					T^2;\ \omega = 0,\\
					\left(\frac{2\sin(\omega T/2)}{\omega}\right)^2;\ \omega\neq 0,
				       \end{cases}\label{eqa6}
\end{align}
and the averages over a finite-time interval $\delta T$ of $\delta T \omega\gg 1$ are
\begin{align}
 \text{Aver}\left[\int_0^Tdte^{i\omega t}\right]=\begin{cases}
						  T;\ \omega=0,\\
						  \frac{i}{\omega};\ \omega\neq 0,
						  \end{cases}\label{eqa7}\\
\text{Aver}\left[\int_0^Tdt_1dt_2e^{i\omega(t_1-t_2)}\right] = \begin{cases}
								T^2;\ \omega=0,\\
								\frac{2}{\omega^2};\ \omega\neq 0.
								\end{cases}\label{eqa8}
\end{align}
For the average probability at finite $T$, the correction is given by
b\begin{align}
  \frac{2}{\omega^2T^2}.\label{eqa9}
 \end{align}

\subsection{Correction by Taylor expansion}\label{appa2}
If $\omega$ is continuous, there exist states of infinitesimal energy differences. The correction becomes non-trivial and is
studied here. The following integral for $\omega_1 < 0 < \omega_2$,
\begin{align}
 I(\omega_1,\omega_2;T) = \int_{\omega_1}^{\omega_2}d\omega g(\omega)\left(\frac{\sin(\omega T/2)}{\omega}\right)^2,
\label{eqa10}
\end{align}
coincides with $2\pi Tg(0)$ if the second equation of Eq. \eqref{eqa3} is used. To find next-order terms in $1/T$, we expand
$g(\omega)$:
\begin{align}
 g(\omega) = g(0) + \sum_{l=1}^\infty \frac{g^{(l)}(0)}{l!}\omega^l,\label{eqa11},
\end{align}
and change the variable to $x=\omega T$:
\begin{align}
 I(\omega_1,\omega_2;T) &= T\int_{\omega_1 T}^{\omega_2T}dx\left(\frac{\sin(x/2)}{x}\right)^2g(x/T)\nonumber\\
&= \sum_l\frac{g^{(l)}(0)}{l!T^{l-1}}\int_{\omega_1T}^{\omega_2T}dx\left(\frac{\sin(x/2)}{x}\right)^2x^l.\label{eqa12}
\end{align}
The integrand in the above equation is finite at $x=0$ for l=0, but those for $l\geq 1$ vanish. At large $x$, $\sin^2(x/2)
\approx 1/2$. So we have 
\begin{align}
 I(\omega_1,\omega_2;T)&\approx 2\pi Tg(0) + \sum_{l\geq q}\frac{g^{(l)}(0)}{l!T^{l-1}}\int_{\omega_1T}^{\omega_2T}dx x^l
\frac{1}{x^2}\frac{1}{2}\nonumber\\
&=2\pi T g(0) + \frac{g^{(1)}(0)}{2}\log\left|\frac{\omega_2}{\omega_1}\right|+\sum_{l\geq2}\frac{g^{(l)}(0)}{2l!}
\frac{(\omega_2^{l-1})-\omega_1^{l-1}}{l-1}.\label{eqa13}
\end{align}
Choosing $\omega_1=-\omega_2$, we have
\begin{align}
 I(-\omega_2,\omega_2;T) = 2\pi Tg(0) + \sum_{l\geq 1}\frac{g^{(l+1)}(0)}{2(l+1)!}\frac{1}{l}(\omega_2^l(1 - (-1)^l)).
\label{eqa14}
\end{align}
The second term in the above equation is the $1/T$ correction. This correction depends on the cut-off frequencies and the
constant term, i.e., the $\omega_2^0$ term vanishes at the symmetric cut-off $\omega_1 = -\omega_2$. So $\displaystyle
{\lim_{\omega_2\to 0}I(-\omega_2,\omega_2;T)}$ agrees with $2\pi Tg(0)$. The finite-size correction is written in the form
\begin{align}
 I(-\omega_2,\omega_2;T) = 2\pi Tg(0)\left(1 + \frac{T_0}{T}\right),\label{eqa15}
\end{align}
where $T_0$ is roughly estimated to be the size of an atom as
\begin{align}
 T_0 = \frac{10^{-10}\text{ m}}{c} = 0.3\times 10^{-18}\text{ s}\label{eqa16}
\end{align}
in atomic physics. Hence, in an experiment of the size 1 m, $T = 1/c = 0.3\times 10^{-8}$ s, and the correction becomes
\begin{align}
 \frac{T_0}{T} = 10^{-10}.\label{eqa17}
\end{align} 
This value is negligible and the finite-size correction vanishes at a macroscopic distance.

\section{Level density and correlation function: quantum mechanics}\label{app2}
We summarize the probability at finite $T$, Eq. \eqref{eq11}, of systems of various level densities. When a level density
$\rho(E)$ is given, the number of states below $E$, $s$, satisfies
\begin{align}
 \frac{ds}{dE} = \rho(E). \label{eqb1}
\end{align}
The correlation function is expressed with $s$ in the form
\begin{align}
 g^{+}(t_1-t_2)&=\int_{E_0}^\infty dE\rho(E)e^{i(E-E_0)(t_1-t_2)}=\int_0^\infty dse^{i(E(e)-E_0)(t_1-t_2)},\label{eqb2}\\
g^{-}(t_1-t_2)&=\int_{E_m}^{E_0}dE\rho(E)e^{i(E-E_0)(t_1-t_2)}=\int_{s_m}^0dse^{i(E(s)-E_0)(t_1-t_2)}.\label{eqb3}
\end{align}
Using $s$, we have the integral over a finite interval of $s$ of Eq. \eqref{eqa3}: $C(T)=I(0,\infty;T),$
\begin{align}
 C(T) &=\int_0^Tdt_1dt_2g(t_1-t_2)=\int_{s_1}^{s_2}ds\int_0^Tdt_1dt_2e^{i\omega(s)(t_1-t_2)},\label{eqb4}\\
\omega(s) &=E(s) - E_0,\ g(t_1-g_2) = g^{(+)}(t_1-t_2) + g^{(-)}(t_1-t_2),\nonumber
\end{align}
which agrees with
\begin{align}
= \int_0^Tds\delta(\omega(s))=2\pi T\frac{1}{\omega'(s_0)}=2\pi T\rho(E_0),\label{eq}
\end{align}
if $\omega(s)=0$ has a simple root, $s_0$ in $s_1<s_0<s_2$, and $\omega'(s_0)$ is not too small. This is equivalent to
Eq. \eqref{eqa3}. $C(T)$ is also given in the expression
\begin{align}
 C(T)=T\int_{-T}^Td\xi g(\xi) +\int_{-T}^0d\xi \xi g(\xi) - \int_0^Td\xi xi g(\xi).\label{eqb6}
\end{align}

\subsection{Regular level density}
For the level densities that are regular at $E=E_0$,
\begin{align}
\rho(E) = c_0,\ \frac{1}{E-E_0-i\Gamma},\ e^{-c|E-E_0|^2},\label{eqb7} 
\end{align}
we have
\begin{table}[h]
\centering{
\caption{} 
\begin{tabular}[t]{|l|c|c|c|}
 $\rho(E)$& $s$& $g(t_1-t_2)$& $C(T)$\\\hline
$c_0$ & $c_0(E_-E_0)$& $2\pi c_0\delta(t_1-t_2)$&$2\pi c_0 T$\\\hline
$\displaystyle{\frac{1}{E-E_0-i\Gamma}}$ & $\displaystyle{\int_{E_0}^E\frac{dE'}{E'-E_0-i\Gamma}}$& $e^{-\Gamma(t_1-t_2)}$
&$\displaystyle{\frac{2T}{\Gamma}\left(1 - \frac{1-e^{\Gamma T}}{\Gamma T}\right)}$\\\hline
$e^{-c|E-E_0|}$ & $\displaystyle{\int_{E_0}^EdE'e^{-c|E'-E_0|^2}}$ & $\displaystyle{e^{-\frac{(t_1-t_2)^2}{2c}}}$
& $\displaystyle{\sqrt{\frac{c\pi}{2}}\left(T-\sqrt{\frac{8c}{\pi}}\right)}$
\end{tabular}
}\label{tableb1}
\end{table}
The correlation functions $g(t_1-t_2)$ in Table \ref{tableb1} are short range and Eq. \eqref{eqa3} is applicable. $C(T)$ are
proportional to $T$ and the corrections are proportional to $1/T$.

\subsection{Weakly singular level density}
For the level densities,
\begin{table}[h]
\centering{
\caption{} 
 \begin{tabular}[t]{|l|c|c|c|c|c|}
 $\rho(E)$ & $\rho(0)$ & $\rho'(0)$& $s$& $g(t_1-t_2)$&$C(T)$\\\hline
$e^{C|E-E_0}$ & 1 & $\pm$ & $\displaystyle{C\int_{E_0}^EdE'e^{-C|E'-E_0|}}$ & $\displaystyle{\frac{1}{C-i(t_1-t_2)}}$
& $\left(T\pi - 2C\log\displaystyle{\frac{T}{C}}\right)$\\\hline
$C(E-E_0)^p(0<p<1)$ & 0 & $\infty$ & $\displaystyle{\frac{C(E-E_0)^{p+1}}{p+1}}$ & $\displaystyle{\frac{\Gamma(\frac{p+1}{2})
2^{p-\frac{1}{2}}}{\Gamma(-\frac{p}{2}|t_1-t_2|^{p+1})}}$ & $T\times \infty$
 \end{tabular}
}
\end{table}
 The correlation functions $g(t_1-t_2)$ are long range Nd Eq. \eqref{eqa3} is applicable in the first case but not in the 
second. $C(T)$ are proportional to $T$; the correction is proportional to $(\log T)/T$ in the former case and the proportional 
constant diverges in the latter. In the latter case in particular, the level density satisfies $\rho(E_0)=0$, but the coefficient
$C(T)/T$ diverges.

\subsection{Singular level density}
For the singular level densities,
\begin{table}[h]
 \centering{
\caption{} 
\begin{tabular}[t]{|l|c|c|c|c|c|}
 $\rho(E)$& $\rho(0)$ & $\rho'(0)$& $s$& $g(t_1-t_2)$& $C(T)$\\\hline
$C\delta(E-E_0)$ & $\infty$ & $\infty$ & $C\theta(E-E_0)$ & $C$ & $CT^2$\\\hline
$\displaystyle{\frac{C}{|E-E_0|^p}};\ 0<p<1$ & $\infty$ & $\infty$ & $\displaystyle{\frac{C|E-E_0|^{1-p}}{1-p}}$ &
$C\displaystyle{\sqrt{\frac{2}{\pi}}\frac{p\pi\Gamma(1-p)}{2|t_1-t_2|^{1-p}}}$ & $\displaystyle{\frac{2T^{1+p}}{p}}$
\end{tabular}
}\label{tableb3}
\end{table}
Thus Eq. \eqref{eqa3} is not applicable in the level densities of Table \ref{tableb3}. The level density satisfies 
$\rho(E)=\infty$ and the decay rates $C(T)/T$ are proportional to a positive power of $T$. Thus this system should reveal the
unusual non-Markov property.

\subsection{Relativistic wave packet}
The spectrum density $\rho(\omega)$ for a two-body decay $A\to B+C$ of relativistic particles, Eq. \eqref{eq128}, integrated 
over $\vec{p}_C$ is
\begin{align}
 \rho(\omega) = \int\frac{d\vec{p}_C}{(2\pi)^3}e^{-\sigma_B(\vec{p}_A-\vec{p}_B-\vec{p}_C)^2}
\delta(\omega - E_A(\vec{p}_A) + E_C(\vec{p}_C) + E_B(\vec{p}_A-\vec{p}_C)).\label{eqb8}
\end{align}
$\rho(0)$ is finite because the root of $\omega=0$ exists and $\rho(\omega) satisfies$
\begin{align}
 \rho(\omega)=\begin{cases}
	       0,\ \omega>E_A,\\
	       e^{-\frac{\sigma_B}{4}\omega^2},\ \omega\to -\infty.
	      \end{cases}
\label{eqb9}
\end{align}
Due to the smooth asymptotic behavior of Eq. \eqref{eqb9}, the following integral converges even at finite $T$:
\begin{align}
 \int d\omega 
\left(
\frac{\sin(\omega T/2)}{\omega}
\right)^2 
\rho(\omega)
= T\left(2\rho(0)\int dx \left(\frac{\sin x}{x}\right)^2
+ \frac{1}{T}\eta\right),\label{eqb10}
\end{align}
when $\eta$ is given by 
\begin{align}
 \eta = \int d\omega\left(\frac{\sin(\omega T/2)}{\omega}\right)^2\tilde{\rho}(\omega)>0,\ \tilde{\rho}(\omega) = \rho(\omega)
-\rho(0),\label{eqb11}
\end{align}
and the positivity Eq. \eqref{eq35} is fulfilled.

\section{Light-cone singularity for general composite systems}\label{app3}
\subsection{Light-cone singularity and form factor}
Composite particles such as hadrons, nuclei, atoms, and molecules have internal structures and have modified light-cone
singularities. The form factor $F((p_A-p_C)^2)$ in
\begin{align}
 \langle C;\vec{p}_C|J(x)|A;\vec{p}_A\rangle = e^{-i(p_A-p_C)\cdot x}\Gamma F((p_A-p_C)^2)\label{eqc1}
\end{align}
depends on the Lorentz scalar $(p_A-p_C)^2$, where $J(x)$ is the source operator of $\gamma$, $\nu$, or others that are
detected. Here, Lorentz indices are ignored. The correlation function is
\begin{align}
 \Delta{A,C}(\delta x) &=\frac{1}{(2\pi)^3}\int\frac{d\vec{p}_C}{E(\vec{p}_C)}|F((p_A-p_C)^2)|^2G(p_A,p_C)e^{-i(p_A-p_C)\cdot
\delta x}\nonumber\\
&=G(p_A,-i\frac{\partial}{\partial\delta x}-p_A)\frac{1}{(2\pi)^3}\int\frac{d\vec{p}_C}{E(\vec{p}_C)}|F((p_A-p_C)^2)|^2
e^{-i(p_A-p_C)\cdot\delta x},\nonumber\\
&G(p_A,p_C) = \sum(\Gamma\Gamma^*)_{A,C},\label{eqc2}
\end{align}
and we have the light-cone singularity
\begin{align}
& \frac{1}{(2\pi)^3}\int\frac{d\vec{p}_C}{E(\vec{p}_C)}|F((p_A-p_C)^2)|^2e^{-i(p_A-p_C)\cdot\delta x}\nonumber\\
&\ =\frac{2}{i\pi(2\pi)^3}\int d^4q\text{Im}\left[
\frac{1}{q^2+m_A^2-m_C^2+2p_A\cdot q - i\epsilon}\right]|F(q^2)e^{-iq\cdot \delta x}\nonumber\\
&\ = |F(m_C^2-m_A^2)|^2\left(\frac{i}{2\pi}\delta(\lambda)\epsilon(\delta t) + \text{``others''}\right).\label{eqc3}
\end{align}
Thus we have
\begin{align}
 \Delta_{A,C} (\delta x) = G(p_A,-i\frac{\partial}{\partial\delta x}-p_A)|F(m_C^2-m_A^2)|^2
\left(\frac{i}{2\pi}\delta(\lambda)\epsilon(\delta t) + \text{``others''}\right).\nonumber
\end{align}
The magnitude is renormalized by the form factor $|F(m_C^2-m_A^2)|^2$ and the form is kept intact.

\subsection{Strength}
$|F(m_C^2-m_A^2)|^2$ of various systems, using the energy gap, $\Delta E$, and size $R$,
\begin{align}
 m_A = m_C + \delta E, \ F(q^2) = F(0)e^{-\frac{R}{2}\sqrt{-q^2},\label{eqc4}}
\end{align}
is written as 
\begin{align}
 F(m_C^2-m_A^2)^2 = F^2(0)e^{-R\sqrt{2m_C\delta E}},\label{eqc5}
\end{align}
and determined by $R\sqrt{2m_C\delta E}$.

The typical values for bound states composed of electrons, nucleons, quarks, and $\mu$, $eA$, $e^+e^-$, $NN$, $q\bar{q}$, $qqq$,
$\mu N$, $\mu^+ e^-$, and $e)\text{K-shell} A$ are
\begin{align}
 R=\begin{cases}
    \frac{\hbar c}{m_e c^2}\frac{1}{\alpha};\ (\text{atom: }eA),\\
    \frac{2\hbar c}{m_e c^2}\frac{1}{\alpha};\ (\text{positronium: } e^+e^-),\\
    \frac{\hbar}{m_\pi c};\ (\text{nucleus: }NN),\\
    \frac{\hbar}{m_q c};\ (\text{hadron: }q\bar{q},\ qqq),\\
    \frac{\hbar c}{m_\mu c^2}\frac{1}{\alpha};\ (\mu\text{ atom: }\mu A),\\
    \frac{\hbar c}{m_e c^2}\frac{1}{\alpha};\ (\mu\text{ atom: } \mu e),\\
\frac{\hbar c}{m_e c^2}\frac{1}{N_K\alpha};\ (\text{K-shell atom: }eA),
   \end{cases}
\Delta E = \begin{cases}
	    \frac{m_ec^2\alpha^2}{2};\ (\text{atom: }eA),\\
	    \frac{m_ec^e\alpha^2}{2};\ (\text{positronium: }e^+e^-),\\
	    \frac{m_\pi c^2}{100};\ (\text{nucleus: }NN),\\
	    m_qc^2;\ (\text{hadron; }q\bar{q},\ qqq),\\
	    \frac{m_\mu c^2\alpha^2}{2};\ (\mu\text{ atom: }\mu A),\\
	    \frac{m_2c^2\alpha^2}{2};\ (\mu \text{ atom: }\mu e),\\
	    \frac{m_ec^2\alpha^2}{2};\ (\text{K-shell atom: }eA),
	   \end{cases}
\label{eqc6}
\end{align}
and 
\begin{align}
 m_C =\begin{cases}
       m_N;\ (\text{atom: }eA),\\
       2m_2;\ (\text{positronium: }e^+e^-),\\
       Am_N;\ (\text{nucleus: }NN),
       Bm_qc^2;\ (\text{hadron: }q\bar{q},\ qqq),\\
       m_N;\ (\mu\text{ atom: }\mu A),\\
       m_\mu;\ (\mu\text{ atom: }\mu e),\\
       m_N;\ (\text{K-shell atom: }eA).
       \end{cases}
 \label{eqc7}
\end{align}
We have 
\begin{align}
 \frac{R}{\hbar}\sqrt{2m_C\Delta E} = 
\begin{cases}
\sqrt{\frac{m_N}{m_e}}\geq 50;\ (\text{atom: }eA),\\
 1;\ (\text{positronium: }e^+e^-),\\
 \sqrt{\frac{Am_N}{100m_\pi}}\leq 1;\ (\text{light nucleus}),\\
 \sqrt{\frac{Am_N}{100m_\pi}} \approx 3\ (A=100);\ (\text{nucleus: }),\\
 \sqrt{B}\approx 1;\ (\text{hadron: }q\bar{q},\ qqq),\\
 \sqrt{\frac{m_N}{m_\mu}} = 3;\ (\mu\text{ atom: }\mu A),\\
 \sqrt{\frac{m_\mu}{m_e}}=14;\ (\mu\text{ atom: }\mu e),\\
 \sqrt{\frac{m_N}{N_Km_e}}=10;\ (\text{K-shell atom: }eA),
\end{cases}
\label{eqc8}
\end{align}
and
\begin{align}
 F^2(m_C^2-m_A^2)=
\begin{cases}
F^2(0)\times O(1);\ \text{hadron, positronium, light nucleus,}\\
F^2(0)\times O(10^{-1});\ \mu N-\text{atom, heavy nucleus,}\\
F^2(0)\times O(10^{-5});\ \mu e-\text{atom, K-electron},\\
F^2(0)\times O(10^{-10});\ \text{atom.}
\end{cases}
\label{eqc9}
\end{align}
Thus the magnitude of the light-cone singularity is about 1 of $F^2(0)$ for positronium, light nuclei, and hadrons, $10^{-1}$
for $\mu N-$atoms and heavy nuclei, $10^{-5}$ for $\mu e-$atoms and K-electrons, and $10^{-10}$ for atoms. An atom is
 composed of a heavy nucleus and electrons and its size and energy gap are determined by the electron mass. Hence the large
ratio $\sqrt{m_N/m_e}$ determines the overlap of an atom $C$ with an atom $A$, and the strength of the light-cone singularity
becomes extremely weak, accordingly. For other particles composed of equal masses, $F(m_C^2-m_A^2)\approx F(0)$.


\begin{thebibliography}{}
 \bibitem{1}
H. Lehman, K. Symanzik, and W. Zimmermann, Nuovo Cim. \textbf{1}, 205 (1955).
\bibitem{2}
F. Low, Phys Rev. \textbf{97}, 1392 (1955).
\bibitem{3}
M. L. Goldberger and K. M. Watson, \textit{Collision Theory} (Wiley, New York, 1965).
\bibitem{4}
R. G. Newton, \textit{Scattering Theory of Waves and Particles} (Springer, New York, 1982).
\bibitem{5}
J. R. Taylor, \textit{Scattering Theory: The quantum Theory of Non-relativistic Collisions} (Dover, New York, 2006).
\bibitem{6}
M. Gell-mann, \textit{The Quark and The Jaguar: Adventures in the Simple and th Complex} (St. Martin's Griffin, London, 1995), 
ILL ed.
\bibitem{7}
R. G. Winter, Phys. Rev. \textbf{123}, 1503 (1961).
\bibitem{8}
M. L. Goldberger and K. M. Watson, Phys. Rev. \textbf{136}, 1472 (1964).
\bibitem{9}
H. Ekstein and A. J. F. Siegert, Ann. Phys. \textbf{68}, 509 (1971).
\bibitem{10}
K. J. F. Gaemers and T. D. Visser, Physica A \textbf{153}, 234 (1988).
\bibitem{11}
A. Einstein, B. Podlsky, and N. Rosen, Phys. Rev. \textbf{47}, 777 (1935).
\bibitem{12}
L. Schiff, \textit{Quantum Mechanics} (McGraw-Hill, New York, 1955), p. 197.
\bibitem{13}
L. D. Landau and E. M. Lifshitz, \textit{Quantum Mechanics} (Butterworth Heinemann, New York, 2003). p. 157.
\bibitem{14}
P. A. M. Dirac, Proc. R. Soc. London, Ser A \textbf{114}, 243 (1927).
\bibitem{15}
V. Weisskopf and E. Wigner, Z. Phys. \textbf{63}, 54 (1930).
\bibitem{16}
M. E. Peskin and D. V. Schroeder, \textit{An introduction to Quantum Field Theory} 
(Westview Press, Boulder, Colorado), Sect. 7.2, p. 222.
\bibitem{17}
S. Weinberg, \textit{The Quantum Theory of Fields I} (Cambridge University Press, Cambridgs, UK, 1995), p. 109.
\bibitem{18}
M. Srednicki, \textit{Quantum Filed Theory} (Cambridge University Press, Cambridge, UK, 2007), p. 37.
\bibitem{19}
N. N. Bogolubov, et al., \textit{General Principles of Quantum Field Theory} (Kluwer, Dordrecht, Netherlands, 1990).
\bibitem{20}
R. Haag, \textit{Local Quantum Physics} (Springer, Berlin, 1992).
\bibitem{21}
H. Araki, \textit{Mathematical Theory of Quantum Field} (Iwanami, Tokyo, 2002).
\bibitem{22}
K. Ishikawa and T. Shimomura, Prog. Theor. Phys. \textbf{114}, 1201 (2005).
\bibitem{23}
K. Ishikawa and Y. Tobita, arXiv:1106.4968 [hep-ph].
\bibitem{24}
K. Ishikawa and Y. Tobita, arXiv:1109.3105 [hep-ph].
\bibitem{25}
K. Ishikawa and Y. Tobita, Prog. Theor. Phys. \textbf{122}, 1111 (2009).
\bibitem{26}
K. Ishikawa and Y. Tobita, arXiv:0801.3124 [hep-ph].
\bibitem{27}
K. Ishikawa and Y. Tobita, AIP Conf. Proc. \textbf{1016}, 329 (2008).
\bibitem{28}
B. Kayser, Phys. Rev. D \textbf{24}, 110 (1981). 
\bibitem{29}
B. Kayser, Nucl. Phys. B \textbf{19}, 177 (1991).
\bibitem{30}
C. Giunti, C. W. Kim, and U. W. Lee, Phys. Rev. D \textbf{44}, 3635 (1991).
\bibitem{31}
S. Nussinov, Phys. Lett. B \textbf{63}, 201 (1976).
\bibitem{32}
K. Kiers, S. Nussinov, and N. Weiss, Phys. Rev. D \textbf{53}, 537 (1996).
\bibitem{33}
L. Stodolsky, Phys. Rev. D \textbf{58}, 036006 (1998).
\bibitem{34}
M. Beuthe, Phys. Rep. \textbf{375}, 105 (2003).
\bibitem{35}
H. J. Lipkin, Phys. Lett. B \textbf{642} 366 (2006).
\bibitem{36}
E. K. Akhmedov, J. High Energy Phys. \textbf{0709}, 116 (2007).
\bibitem{37}
A. Asahara, K. Ishikawa, T. Shimomura, and T. Yabuki, Prog. Theor. Phys. \textbf{113}, 385 (2005).
\bibitem{38}
T. Yabuki and K. Ishikawa, Prog. Theor. Phys. \textbf{108}, 347 (2002).
\bibitem{39}
A. H. Mueller, Phys. Rev. D \textbf{12}, 2963 (1970).
\bibitem{40}
K. Wilson, Proc. 5th Int. Symp. Electron and Photon Interactions at High Energies, p. 115 (1971).
\bibitem{41}
N. N. Bogoliubov and D. V. Shirkov, \textit{Introduction to the Theory of Quantized Fields} (Wiley, New York, 1976).
\bibitem{42}
D. J. Flannigan and K. S. Suslick, Nature \textbf{434}, 52 (2005).
\bibitem{43}
B. P. Barber and S. J. Putterman, Phys. Rev. Lett. \textbf{69}, 3839 (1992).
\bibitem{44}
C. G. Camara et al., Nature \textbf{455}, 1089 (2008).
\bibitem{45}
H. Tsuchiya et al., Phys. Rev. Lett. \textbf{99}, 165002 (2007).
\bibitem{46}
A. Chilingarian et al., Phys. Rev. D \textbf{82}, 043009 (2010).
\bibitem{47}
T. Torii et al., Geophys. Res. Lett. \textbf{38}, L24801 (2011).
\bibitem{48}
M. Tavani et al., Phys. Rev. Lett. \textbf{106}, 018501 (2011).
\bibitem{49}
J. E. Gaiser, SLAC-255, UC-34D (1982).
\bibitem{50}
J. Beringer et al. [Particle Data Group], Phys. Rev. D \textbf{86}, 010001 (2012).
\bibitem{51}
K. Ishikawa, Phys. Rev. Lett. \textbf{46}, 978 (1981).
\bibitem{52}
M. Chanozitz, Phys. Rev. Lett. \textbf{46}, 981 (1981).
\bibitem{53}
J. Adam et al., arXiv:1303.0754 [hep-ex].
\bibitem{54}
K. Ishikawa and Y. Tobita, arXiv:1206.2593 [hep-ph].
\bibitem{55}
K. Ishikawa and Y. Tobita, arXiv:1209.5585 [hep-ph].
\bibitem{56}
K. Ishikawa and Y. Tobita, arXiv:1209.5586 [hep-ph].




\end{thebibliography}
\end{document}